\documentclass[preprint,prd,aps,showpacs,showkeys,nofootinbib]{revtex4}
\usepackage{graphicx}
\begin{document}

%\preprint{hep-ph/0412147}

\title{$B\rightarrow X_{_S}l^+l^-$ in the minimal gauged $(B-L)$ supersymmetry}

\author{Tai-Fu Feng\footnote{email:fengtf@hbu.edu.cn}, Jin-Lei Yang,
Hai-Bin Zhang\footnote{email:hbzhang@hbu.edu.cn}, Shu-Min Zhao, Rong-Fei Zhu}

\affiliation{Department of Physics, Hebei University, Baoding, 071002, China}

\begin{abstract}
Applying the effective Hamilton for $b\rightarrow sl^+l^-,\;(l=e,\;\mu)$ in the framework
of minimal supersymmetric extension of the standard model with local $B-L$ gauge symmetry,
we investigate branching ratios and forward-backward asymmetries of rare decay
$B\rightarrow X_{_S}l^+l^-$ in low and high $q^2$ regions, respectively.
In addition we also study the $CP$ asymmetries depending on new $CP$ phases
from soft breaking terms in low and high $q^2$ regions.
With some assumptions on parameter space of the model, the numerical analyses of the supersymmetric
contributions to the branching ratios, forward-backward and $CP$ asymmetries of
$B\rightarrow X_{_S}l^+l^-$ are presented in low and high $q^2$ regions, respectively.
\end{abstract}


\maketitle

\section{Introduction\label{sec1}}
\indent\indent
The study on rare $B$ decays can detect new physics beyond the standard model (SM) since the
theoretical evaluations of corresponding observations are not
seriously affected by the uncertainties originating from unperturbative QCD  effects.
So far the Charmless semileptonic $B$ decays are studied extensively in the SM,
the authors of Ref.\cite{Buchalla93,Misiak93,Buras95,Buchalla96} analyze
the QCD corrections to the branching ratios of rare $B$-decay, and the authors
of Ref.\cite{Deshpande89,Lim89,Donnell91} present corrections from $c\bar{c}$ resonances to
the branching ratio of $B\rightarrow X_{_S}l^+l^-$. In order to obtain
the QCD evolution effects precisely in rare $B$-decay,
the relevant two-loop QCD anomalous dimension matrix (ADM)
for all the flavour-changing four-quark dimension-six operators is given in Ref.\cite{Buras00}.

Considering those corrections mentioned above, one obtains the theoretical evaluations in the
SM as\cite{SMBtoXsll,Hurth09,Huber06}
\begin{eqnarray}
%%%%%%%%%%%%%%%%%%%%%%%%%%%%%%%%%%%%%%%%%%%%%%%%%%%
&&BR(B\rightarrow X_{_s}l^+l^-)_{_{q^2\in[1,6]{\rm GeV}^2}}^{SM}=(1.59\pm0.11)\times10^{-6},\;
\nonumber\\
&&BR(B\rightarrow X_{_s}l^+l^-)_{_{q^2\in[14.4,25]{\rm GeV}^2}}^{SM}=(2.3\pm0.7)\times10^{-7}\;,
(l=e,\;\mu)\;.
%%%%%%%%%%%%%%%%%%%%%%%%%%%%%%%%%%%%%%%%%%%%%%%%%%%
\label{SM-BR-BtoXsll}
\end{eqnarray}
Only the contributions to the branching ratios in the low $q^2$ region with
$1\;{\rm GeV}^2\le q^2\le6\;{\rm GeV}^2$
and the high $q^2$ region with $14.4\;{\rm GeV}^2\le q^2\le25\;{\rm GeV}^2$ are
evaluated respectively.
Here $q^2=(p_{_{l^+}}+p_{_{l^-}})^2$ denotes the dilepton invariant mass squared.
Averaging available experimental data from BaBar\cite{BABAR04} and Belle\cite{BELLE05},
we obtain the experimental averages for the branching ratios in two regions as follows\cite{HFAG}
\begin{eqnarray}
%%%%%%%%%%%%%%%%%%%%%%%%%%%%%%%%%%%%%%%%%%%%%%%%%%%
&&BR(B\rightarrow X_{_s}l^+l^-)_{_{q^2\in[1,6]{\rm GeV}^2}}^{exp}=(1.63\pm0.50)\times10^{-6},\;
\nonumber\\
&&BR(B\rightarrow X_{_s}l^+l^-)_{_{q^2\in[14.4,25]{\rm GeV}^2}}^{exp}=(4.3\pm1.2)\times10^{-7}
\;,(l=e,\;\mu)\;.
%%%%%%%%%%%%%%%%%%%%%%%%%%%%%%%%%%%%%%%%%%%%%%%%%%%
\label{EXP-BR-BtoXsll}
\end{eqnarray}
Obviously the SM theoretical evaluations on those branching ratios
coincide with the experimental data in three standard deviations,
and the coming precise measurements on the rare $B$-decay processes will set more strong
constraints on the new physics beyond SM. The main purpose of investigation of $B$-decays
is to search for traces of new physics and determine its parameter space.
Besides the branching ratios, the forward-backward asymmetries
in the process $B\rightarrow X_{_S}l^+l^-\;(l=e,\;\mu)$ are
the physics quantities to detect new physics beyond the SM.
The updated experimental data from Belle Collaboration\cite{BELLE16} are
\begin{eqnarray}
%%%%%%%%%%%%%%%%%%%%%%%%%%%%%%%%%%%%%%%%%%%%%%%%%%%
&&A_{_{FB}}(B\rightarrow X_{_s}l^+l^-)\Big|_{q^2\in[1,\;6]\;{\rm GeV}^2}^{exp}
=0.30\pm0.24\pm0.03\;,
\nonumber\\
&&A_{_{FB}}(B\rightarrow X_{_s}l^+l^-)\Big|_{q^2\in[14.4,\;25]\;{\rm GeV}^2}^{exp}
=0.28\pm0.15\pm0.01\;,
%%%%%%%%%%%%%%%%%%%%%%%%%%%%%%%%%%%%%%%%%%%%%%%%%%%
\label{exp-AFB0}
\end{eqnarray}
where the first uncertainty is statistical and the second uncertainty
is systematic. The corresponding SM predictions are given in Ref.\cite{SMBtoXsllAFB} as:
\begin{eqnarray}
%%%%%%%%%%%%%%%%%%%%%%%%%%%%%%%%%%%%%%%%%%%%%%%%%%%
&&A_{_{FB}}(B\rightarrow X_{_s}l^+l^-)\Big|_{q^2\in[1,\;6]\;{\rm GeV}^2}^{SM}
=-0.07\pm0.04\;,
\nonumber\\
&&A_{_{FB}}(B\rightarrow X_{_s}l^+l^-)\Big|_{q^2\in[14.4,\;25]\;{\rm GeV}^2}^{SM}
=0.40\pm0.04\;.
%%%%%%%%%%%%%%%%%%%%%%%%%%%%%%%%%%%%%%%%%%%%%%%%%%%
\label{SM-AFB0}
\end{eqnarray}
For the present experimental uncertainty is large, we cannot yet apply the experimental
data of the physics quantity to test the SM precisely. Nevertheless, the future experimental
data will constrain the parameter space of new physics strongly with accumulating of data sample.

Meanwhile the SM evaluation of the $CP$ asymmetry in
the entire $q^2$ region is very small \cite{SMBtoXsllACP}, the theoretical prediction
on the $CP$ asymmetry may be enhanced significantly \cite{NPBtoXsllACP}. The updated
experimental data on the $CP$ asymmetry from Babar Collaboration are\cite{BABAR14}
\begin{eqnarray}
%%%%%%%%%%%%%%%%%%%%%%%%%%%%%%%%%%%%%%%%%%%%%%%%%%%
&&A_{_{CP}}(B\rightarrow X_{_s}l^+l^-)\Big|_{q^2\in[1,\;6]\;{\rm GeV}^2}^{exp}
=-0.06\pm0.22\pm0.01\;,
\nonumber\\
&&A_{_{CP}}(B\rightarrow X_{_s}l^+l^-)\Big|_{q^2\in[14.4,\;25]\;{\rm GeV}^2}^{exp}
=0.19^{+0.18}_{-0.17}\pm0.01\;.
%%%%%%%%%%%%%%%%%%%%%%%%%%%%%%%%%%%%%%%%%%%%%%%%%%%
\label{exp-ACP0}
\end{eqnarray}

In supersymmetric extensions of the SM, new sources of flavor and CP
violations may appear in those soft breaking terms\cite{Gabrielli}.
Actually the analyses of constraints on extensions of the SM are extensively
discussed in literature. The calculation of the branching ratio of inclusive decay $B\rightarrow X_s\gamma$ is
presented by authors of \cite{Ciuchini1,Ciafaloni,Borzumati1} in the two-Higgs doublet model (2HDM).
The supersymmetric effect on $B\rightarrow X_s\gamma$ is discussed in
\cite{Bertolini1,Barbieri,Borzumati2,Causse,Prelovsek}
and the next-to-leading order (NLO) QCD corrections are given in \cite{Ciuchini2}.
The transition $b\rightarrow s\gamma\gamma$ in the supersymmetric extensions of
the SM is computed in \cite{Bertolini2}. The hadronic $B$ decays\cite{Cottingham} and
CP-violation in those processes\cite{Barenboim} have been discussed also.
The authors of \cite{Hewett} have discussed possibility to observe supersymmetric effects
in rare decays $B\rightarrow X_s\gamma$ and $B\rightarrow X_se^+e^-$ at the $B$-factory.
Studies on decays $B\rightarrow (K,K^*)\mu^+\mu^-$ in the SM and
supersymmetric models have been carried out in \cite{Ali},
a relevant review can be found in Ref.\cite{Masiero1} also. The theoretical analyses on
oscillations of $B_0-\bar{B}_0$ ($K_0-\bar{K}_0$)
have been done in the SM and 2HDM. In supersymmetric extensions of the SM,
the calculation involving the
gluino contributions should be re-studied carefully for gluino has a nonzero mass. At the NLO approximation,
the QCD corrections to the $B_0-\bar{B}_0$ mixing in the supersymmetry extensions have been discussed also.
Adopting the mass-insertion approximation (MIA) method the
authors of \cite{Ciuchini3,Contion,Krauss} estimate QCD corrections
to the $B_0-\bar{B}_0$ mixing, and later we have re-derived the formulation
by including the contribution of gluinos \cite{Feng1}.

The discovery of Higgs on the Large Hadron Collider (LHC) implies that the searching of
particle spectrum predicted by the SM is finished now \cite{CMS,ATLAS}.
One main target of particle physics is testing the SM precisely and searching for
the new physics (NP) beyond it. The updated bound from ATLAS collaboration on the gluino mass
is $m_{_{\tilde g}}\ge1460\;{\rm GeV}$, and the bound on
the mass of scalar top is $m_{_{\tilde t}}\ge780\;{\rm GeV}$\cite{ATLAS2016}.
Additionally the LHCb experiment can measure the quantities
of exclusive hadronic,semi-leptonic, and leptonic $B$ and $B_s$ decays at a high sensitivity\cite{LHCB}.
The measurements on inclusive rare $B$ decay and decays with neutrino
final states will be performed also in two next generation $B$ factories
in near future \cite{B-factory1,B-factory2}.

The discrete symmetry R-parity in supersymmetry is defined through
$R=(-1)^{3(B-L)+2S}$, where $B$, $L$ and $S$ are baryon number,
lepton number and spin respectively for a concerned
field\cite{R-parity}. In the minimal supersymmetric extension of SM (MSSM) with
local $U(1)_{B-L}$ symmetry, R-parity is spontaneously
broken when left- and right-handed sneutrinos acquire nonzero
vacuum expectation values (VEVs)\cite{Perez1,Perez2,Perez3,Perez4}.
Meanwhile, the nonzero VEVs of left- and right-handed sneutrinos induce the mixing
between neutralinos (charginos) and neutrinos (charged leptons). Furthermore,
the MSSM with  local $U(1)_{B-L}$ symmetry naturally predicates two sterile neutrinos
\cite{Perez6,Senjanovic1,Chang-Feng}, which are favored by the Big-bang
nucleosynthesis (BBN) in cosmology\cite{Hamann}. In other words, there are
exotic sources to mediate flavor changing neutral current processes (FCNC)
in this model.

Here we investigate some interesting physical quantities in
the FCNC processes $B\rightarrow X_sl^+l^-,\;(l=e,\;\mu)$ in the
MSSM with local $U(1)_{B-L}$ symmetry, our presentation is organized as follows. In section \ref{sec2}, we
briefly summarize the main ingredients of the MSSM with
local $U(1)_{B-L}$ symmetry, then present effective Hamilton for  $b\rightarrow sl^+l^-$
in section \ref{sec3}. The formulae of decay widths,
forward-backward asymmetries, $CP$ asymmetries at hadronic scale are given in \ref{sec4}, respectively.
The numerical analyses are given in section \ref{sec5}, and our
conclusions are summarized in section \ref{sec6} finally.

\section{The MSSM with local $U(1)_{B-L}$ symmetry\label{sec2}}
\indent\indent
When $U(1)_{B-L}$ is a local gauge symmetry, one can
enlarge the local gauge group of the SM to $SU(3)_{_C}\otimes
SU(2)_{_L}\otimes U(1)_{_Y}\otimes U(1)_{_{(B-L)}}$. In the
model proposed in Ref.\cite{Perez1,Perez2,Perez3,Perez4}, the
exotic superfields are three generation right-handed neutrinos
$\hat{N}_{_i}^c\sim(1,\;1,\;0,\;1)$. Meanwhile, quantum numbers of
the matter chiral superfields for quarks and leptons are given by
\begin{eqnarray}
&&\hat{Q}_{_I}=\left(\begin{array}{l}\hat{U}_{_I}\\
\hat{D}_{_I}\end{array}\right)\sim(3,\;2,\;{1\over3},\;{1\over3})\;,\;\;
\hat{L}_{_I}=\left(\begin{array}{l}\hat{\nu}_{_I}\\
\hat{E}_{_I}\end{array}\right)\sim(1,\;2,\;-1,\;-1)\;,
\nonumber\\
&&\hat{U}_{_I}^c\sim(3,\;1,\;-{4\over3},\;-{1\over3})\;,\;\;
\hat{D}_{_I}^c\sim(3,\;1,\;{2\over3},\;-{1\over3})\;,\;\;
\hat{E}_{_I}^c\sim(1,\;1,\;2,\;1)\;,
\label{quantum-number1}
\end{eqnarray}
with $I=1,\;2,\;3$ denoting the index of generation. In addition,
the quantum numbers of two Higgs doublets
are assigned as
\begin{eqnarray}
&&\hat{H}_{_u}=\left(\begin{array}{l}\hat{H}_{_u}^+\\
\hat{H}_{_u}^0\end{array}\right)\sim(1,\;2,\;1,\;0)\;,\;\;
\hat{H}_{_d}=\left(\begin{array}{l}\hat{H}_{_d}^0\\
\hat{H}_{_d}^-\end{array}\right)\sim(1,\;2,\;-1,\;0)\;.
\label{quantum-number2}
\end{eqnarray}
The superpotential of the MSSM with local $U(1)_{B-L}$ symmetry is written as
\begin{eqnarray}
&&{\cal W}={\cal W}_{_{MSSM}}+{\cal W}_{_{(B-L)}}^{(1)}\;.
\label{superpotential1}
\end{eqnarray}
Here ${\cal W}_{_{MSSM}}$ is superpotential of the MSSM, and
\begin{eqnarray}
&&{\cal W}_{_{(B-L)}}^{(1)}=\Big(Y_{_N}\Big)_{_{IJ}}\hat{H}_{_u}^Ti\sigma_2\hat{L}_{_I}\hat{N}_{_J}^c\;.
\label{superpotential-BL}
\end{eqnarray}
Correspondingly, the soft breaking terms for the MSSM with local $U(1)_{B-L}$
symmetry are generally given as
\begin{eqnarray}
&&{\cal L}_{_{soft}}={\cal L}_{_{soft}}^{MSSM}+{\cal L}_{_{soft}}^{(1)}\;.
\label{soft-breaking1}
\end{eqnarray}
Here ${\cal L}_{_{soft}}^{MSSM}$ is soft breaking terms of the MSSM, and
\begin{eqnarray}
&&{\cal L}_{_{soft}}^{(1)}=
-(m_{_{\tilde{N}^c}}^2)_{_{IJ}}\tilde{N}_{_I}^{c*}\tilde{N}_{_J}^c
-\Big(m_{_{BL}}\lambda_{_{BL}}\lambda_{_{BL}}+m_{_{1BL}}\lambda_{_{1}}\lambda_{_{BL}}+h.c.\Big)
\nonumber\\
&&\hspace{1.5cm}
+\Big\{\Big(A_{_N}\Big)_{_{IJ}}H_{_u}^Ti\sigma_2\tilde{L}_{_I}\tilde{N}_{_J}^c+h.c.\Big\}\;,
\label{soft-breaking3}
\end{eqnarray}
with $\lambda_{_{BL}}$ denoting the gaugino of $U(1)_{_{B-L}}$, $m_{_{1BL}}$ denoting
the mixing mass parameter between the $U(1)_{_Y}$ gaugino and $U(1)_{_{B-L}}$ gaugino, respectively.
After the $SU(2)_L$ doublets $H_{_u},\;H_{_d},\;\tilde{L}_{_I}$ and $SU(2)_L$ singlets $\tilde{N}_{_I}^c$
acquire the nonzero VEVs,
\begin{eqnarray}
&&H_{_u}=\left(\begin{array}{c}H_{_u}^+\\{1\over\sqrt{2}}\Big(\upsilon_{_u}+H_{_u}^0+iP_{_u}\Big)\end{array}\right)\;,
\nonumber\\
&&H_{_d}=\left(\begin{array}{c}{1\over\sqrt{2}}\Big(\upsilon_{_d}+H_{_d}^0+iP_{_d}\Big)\\H_{_d}^-\end{array}\right)\;,
\nonumber\\
&&\tilde{L}_{_I}=\left(\begin{array}{c}{1\over\sqrt{2}}\Big(\upsilon_{_{L_I}}+\tilde{\nu}_{_{L_I}}+iP_{_{L_I}}\Big)\\
\tilde{L}_{_I}^-\end{array}\right)\;,\nonumber\\
&&\tilde{N}_{_I}^c={1\over\sqrt{2}}\Big(\upsilon_{_{N_I}}+\tilde{\nu}_{_{R_I}}+iP_{_{N_I}}\Big)\;,
\label{VEVs}
\end{eqnarray}
the R-parity is broken spontaneously, and the local gauge symmetry $SU(2)_{_L}\otimes U(1)_{_Y}\otimes U(1)_{_{(B-L)}}$
is broken down to the electromagnetic symmetry $U(1)_{_e}$, and the neutral and
charged gauge bosons acquire the nonzero masses as
\begin{eqnarray}
&&m_{_{\rm Z}}^2={1\over4}(g_1^2+g_2^2)\upsilon_{_{\rm EW}}^2\;,\nonumber\\
&&m_{_{\rm W}}^2={1\over4}g_2^2\upsilon_{_{\rm EW}}^2\;,\nonumber\\
&&m_{_{Z_{BL}}}^2=g_{_{BL}}^2\Big(\upsilon_{_N}^2+\upsilon_{_{\rm EW}}^2-\upsilon_{_{\rm SM}}^2\Big)\;.
\label{gauge-masses}
\end{eqnarray}
Where $\upsilon_{_{\rm SM}}^2=\upsilon_{_u}^2+\upsilon_{_d}^2$,
$\upsilon_{_{\rm EW}}^2=\upsilon_{_u}^2+\upsilon_{_d}^2+\sum\limits_{\alpha=1}^3\upsilon_{_{L_\alpha}}^2$,
$\upsilon_{_N}^2=\sum\limits_{\alpha=1}^3\upsilon_{_{N_\alpha}}^2$,
and $g_2,\;g_1$, $g_{_{BL}}$ denote the gauge couplings of
$SU(2)_{_L},\;\;U(1)_{_Y}$ and $U(1)_{_{(B-L)}}$, respectively.

%%%%%%%%%%%%%%%%%%%%%%%%%%%%%%%%%%Begin 1st Revision%%%%%%%%%%%%%%%%%%%%%%%%%%%%%%%%%%%%
To satisfy present electroweak precision observations we assume
the mass of neutral $U(1)_{_{(B-L)}}$ gauge boson $m_{_{Z_{BL}}}>1\;{\rm TeV}$ which implies
$\upsilon_{_N}>1\;{\rm TeV}$ when $g_{_{BL}}<1$, then we derive $\max((Y_{_{N}})_{ij})\le10^{-6}$
and $\max(\upsilon_{_{L_I}})\le10^{-3}\;{\rm GeV}$\cite{Perez4}
to explain experimental data on neutrino oscillation.
Considering the minimization conditions at one-loop level,
we formulate the $3\times3$ mass-squared matrix for right-handed sneutrinos as
\begin{eqnarray}
&&m_{_{{\tilde N}^c}}^2\simeq\left(\begin{array}{ccc}
\Lambda_{_{\tilde{N}_1^c}}^2-\Lambda_{_{BL}}^2\;,\;\;&0\;,\;\;
&-{\upsilon_{_{N_1}}\over\upsilon_{_{N_3}}}\Lambda_{_{\tilde{N}_1^c}}^2\\
0\;,\;\;&\Lambda_{_{\tilde{N}_2^c}}^2-\Lambda_{_{BL}}^2\;,\;\;
&-{\upsilon_{_{N_2}}\over\upsilon_{_{N_3}}}\Lambda_{_{\tilde{N}_2^c}}^2\\
-{\upsilon_{_{N_1}}\over\upsilon_{_{N_3}}}\Lambda_{_{\tilde{N}_1^c}}^2\;,\;\;
&-{\upsilon_{_{N_2}}\over\upsilon_{_{N_3}}}\Lambda_{_{\tilde{N}_2^c}}^2\;,\;\;
&{\upsilon_{_{N_1}}^2\Lambda_{_{\tilde{N}_1^c}}^2+\upsilon_{_{N_2}}^2\Lambda_{_{\tilde{N}_2^c}}^2
\over\upsilon_{_{N_3}}^2}-\Lambda_{_{BL}}^2
\end{array}\right)
\label{R-sneutrino-mass}
\end{eqnarray}
with $\Lambda_{_{BL}}^2=m_{_{Z_{BL}}}^2/2+\Delta T_{_{\tilde N}}$.
Where $\Delta T_{_{\tilde N}}$ denotes one-loop radiative corrections
to the mass matrix of right-handed sneutrinos from top, bottom, tau and their supersymmetric
partners \cite{Chang-Feng}.

\section{Effective Hamilton for  $b\rightarrow sl^+l^-,\;(l=e,\;\mu,\;\tau)$\label{sec3}}
\indent\indent
The transition $b\rightarrow sl^+l^-$ is attributed to the effective Hamilton at hadronic scale
\begin{eqnarray}
&&{\cal H}_{_{eff}}=-{4G_{_F}\over\sqrt{2}}V_{_{tb}}V_{_{ts}}^*\Big[C_{_1}{\cal O}_{_1}^c
+C_{_2}{\cal O}_{_2}^c+\sum\limits_{i=3}^6C_{_i}{\cal O}_{_i}
+\sum\limits_{i=7}^{10}C_{_i}{\cal O}_{_i}
+\sum\limits_{i=S,P}C_{_i}{\cal O}_{_i}\Big]\;,
\label{effective-Hamilton}
\end{eqnarray}
where ${\cal O}_{_i}\;(i=1,\;2,\;\cdots,\;10,\;S,\;P)$
are defined as \cite{Buras1}
\begin{eqnarray}
&&{\cal O}_{_1}^u=(\bar{s}_{_L}\gamma_\mu T^au_{_L})(\bar{u}_{_L}\gamma^\mu T^ab_{_L})\;,\;\;
{\cal O}_{_2}^u=(\bar{s}_{_L}\gamma_\mu u_{_L})(\bar{u}_{_L}\gamma^\mu b_{_L})\;,
\nonumber\\
&&{\cal O}_{_3}=(\bar{s}_{_L}\gamma_\mu b_{_L})\sum\limits_q(\bar{q}\gamma^\mu q)\;,\;\;
{\cal O}_{_4}=(\bar{s}_{_L}\gamma_\mu T^ab_{_L})\sum\limits_q(\bar{q}\gamma^\mu T^aq)\;,
\nonumber\\
&&{\cal O}_{_5}=(\bar{s}_{_L}\gamma_\mu\gamma_\nu\gamma_\rho b_{_L})\sum\limits_q(\bar{q}\gamma^\mu
\gamma^\nu\gamma^\rho q)\;,\;\;
{\cal O}_{_6}=(\bar{s}_{_L}\gamma_\mu\gamma_\nu\gamma_\rho T^ab_{_L})\sum\limits_q(\bar{q}\gamma^\mu
\gamma^\nu\gamma^\rho T^aq)\;,
\nonumber\\
&&{\cal O}_{_7}={e\over g_{_s}^2}m_{_b}(\bar{s}_{_L}\sigma_{_{\mu\nu}}b_{_R})F^{\mu\nu}\;,\;\;
{\cal O}_{_8}={1\over g_{_s}}m_{_b}(\bar{s}_{_L}\sigma_{_{\mu\nu}}T^ab_{_R})G^{a,\mu\nu}\;,\;\;
\nonumber\\
&&{\cal O}_{_9}={e^2\over g_{_s}^2}(\bar{s}_{_L}\gamma_\mu b_{_L})\bar{l}\gamma^\mu l\;,\;\;
{\cal O}_{_{10}}={e^2\over g_{_s}^2}(\bar{s}_{_L}\gamma_\mu b_{_L})\bar{l}\gamma^\mu\gamma_5 l\;,\;\;
\nonumber\\
&&{\cal O}_{_S}={e^2\over16\pi^2}m_{_b}(\bar{s}_{_L}b_{_R})\bar{l}l\;,\;\;
{\cal O}_{_P}={e^2\over16\pi^2}m_{_b}(\bar{s}_{_L}b_{_R})\bar{l}\gamma_5l\;.
\label{operators}
\end{eqnarray}
Here we adopt the MIA to get the corrections to relevant Wilson
coefficients from supersymmetric particles because the updated experiment data
push the energy scale of supersymmetry far above the electroweak energy scale.
Furthermore we can formulate the relevant Wilson coefficients depending
on the flavor changing sources from scalar quark sectors transparently by the MIA method.
At the electroweak energy scale $\mu_{_{\rm EW}}$, the Wilson coefficients
$C_{_{7,NP}}(\mu_{_{\rm EW}}),\;C_{_{8,NP}}(\mu_{_{\rm EW}})$
from the new physics beyond SM can be found in
our previous work \cite{Feng2016}, other relevant Wilson coefficients
are split as following
\begin{eqnarray}
%%%%%%%%%%%%%%%%%%%%%%%%%%%%%%%%%%%%%%%%%%%%%%%%%%%
&&C_{_{9,NP}}(\mu_{_{\rm EW}})=C_{_{9,NP}}^\gamma(\mu_{_{\rm EW}})
+C_{_{9,NP}}^Z(\mu_{_{\rm EW}})+C_{_{9,NP}}^{Z_{_{BL}}}(\mu_{_{\rm EW}})
+C_{_{9,NP}}^{box}(\mu_{_{\rm EW}})\;,\nonumber\\
%%%%%%%%%%%%%%%%%%%%%%%%%%%%%%%%%%%%%%%%%%%%%%%%%%%
&&C_{_{10,NP}}(\mu_{_{\rm EW}})=C_{_{10,NP}}^\gamma(\mu_{_{\rm EW}})
+C_{_{10,NP}}^Z(\mu_{_{\rm EW}})+C_{_{10,NP}}^{Z_{_{BL}}}(\mu_{_{\rm EW}})
+C_{_{10,NP}}^{box}(\mu_{_{\rm EW}})\;,\nonumber\\
%%%%%%%%%%%%%%%%%%%%%%%%%%%%%%%%%%%%%%%%%%%%%%%%%%%
&&C_{_{S,NP}}(\mu_{_{\rm EW}})=\sum\limits_{i=1}^2
C_{_{S,NP}}^{H_i^0}(\mu_{_{\rm EW}})+C_{_{S,NP}}^{box}(\mu_{_{\rm EW}})
\;,\nonumber\\
%%%%%%%%%%%%%%%%%%%%%%%%%%%%%%%%%%%%%%%%%%%%%%%%%%%
&&C_{_{P,NP}}(\mu_{_{\rm EW}})=C_{_{P,NP}}^{A^0}(\mu_{_{\rm EW}})
+C_{_{P,NP}}^{box}(\mu_{_{\rm EW}})\;,
%%%%%%%%%%%%%%%%%%%%%%%%%%%%%%%%%%%%%%%%%%%%%%%%%%%
\label{Wilson-Coefficients1}
\end{eqnarray}
where the superscripts $\gamma,\;Z,\;Z_{_{BL}},\;H_i^0,\;A^0,\;box$
denote that new physics corrections to relevant Wilson coefficients
originate from $\gamma-,\;Z-,\;Z_{_{BL}}-,\;H_i^0-,\;A^0-$ penguins and box diagrams,
respectively. In order to formulate the corrections transparently, we
split those pieces further as
\begin{eqnarray}
%%%%%%%%%%%%%%%%%%%%%%%%%%%%%%%%%%%%%%%%%%%%%%%%%%%
&&C_{_{9,NP}}^\gamma(\mu_{_{\rm EW}})=C_{_{9,H^\pm}}^\gamma(\mu_{_{\rm EW}})
+C_{_{9,\chi^\pm}}^\gamma(\mu_{_{\rm EW}})+C_{_{9,\chi^0}}^\gamma(\mu_{_{\rm EW}})
+C_{_{9,\tilde{g}}}^\gamma(\mu_{_{\rm EW}})+C_{_{9,{\tilde Z}_{_{BL}}}}^\gamma(\mu_{_{\rm EW}})
\;,\nonumber\\
%%%%%%%%%%%%%%%%%%%%%%%%%%%%%%%%%%%%%%%%%%%%%%%%%%%
&&C_{_{9,NP}}^Z(\mu_{_{\rm EW}})=(4s_{_{\rm W}}^2-1)C_{_{10,NP}}^Z(\mu_{_{\rm EW}})
\;,\nonumber\\
%%%%%%%%%%%%%%%%%%%%%%%%%%%%%%%%%%%%%%%%%%%%%%%%%%%
&&C_{_{10,NP}}^Z(\mu_{_{\rm EW}})=C_{_{10,H^\pm}}^Z(\mu_{_{\rm EW}})
+C_{_{10,\chi^\pm}}^Z(\mu_{_{\rm EW}})+C_{_{10,\chi^0}}^Z(\mu_{_{\rm EW}})
+C_{_{10,\tilde{g}}}^Z(\mu_{_{\rm EW}})+C_{_{10,{\tilde Z}_{_{BL}}}}^Z(\mu_{_{\rm EW}})
\;,\nonumber\\
%%%%%%%%%%%%%%%%%%%%%%%%%%%%%%%%%%%%%%%%%%%%%%%%%%%
&&C_{_{9,NP}}^{Z_{_{BL}}}(\mu_{_{\rm EW}})=C_{_{9,H^\pm}}^{Z_{_{BL}}}(\mu_{_{\rm EW}})
+C_{_{9,\chi^\pm}}^{Z_{_{BL}}}(\mu_{_{\rm EW}})+C_{_{9,\chi^0}}^{Z_{_{BL}}}(\mu_{_{\rm EW}})
+C_{_{9,\tilde{g}}}^{Z_{_{BL}}}(\mu_{_{\rm EW}})+C_{_{9,{\tilde Z}_{_{BL}}}}^{Z_{_{BL}}}(\mu_{_{\rm EW}})
\;,\nonumber\\
%%%%%%%%%%%%%%%%%%%%%%%%%%%%%%%%%%%%%%%%%%%%%%%%%%%
&&C_{_{S,NP}}^{H_i^0}(\mu_{_{\rm EW}})=
C_{_{S,H^\pm}}^{H_i^0}(\mu_{_{\rm EW}})
+C_{_{S,\chi^\pm}}^{H_i^0}(\mu_{_{\rm EW}})
+C_{_{S,\chi^0}}^{H_i^0}(\mu_{_{\rm EW}})
+C_{_{S,\tilde{g}}}^{H_i^0}(\mu_{_{\rm EW}})
+C_{_{S,{\tilde Z}_{_{BL}}}}^{H_i^0}(\mu_{_{\rm EW}})
\;,\nonumber\\
%%%%%%%%%%%%%%%%%%%%%%%%%%%%%%%%%%%%%%%%%%%%%%%%%%%
&&C_{_{P,NP}}^{A^0}(\mu_{_{\rm EW}})=
C_{_{P,H^\pm}}^{A^0}(\mu_{_{\rm EW}})
+C_{_{P,\chi^\pm}}^{A^0}(\mu_{_{\rm EW}})
+C_{_{P,\chi^0}}^{A^0}(\mu_{_{\rm EW}})
+C_{_{P,\tilde{g}}}^{A^0}(\mu_{_{\rm EW}})
+C_{_{P,{\tilde Z}_{_{BL}}}}^{A^0}(\mu_{_{\rm EW}})
\;,\nonumber\\
%%%%%%%%%%%%%%%%%%%%%%%%%%%%%%%%%%%%%%%%%%%%%%%%%%%
&&C_{_{9,NP}}^{box}(\mu_{_{\rm EW}})=C_{_{9,H^\pm}}^{box}(\mu_{_{\rm EW}})
+C_{_{9,\chi^\pm}}^{box}(\mu_{_{\rm EW}})+C_{_{9,\chi^0}}^{box}(\mu_{_{\rm EW}})
+C_{_{9,{\tilde Z}_{_{BL}}}}^{box}(\mu_{_{\rm EW}})
+C_{_{9,{\chi^0\tilde Z}_{_{BL}}}}^{box}(\mu_{_{\rm EW}})
\;,\nonumber\\
%%%%%%%%%%%%%%%%%%%%%%%%%%%%%%%%%%%%%%%%%%%%%%%%%%%
&&C_{_{10,NP}}^{box}(\mu_{_{\rm EW}})=C_{_{10,H^\pm}}^{box}(\mu_{_{\rm EW}})
+C_{_{10,\chi^\pm}}^{box}(\mu_{_{\rm EW}})+C_{_{10,\chi^0}}^{box}(\mu_{_{\rm EW}})
+C_{_{10,{\tilde Z}_{_{BL}}}}^{box}(\mu_{_{\rm EW}})
+C_{_{10,{\chi^0\tilde Z}_{_{BL}}}}^{box}(\mu_{_{\rm EW}})
\;,\nonumber\\
%%%%%%%%%%%%%%%%%%%%%%%%%%%%%%%%%%%%%%%%%%%%%%%%%%%
&&C_{_{S,NP}}^{box}(\mu_{_{\rm EW}})=C_{_{S,H^\pm}}^{box}(\mu_{_{\rm EW}})
+C_{_{S,\chi^\pm}}^{box}(\mu_{_{\rm EW}})+C_{_{S,\chi^0}}^{box}(\mu_{_{\rm EW}})
+C_{_{S,\chi^0{\tilde Z}_{_{BL}}}}^{box}(\mu_{_{\rm EW}})
\;,\nonumber\\
%%%%%%%%%%%%%%%%%%%%%%%%%%%%%%%%%%%%%%%%%%%%%%%%%%%
&&C_{_{P,NP}}^{box}(\mu_{_{\rm EW}})=C_{_{P,H^\pm}}^{box}(\mu_{_{\rm EW}})
+C_{_{P,\chi^\pm}}^{box}(\mu_{_{\rm EW}})+C_{_{P,\chi^0}}^{box}(\mu_{_{\rm EW}})
+C_{_{P,\chi^0{\tilde Z}_{_{BL}}}}^{box}(\mu_{_{\rm EW}})\;,
%%%%%%%%%%%%%%%%%%%%%%%%%%%%%%%%%%%%%%%%%%%%%%%%%%%
\label{Wilson-Coefficients2}
\end{eqnarray}
where the concrete expressions for the corrections involving $U(1)_{_{B-L}}$
interaction are presented in Eq.(\ref{appB-2}).

The Wilson coefficients in Eq.(\ref{Wilson-Coefficients1}) are calculated at the matching
scale $\mu_{_{\rm EW}}$, then evolved down to hadronic scale $\mu\sim m_{_b}$ by the renormalization
group equations. In order to obtain hadronic matrix elements conveniently, we
define effective coefficients \cite{Buras1}
\begin{eqnarray}
%%%%%%%%%%%%%%%%%%%%%%%%%%%%%%%%%%%%%%%%%%%%%%%%%%%
&&C_{_7}^{eff}={4\pi\over\alpha_{_s}}C_{_7}-{1\over3}C_{_3}-{4\over9}C_{_4}
-{20\over3}C_{_5}-{80\over9}C_{_6}
\;,\nonumber\\
%%%%%%%%%%%%%%%%%%%%%%%%%%%%%%%%%%%%%%%%%%%%%%%%%%%
&&C_{_8}^{eff}={4\pi\over\alpha_{_s}}C_{_8}+C_{_3}-{1\over6}C_{_4}
+20C_{_5}-{10\over3}C_{_6}
\;,\nonumber\\
%%%%%%%%%%%%%%%%%%%%%%%%%%%%%%%%%%%%%%%%%%%%%%%%%%%
&&C_{_9}^{eff}=C_{_9}^{eff,SM}+{4\pi\over\alpha_{_s}}C_{_9}^{NP}
\;,\nonumber\\
%%%%%%%%%%%%%%%%%%%%%%%%%%%%%%%%%%%%%%%%%%%%%%%%%%%
&&C_{_{10}}^{eff}={4\pi\over\alpha_{_s}}C_{_{10}}\;,\;\;\;
C_{_{7,8,9,10}}^{\prime eff}={4\pi\over\alpha_{_s}}C_{_{7,8,9,10}}^\prime\;.
%%%%%%%%%%%%%%%%%%%%%%%%%%%%%%%%%%%%%%%%%%%%%%%%%%%
\label{Wilson-Coefficients}
\end{eqnarray}
In the SM, the Wilson coefficient $C_{_7}^{eff,SM}$ and $C_{_{10}}^{eff,SM}$
are real. Nevertheless, the Wilson coefficient $C_{_{9}}^{eff,SM}$ contains slightly
complex $CP$ phase originating from the continuum part of $u\bar{u}$ and $c\bar{c}$
loop which is proportional to $V_{_{ub}}V_{_{us}}^*$:
\begin{eqnarray}
%%%%%%%%%%%%%%%%%%%%%%%%%%%%%%%%%%%%%%%%%%%%%%%%%%%
&&C_{_9}^{eff,SM}={4\pi\over\alpha_{_s}}C_{_9}^{SM}+\xi_{_1}(q^2)
+{V_{_{ub}}V_{_{us}}^*\over V_{_{tb}}V_{_{ts}}^*}\xi_{_2}(q^2)\;,
%%%%%%%%%%%%%%%%%%%%%%%%%%%%%%%%%%%%%%%%%%%%%%%%%%%
\label{C9-SM}
\end{eqnarray}
with
\begin{eqnarray}
%%%%%%%%%%%%%%%%%%%%%%%%%%%%%%%%%%%%%%%%%%%%%%%%%%%
&&\xi_{_1}(q^2)=0.138\omega({q^2\over m_{_b}^2})
+g({m_{_c}\over m_{_b}},{q^2\over m_{_b}^2})\Big({4\over3}C_{_1}(\mu_{_b})
+C_{_2}(\mu_{_b})+3C_{_3}(\mu_{_b})+3C_{_5}(\mu_{_b})\Big)
\nonumber\\
&&\hspace{1.5cm}
+{2\over3}\Big(C_{_3}(\mu_{_b})+C_{_5}(\mu_{_b})\Big)
-{1\over2}g({m_{_d}\over m_{_b}},{q^2\over m_{_b}^2})\Big(C_{_3}(\mu_{_b})
+{4\over3}C_{_4}(\mu_{_b})\Big)
\nonumber\\
&&\hspace{1.5cm}
-{1\over2}g(1,{q^2\over m_{_b}^2})\Big(4C_{_3}(\mu_{_b})
+{4\over3}C_{_4}(\mu_{_b})+3C_{_5}(\mu_{_b})\Big)\;,
\nonumber\\
%%%%%%%%%%%%%%%%%%%%%%%%%%%%%%%%%%%%%%%%%%%%%%%%%%%
&&\xi_{_2}(q^2)=\Big[g({m_{_c}\over m_{_b}},{q^2\over m_{_b}^2})
-g({m_{_u}\over m_{_b}},{q^2\over m_{_b}^2})\Big]
\Big({4\over3}C_{_1}(\mu_{_b})+C_{_2}(\mu_{_b})\Big)\;.
%%%%%%%%%%%%%%%%%%%%%%%%%%%%%%%%%%%%%%%%%%%%%%%%%%%
\label{xi-1-2}
\end{eqnarray}
Where the concrete expressions for $\omega(z),g(x,y)$ are written respectively as \cite{Buras1}:
\begin{eqnarray}
%%%%%%%%%%%%%%%%%%%%%%%%%%%%%%%%%%%%%%%%%%%%%%%%%%%
&&\omega(z)=-{2\over9}\pi^2-{4\over3}Li_{_2}(z)-{2\over3}\ln z\ln(1-z)
-{5+4z\over3(1+2z)}\ln(1-z)
\nonumber\\
&&\hspace{1.4cm}
-{2z(1+z)(1-2z)\over3(1-z)^2(1+2z)}\ln z+{5+9z-6z^2\over6(1-z)(1+2z)}\;,
\nonumber\\
&&g(x,y)={8\over27}-{8\over9}\ln{m_{_b}\over\mu_{_b}}-{8\over9}\ln x+{16x^2\over9y}
\nonumber\\
&&\hspace{1.4cm}
-{4\over9}(1+{2x^2\over y})\sqrt{1-{4x^2\over y}}
\left\{\begin{array}{ll}\ln\Big|{\sqrt{y-4x^2}+\sqrt{y}\over\sqrt{y-4x^2}-\sqrt{y}}\Big|-i\pi,
&{\rm if}\: y>4x^2\\
2\arctan{\sqrt{y}\over\sqrt{4x^2-y}},&{\rm if}\: y<4x^2\end{array}\right.
%%%%%%%%%%%%%%%%%%%%%%%%%%%%%%%%%%%%%%%%%%%%%%%%%%%
\label{omega-g}
\end{eqnarray}
In the limit of $x=0$, $g(0,y)={8\over27}-{8\over9}\ln{m_{_b}\over\mu_{_b}}-{8\over9}\ln y+i{4\over9}\pi$.
In the following numerical analysis, we take $\mu_{_b}=m_{_b}$ for simplification.

In our numerical analyses, we evaluate the Wilson coefficients from the SM to
next-to-next-to-logarithmic (NNLL) accuracy in Table.\ref{tab1} at hadronic energy scale.
\begin{table}
\begin{tabular}{|c|c|c|c|}
\hline
\hline
$C_{_7}^{eff,SM}$    & $C_{_8}^{eff,SM}$    & $C_{_9}^{eff,SM}$ & $C_{_{10}}^{eff,SM}$\\
\hline
$-0.304$   & $-0.167$  & $4.211$ & $-4.103$\\
\hline
\hline
\end{tabular}
\caption{At hadronic scale $\mu=m_{_b}\simeq4.8$GeV, SM Wilson coefficients to NNLL accuracy. \label{tab1}}
\end{table}
On the other hand, the corrections to the Wilson coefficients from new physics
are only included to one-loop accuracy:
\begin{eqnarray}
%%%%%%%%%%%%%%%%%%%%%%%%%%%%%%%%%%%%%%%%%%%%%%%%%%%
&&\overrightarrow{C}_{_{NP}}(\mu)=\widehat{U}(\mu,\mu_0)\overrightarrow{C}_{_{NP}}(\mu_0)
\;,\nonumber\\
%%%%%%%%%%%%%%%%%%%%%%%%%%%%%%%%%%%%%%%%%%%%%%%%%%%
&&\overrightarrow{C^\prime}_{_{NP}}(\mu)=\widehat{U^\prime}(\mu,\mu_0)
\overrightarrow{C^\prime}_{_{NP}}(\mu_0)
%%%%%%%%%%%%%%%%%%%%%%%%%%%%%%%%%%%%%%%%%%%%%%%%%%%
\label{evaluation1}
\end{eqnarray}
with
\begin{eqnarray}
%%%%%%%%%%%%%%%%%%%%%%%%%%%%%%%%%%%%%%%%%%%%%%%%%%%
&&\overrightarrow{C}_{_{NP}}^{T}=\Big(C_{_{1,NP}},\;\cdots,\;C_{_{6,NP}},
C_{_{7,NP}}^{eff},\;C_{_{8,NP}}^{eff},\;C_{_{9,NP}}^{eff}-Y(q^2),\;
C_{_{10,NP}}^{eff}\Big)
\;,\nonumber\\
%%%%%%%%%%%%%%%%%%%%%%%%%%%%%%%%%%%%%%%%%%%%%%%%%%%
&&\overrightarrow{C}_{_{NP}}^{\prime,\;T}=\Big(C_{_{7,NP}}^{\prime,\;eff},\;
C_{_{8,NP}}^{\prime,\;eff},\;C_{_{9,NP}}^{\prime,\;eff},\;
C_{_{10,NP}}^{\prime,\;eff}\Big)\;.
%%%%%%%%%%%%%%%%%%%%%%%%%%%%%%%%%%%%%%%%%%%%%%%%%%%
\label{evaluation2}
\end{eqnarray}
Correspondingly the evolving matrices are approached as
\begin{eqnarray}
%%%%%%%%%%%%%%%%%%%%%%%%%%%%%%%%%%%%%%%%%%%%%%%%%%%
&&\widehat{U}(\mu,\mu_0)\simeq1-\Big[{1\over2\beta_0}\ln{\alpha_{_s}(\mu)\over
\alpha_{_s}(\mu_0)}\Big]\widehat{\gamma}^{(0)T}
\;,\nonumber\\
%%%%%%%%%%%%%%%%%%%%%%%%%%%%%%%%%%%%%%%%%%%%%%%%%%%
&&\widehat{U^\prime}(\mu,\mu_0)\simeq1-\Big[{1\over2\beta_0}\ln{\alpha_{_s}(\mu)\over
\alpha_{_s}(\mu_0)}\Big]\widehat{\gamma^\prime}^{(0)T}\;,
%%%%%%%%%%%%%%%%%%%%%%%%%%%%%%%%%%%%%%%%%%%%%%%%%%%
\label{evaluation3}
\end{eqnarray}
where the anomalous dimension matrices can be read from Ref. \cite{Gambino1} as
\begin{eqnarray}
%%%%%%%%%%%%%%%%%%%%%%%%%%%%%%%%%%%%%%%%%%%%%%%%%%%
&&\widehat{\gamma}^{(0)}=\left(\begin{array}{cccccccccc}
-4&{8\over3}&0&-{2\over9}&0&0&-{208\over243}&{173\over162}&-{2272\over729}&0\\
12&0&0&{4\over3}&0&0&{416\over81}&{70\over27}&{1952\over243}&0\\
0&0&0&-{52\over3}&0&2&-{176\over81}&{14\over27}&-{6752\over243}&0\\
0&0&-{40\over9}&-{100\over9}&{4\over9}&{5\over6}&-{152\over243}&-{587\over162}&-{2192\over729}&0\\
0&0&0&-{256\over3}&0&20&-{6272\over81}&{6596\over27}&-{84032\over243}&0\\
0&0&-{256\over9}&{56\over9}&{40\over9}&-{2\over3}&{4624\over243}&{4772\over81}&-{37856\over729}&0\\
0&0&0&0&0&0&{32\over3}&0&0&0\\
0&0&0&0&0&0&-{32\over9}&{28\over3}&0&0\\
0&0&0&0&0&0&0&0&0&0\\
0&0&0&0&0&0&0&0&0&0\\
\end{array}\right)
\;,\nonumber\\
%%%%%%%%%%%%%%%%%%%%%%%%%%%%%%%%%%%%%%%%%%%%%%%%%%%
&&\widehat{\gamma^\prime}^{(0)}=\left(\begin{array}{cccc}
{32\over3}&0&0&0\\
-{32\over9}&{28\over3}&0&0\\
0&0&0&0\\0&0&0&0\\
\end{array}\right)\;.
%%%%%%%%%%%%%%%%%%%%%%%%%%%%%%%%%%%%%%%%%%%%%%%%%%%
\label{ADM1}
\end{eqnarray}
In addition, the operators ${\cal O}_{_{S,P}}^{(\prime)}$ do not mix with other operators
and their Wilson coefficients are given by the corresponding coefficients at matching
scale.

\section{Differential decay branching ratios, forward-backward and $CP$ asymmetries\label{sec4}}
\indent\indent
Keeping full dependence on the lepton mass while neglecting the strange quark mass,
we write the unnormalized double differential decay width for $B\rightarrow X_{_s}l^+l^-\;(l=e,\;\mu,\;\tau)$
as
\begin{eqnarray}
%%%%%%%%%%%%%%%%%%%%%%%%%%%%%%%%%%%%%%%%%%%%%%%%%%%
&&{d^2\Gamma\over dq^2d\cos\theta}={\alpha_{_{\rm EW}}^2\over16\pi^2}
{G_{_F}^2m_{_b}^5|V_{_{tb}}V_{_{ts}}^*|^2\over48\pi^3}
\Big(1-{q^2\over m_{_b}^2}\Big)^2\sqrt{1-{4m_{_l}^2m_{_b}^2\over(q^2)^2}}
\nonumber\\
&&\hspace{2.3cm}\times
\Big\{6\Big(1+{2m_{_l}^2\over q^2}\Big)\Re(C_{_7}^{eff}C_{_9}^{eff*}(q^2))
+{3m_{_l}^2\over m_{_b}^2}\Big[|C_{_9}^{eff}(q^2)|^2-|C_{_{10}}^{eff}|^2\Big]
\nonumber\\
&&\hspace{2.3cm}
+3\Big(1+{2m_{_l}^2\over q^2}\Big)|C_{_7}^{eff}|^2
\Big[{m_{_b}^2\over q^2}(1+\cos^2\theta)+(1-\cos^2\theta)\Big]
\nonumber\\
&&\hspace{2.3cm}
+{3\over4}\Big[|C_{_9}^{eff}(q^2)|^2+|C_{_{10}}^{eff}|^2\Big]
\Big[(1-\cos^2\theta)+({q^2-m_{_l^2}\over m_{_b}^2}+{m_{_l}^2\over q^2})
(1+\cos^2\theta)\Big]
\nonumber\\
&&\hspace{2.3cm}
+{9\over32}q^2\Big(1-{4m_{_l}^2\over q^2}\Big)\Big(|C_{_S}|^2+|C_{_P}|^2\Big)
(1+\cos^2\theta)
\nonumber\\
&&\hspace{2.3cm}
+{9\over4}m_{_l}\Re(C_{_S}C_{_{10}}^{eff*}+C_{_P}C_{_{10}}^{eff*})(1-\cos^2\theta)
\nonumber\\
&&\hspace{2.3cm}
-{3q^2\over m_{_b}^2}\Re(C_{_9}^{eff}(q^2)C_{_{10}}^{eff*})\cos\theta
-6\Re(C_{_7}^{eff}C_{_{10}}^{eff*})\cos\theta
\nonumber\\
&&\hspace{2.0cm}
+{3m_{_l}\over2}\Re((2C_{_7}^{eff}+C_{_9}^{eff}(q^2))(C_{_S}^*+C_{_P}^*))\cos\theta\Big\}\;,
%%%%%%%%%%%%%%%%%%%%%%%%%%%%%%%%%%%%%%%%%%%%%%%%%%%
\label{Double-Differential-DW}
\end{eqnarray}
where $\theta$ denotes the angle between the $l^+$ and $B$ meson three
momenta in the di-lepton rest frame. In order to reduce the uncertainties
originating from the bottom quark mass and $CKM$ matrix elements, one
generally normalize the observables by the semileptonic decay width
of $B$ meson:
\begin{eqnarray}
%%%%%%%%%%%%%%%%%%%%%%%%%%%%%%%%%%%%%%%%%%%%%%%%%%%
&&\Gamma(B\rightarrow X_{_c}e\bar{\nu}_{_e})={G_{_F}^2m_{_b}^5\over192\pi^3}
|V_{_{cb}}|^2f(z)\kappa(z)\;.
%%%%%%%%%%%%%%%%%%%%%%%%%%%%%%%%%%%%%%%%%%%%%%%%%%%
\label{B-semileptonic-DW}
\end{eqnarray}
Here $f(z)=1-8z^2+8z^6-z^8-24z^4\ln z$ is phase space factor, $k(z)=1-2\alpha_{_S}/3\pi
[(\pi^2-31/4)(1-z)^2+3/2]$ is the QCD correction factor
with $z=m_{_c}/m_{_b}$ \cite{phase-QCD}, respectively.
Using Eq.(\ref{Double-Differential-DW}) and Eq.(\ref{B-semileptonic-DW}),
we get the normalized differential decay branching ratio of $B\rightarrow X_{_s}l^+l^-\;(l=e,\;\mu,\;\tau)$
at hadronic scale as
\begin{eqnarray}
%%%%%%%%%%%%%%%%%%%%%%%%%%%%%%%%%%%%%%%%%%%%%%%%%%%
&&R(q^2)={1\over\Gamma(B\rightarrow X_{_c}e\nu)}{d\Gamma(B\rightarrow X_{_s}l^+l^-)\over dq^2}
\nonumber\\
&&\hspace{1.1cm}
={\alpha_{_{\rm EW}}^2\over4\pi^2}\Big|{V_{_{tb}}V_{_{ts}}^*\over V_{_{cb}}}\Big|^2
{1\over f(z)k(z)}\Big(1-{q^2\over m_{_b}^2}\Big)^2\sqrt{1-{4m_{_l}^2m_{_b}^2\over(q^2)^2}}
\nonumber\\
&&\hspace{1.5cm}\times
\Big\{4\Big(1+{2m_{_l}^2\over q^2}\Big)\Big[3\Re(C_{_7}^{eff}C_{_9}^{eff*}(q^2))
+\Big(1+{2m_{_b}^2\over q^2}\Big)|C_{_7}^{eff}|^2\Big]
\nonumber\\
&&\hspace{1.5cm}
+\Big(1+{2q^2\over m_{_b}^2}+{2m_{_l}^2\over q^2}+{4m_{_l}^2\over m_{_b}^2}\Big)
|C_{_9}^{eff}(q^2)|^2
\nonumber\\
&&\hspace{1.5cm}
+\Big(1+{2q^2\over m_{_b}^2}+{2m_{_l}^2\over q^2}-{8m_{_l}^2\over m_{_b}^2}\Big)
|C_{_{10}}^{eff}|^2
\nonumber\\
&&\hspace{1.5cm}
+{3\over4}q^2\Big(1-{4m_{_l}^2\over q^2}\Big)\Big(|C_{_S}|^2+|C_{_P}|^2\Big)
\nonumber\\
&&\hspace{1.5cm}
+3m_{_l}\Re(C_{_S}C_{_{10}}^{eff*}+C_{_P}C_{_{10}}^{eff*})\Big\}\;,
%%%%%%%%%%%%%%%%%%%%%%%%%%%%%%%%%%%%%%%%%%%%%%%%%%%
\label{Differential-BR}
\end{eqnarray}
where $q^2=(p_{_{l^+}}+p_{_{l^-}})^2$ denotes the dilepton invariant mass squared.
Meanwhile the unnormalized forward-backward asymmetry is formulated as
\begin{eqnarray}
%%%%%%%%%%%%%%%%%%%%%%%%%%%%%%%%%%%%%%%%%%%%%%%%%%%
&&\bar{A}_{_{FB}}(q^2)={1\over\Gamma(B\rightarrow X_{_c}e\nu)}
\int_{-1}^1d\cos\theta{d^2\Gamma(B\rightarrow X_{_s}l^+l^-)\over d\cos\theta dq^2}
{\rm Sgn}(\cos\theta)
\nonumber\\
&&\hspace{1.5cm}
=-{3\alpha_{_{\rm EW}}^2\over4\pi^2}\Big|{V_{_{tb}}V_{_{ts}}^*\over V_{_{cb}}}\Big|^2
{1\over f(z)k(z)}\Big(1-{q^2\over m_{_b}^2}\Big)^2\sqrt{1-{4m_{_l}^2m_{_b}^2\over(q^2)^2}}
\nonumber\\
&&\hspace{1.8cm}\times
\Big\{{q^2\over m_{_b}^2}\Re(C_{_9}^{eff}(q^2)C_{_{10}}^{eff*})
+2\Re(C_{_7}^{eff}C_{_{10}}^{eff*})
\nonumber\\
&&\hspace{2.0cm}
-{m_{_l}\over2}\Re((2C_{_7}^{eff}+C_{_9}^{eff}(q^2))(C_{_S}^*+C_{_P}^*))\Big\}\;,
%%%%%%%%%%%%%%%%%%%%%%%%%%%%%%%%%%%%%%%%%%%%%%%%%%%
\label{Unnormalized-AFB}
\end{eqnarray}
and the normalized forward-backward asymmetry is
\begin{eqnarray}
%%%%%%%%%%%%%%%%%%%%%%%%%%%%%%%%%%%%%%%%%%%%%%%%%%%
&&A_{_{FB}}(q^2)={1\over R(q^2)}\bar{A}_{_{FB}}(q^2)\;,
%%%%%%%%%%%%%%%%%%%%%%%%%%%%%%%%%%%%%%%%%%%%%%%%%%%
\label{Differential-AFB}
\end{eqnarray}
respectively. The global forward-backward asymmetry in the region $q^2\in[a,\;b]\;{\rm GeV}^2$
is defined through
\begin{eqnarray}
%%%%%%%%%%%%%%%%%%%%%%%%%%%%%%%%%%%%%%%%%%%%%%%%%%%
&&A_{_{FB}}\Big|_{q^2\in[a,\;b]\;{\rm GeV}^2}
={N(l_\rightarrow^+)-N(l_\leftarrow^+)\over N(l_\rightarrow^+)+N(l_\leftarrow^+)}
\Big|_{q^2\in[a,\;b]\;{\rm GeV}^2}
\nonumber\\
&&\hspace{3.0cm}
={\int_a^b dq^2\bar{A}_{_{FB}}(q^2)\over\int_a^bdq^2R(q^2)}\;.
%%%%%%%%%%%%%%%%%%%%%%%%%%%%%%%%%%%%%%%%%%%%%%%%%%%
\label{global-AFB}
\end{eqnarray}

The direct $CP$ asymmetry in $B\rightarrow X_{_s}l^+l^-$ is defined by
\begin{eqnarray}
%%%%%%%%%%%%%%%%%%%%%%%%%%%%%%%%%%%%%%%%%%%%%%%%%%%
&&A_{_{CP}}(q^2)={d\Gamma(B\rightarrow X_{_s}l^+l^-)/dq^2
-d\Gamma(\bar{B}\rightarrow \bar{X}_{_s}l^+l^-)/dq^2\over
d\Gamma(B\rightarrow X_{_s}l^+l^-)/dq^2
+d\Gamma(\bar{B}\rightarrow \bar{X}_{_s}l^+l^-)/dq^2}
\nonumber\\
&&\hspace{1.5cm}=
{\Delta D(q^2)\over D(q^2)}\;,
%%%%%%%%%%%%%%%%%%%%%%%%%%%%%%%%%%%%%%%%%%%%%%%%%%%
\label{ACP1}
\end{eqnarray}
with
\begin{eqnarray}
%%%%%%%%%%%%%%%%%%%%%%%%%%%%%%%%%%%%%%%%%%%%%%%%%%%
&&\Delta D(q^2)=2\Big(1+{2m_{_l}^2\over q^2}\Big)\Big\{
\Im\Big({V_{_{ub}}V_{_{us}}^*\over V_{_{tb}}V_{_{ts}}^*}\Big)
\Big[2\Big(1+{2q^2\over m_{_b}^2}\Big)\Im(\xi_{_1}\xi_{_2}^*)
-12C_{_7}^{eff,SM}(\mu_{_b})\Im(\xi_{_2})\Big]
\nonumber\\
&&\hspace{1.8cm}
+2\Big(1+{2q^2\over m_{_b}^2}\Big)\Big[\Im(\xi_{_1})\Im\Big(C_{_9}^{eff,NP}(\mu_{_b})\Big)
+\Im(\xi_{_2})\Im\Big({V_{_{ub}}V_{_{us}}^*\over V_{_{tb}}V_{_{ts}}^*}
C_{_9}^{eff,NP}(\mu_{_b})\Big)\Big]
\nonumber\\
&&\hspace{1.8cm}
+12\Big[\Im(\xi_{_1})\Im\Big(C_{_7}^{eff,NP}(\mu_{_b})\Big)
+\Im(\xi_{_2})\Im\Big({V_{_{ub}}V_{_{us}}^*\over V_{_{tb}}V_{_{ts}}^*}
C_{_7}^{eff,NP}(\mu_{_b})\Big)\Big]\Big\}\;,
\nonumber\\
%%%%%%%%%%%%%%%%%%%%%%%%%%%%%%%%%%%%%%%%%%%%%%%%%%%
&&D(q^2)=\Big(1+{2m_{_l}^2\over q^2}\Big)\Big(1+{2q^2\over m_{_b}^2}\Big)
\Big\{B_{_1}+2|C_{_9}^{eff,NP}(\mu_{_b})|^2+4\Re(\xi_{_1})\Re\Big(C_{_9}^{eff,NP}(\mu_{_b})\Big)
\nonumber\\
&&\hspace{1.8cm}
+4\Re(\xi_{_2})\Re\Big({V_{_{ub}}V_{_{us}}^*\over V_{_{tb}}V_{_{ts}}^*}
C_{_9}^{eff,NP}(\mu_{_b})\Big)\Big\}
\nonumber\\
&&\hspace{1.8cm}
+12\Big(1+{2m_{_l}^2\over q^2}\Big)\Big\{B_{_2}+2C_{_7}^{eff,SM}(\mu_{_b})
\Re\Big(C_{_9}^{eff,NP}(\mu_{_b})\Big)
\nonumber\\
&&\hspace{1.8cm}
+2\Re\Big(C_{_7}^{eff,NP}(\mu_{_b})C_{_9}^{eff,NP*}(\mu_{_b})\Big)
+2\Re(\xi_{_1})\Re\Big(C_{_7}^{eff,NP}(\mu_{_b})\Big)
\nonumber\\
&&\hspace{1.8cm}
+2\Re(\xi_{_2})\Re\Big({V_{_{ub}}V_{_{us}}^*\over V_{_{tb}}V_{_{ts}}^*}
C_{_7}^{eff,NP}(\mu_{_b})\Big)\Big\}
\nonumber\\
&&\hspace{1.8cm}
+8\Big(1+{2m_{_l}^2\over q^2}\Big)\Big(1+{2m_{_b}^2\over q^2}\Big)|C_{_7}^{eff}|^2
\nonumber\\
&&\hspace{1.8cm}
+2\Big(1+{2q^2\over m_{_b}^2}+{2m_{_l}^2\over q^2}-{8m_{_l}^2\over m_{_b}^2}\Big)
|C_{_{10}}^{eff}|^2
\nonumber\\
&&\hspace{1.8cm}
+{3\over2}q^2\Big(1-{4m_{_l}^2\over q^2}\Big)\Big(|C_{_S}|^2+|C_{_P}|^2\Big)
\nonumber\\
&&\hspace{1.8cm}
+6m_{_l}\Re(C_{_S}C_{_{10}}^{eff*}+C_{_P}C_{_{10}}^{eff*})\;.
%%%%%%%%%%%%%%%%%%%%%%%%%%%%%%%%%%%%%%%%%%%%%%%%%%%
\label{deltaDl}
\end{eqnarray}
Here
\begin{eqnarray}
%%%%%%%%%%%%%%%%%%%%%%%%%%%%%%%%%%%%%%%%%%%%%%%%%%%
&&B_{_1}=2\Big\{|\xi_{_1}|^2+|{V_{_{ub}}V_{_{us}}^*\over V_{_{tb}}V_{_{ts}}^*}\xi_{_2}|^2
+2\Re({V_{_{ub}}V_{_{us}}^*\over V_{_{tb}}V_{_{ts}}^*})\Re(\xi_{_1}\xi_{_2})\Big\}\;,
\nonumber\\
&&B_{_2}=2C_{_7}^{eff,SM}(\mu_{_b})\Big\{\Re(\xi_{_1})
+\Re({V_{_{ub}}V_{_{us}}^*\over V_{_{tb}}V_{_{ts}}^*})\Re(\xi_{_2})\Big\}\;.
%%%%%%%%%%%%%%%%%%%%%%%%%%%%%%%%%%%%%%%%%%%%%%%%%%%
\label{B12}
\end{eqnarray}

The global $CP$ asymmetry in the region $q^2\in[a,\;b]\;{\rm GeV}^2$
is correspondingly defined through
\begin{eqnarray}
%%%%%%%%%%%%%%%%%%%%%%%%%%%%%%%%%%%%%%%%%%%%%%%%%%%
&&A_{_{CP}}\Big|_{q^2\in[a,\;b]\;{\rm GeV}^2}
={\int_a^b dq^2\Delta D(q^2)\over\int_a^bdq^2D(q^2)}\;.
%%%%%%%%%%%%%%%%%%%%%%%%%%%%%%%%%%%%%%%%%%%%%%%%%%%
\label{global-ACP}
\end{eqnarray}

In the region $1{\rm GeV}^2\le q^2\le6{\rm GeV}^2$ where the theoretical
evaluations are not heavily affected by the photon pole at low $q^2$
and the $c\bar{c}$ resonances at higher $q^2$, the updated theoretical
predictions in this region are\cite{UpdatedSMBtoXsll}:
\begin{eqnarray}
%%%%%%%%%%%%%%%%%%%%%%%%%%%%%%%%%%%%%%%%%%%%%%%%%%%
&&Br(B\rightarrow X_{_s}e^+e^-)\Big|_{q^2\in[1,\;6]\;{\rm GeV}^2}^{SM}
=(1.64\pm0.11)\times10^{-6}\;,
\nonumber\\
&&Br(B\rightarrow X_{_s}\mu^+\mu^-)\Big|_{q^2\in[1,\;6]\;{\rm GeV}^2}^{SM}
=(1.59\pm0.11)\times10^{-6}\;.
%%%%%%%%%%%%%%%%%%%%%%%%%%%%%%%%%%%%%%%%%%%%%%%%%%%
\label{SM-BR-Low}
\end{eqnarray}

In the region $q^2\ge14.4{\rm GeV}^2$, the theoretical uncertainty
is relatively larger than that in the low region $1{\rm GeV}^2\le q^2\le6{\rm GeV}^2$,
and the SM theoretical evaluations are given as\cite{UpdatedSMBtoXsll}:
\begin{eqnarray}
%%%%%%%%%%%%%%%%%%%%%%%%%%%%%%%%%%%%%%%%%%%%%%%%%%%
&&Br(B\rightarrow X_{_s}e^+e^-)\Big|_{q^2\in[14.4,\;25]\;{\rm GeV}^2}^{SM}
=(0.21\pm0.07)\times10^{-6}\;,
\nonumber\\
&&Br(B\rightarrow X_{_s}\mu^+\mu^-)\Big|_{q^2\in[14.4,\;25]\;{\rm GeV}^2}^{SM}
=(0.24\pm0.07)\times10^{-6}\;.
%%%%%%%%%%%%%%%%%%%%%%%%%%%%%%%%%%%%%%%%%%%%%%%%%%%
\label{SM-BR-High}
\end{eqnarray}
In our analysis, the lepton-flavor-averaged branching ratio for
$B\rightarrow X_{_s}l^+l^-$ is averages of the individual
$Br(B\rightarrow X_{_s}e^+e^-)$ and $Br(B\rightarrow X_{_s}\mu^+\mu^-)$.
Furthermore, the updated experimental data on the forward-backward and $CP$
asymmetries constrain the parameter space of new physics concretely.

\section{Numerical analyses\label{sec5}}
\indent\indent
%-----------------------------------------------------------------------------
\begin{table}[t]
\renewcommand{\arraystretch}{1.3}
\centering
\begin{tabular}{|c|c|}
\hline
Input & Input \\
\hline
$m_{_B}=5.280$ GeV & $m_{_{K^*}}=0.896$ GeV\\
$m_{_{B_s}}=5.367$ GeV & $m_{_\mu}=0.106$ GeV\\
$m_{_{\rm W}}=80.40$ GeV & $m_{_{\rm Z}}=91.19$ GeV\\
$\tau_{_B}=2.307\times10^{12}$ GeV & $f_{_B}=0.190\pm0.004$\\
$\alpha_{_s}(m_{_{\rm Z}})=0.118\pm0.002$ & $\alpha_{_{ em}}(m_{_{\rm Z}})=1/128.9$\\
$m_c(m_c)=1.27\pm0.11$ GeV & $m_b(m_b)=4.18\pm0.17$ GeV\\
$m_t^{pole}=173.1\pm1.3$ GeV &  \\
$\lambda_{_{CKM}}=0.225\pm0.001$  & $A_{_{CKM}}=0.811\pm0.022$\\
$\bar{\rho}=0.131\pm0.026$  & $\bar{\eta}=0.345\pm0.014$\\

\hline
\end{tabular}
\caption{Input parameters\cite{PDG} of the SM used in the numerical analysis\label{tab2}}
\label{InputSM}
\end{table}
%-----------------------------------------------------------------------------

%%%%%%%%%%%%%%%%%%%%%%%%%%%%%%%%%%%%%%%%%%%%%%%%%%%%%
\begin{figure}[h]
\setlength{\unitlength}{1cm}
\centering
\vspace{0.0cm}\hspace{-1.5cm}
\includegraphics[height=16cm,width=18.0cm]{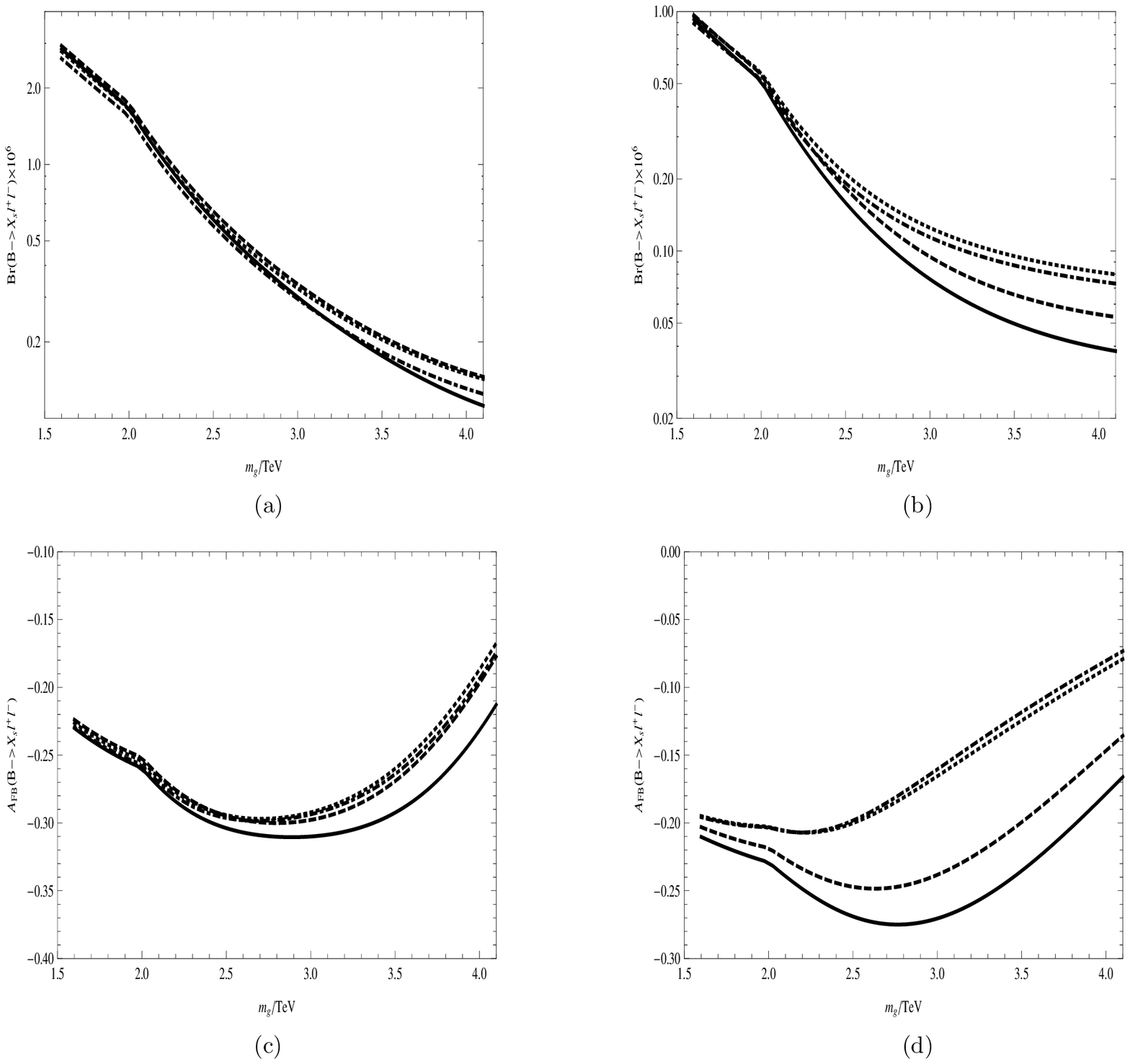}
\vspace{-7cm}
\caption[]{Taking $(\delta_{D}^{LL})_{23}=(\delta_{D}^{RR})_{23}=(\delta_{D}^{LR})_{23}=0.125$,
we plot (a) $BR(B\rightarrow X_{_s}l^+l^-)_{_{q^2\in[1,6]{\rm GeV}^2}}\times10^6$,
(b) $BR(B\rightarrow X_{_s}l^+l^-)_{_{q^2\in[14.4,25.0]{\rm GeV}^2}}\times10^6$
(c) $A_{_{FB}}(B\rightarrow X_{_s}l^+l^-)_{_{q^2\in[1,6]{\rm GeV}^2}}$,
(d) $A_{_{FB}}(B\rightarrow X_{_s}l^+l^-)_{_{q^2\in[14.4,25.0]{\rm GeV}^2}}$,
varying with the gluino mass. Where the solid lines denote $\tan\beta=5$,
dashed lines denote $\tan\beta=10$, dotted lines denote $\tan\beta=30$,
dashed-dotted lines denote $\tan\beta=50$, respectively.}
\label{fig1}
\end{figure}
%%%%%%%%%%%%%%%%%%%%%%%%%%%%%%%%%%%%%%%%%%%%%%%%%%%%%
In order to perform our numerical analyses, we present
the relevant SM inputs from \cite{PDG} in table.\ref{tab2}.
The supersymmetric parameters involved here are soft breaking masses
of the 2nd and 3rd generation squarks, $m_{_{\tilde{Q}_{2,3}}}^2,\;m_{_{\tilde{U}_{2,3}}}^2,\;
m_{_{\tilde{D}_{2,3}}}^2$, soft breaking masses of the 1st and 2nd
generation sleptons $m_{_{{\tilde L}_{1,2}}},\;m_{_{{\tilde E}_{1,2}}},
\;m_{_{{\tilde N}_{1,2}}}$, neutralino and chargino masses $m_{_{\chi_\alpha^0}},\;
m_{_{\chi_\beta^\pm}},\;(\alpha=1,\;\cdots,\;4,\;\beta=1,\;2)$ and their mixing matrices.
Additionally the masses and mixing matrix of $B-L$ gaugino/right-handed neutrinos are mainly determined
from nonzero VEVs of right-handed sneutrinos, local $B-L$ gauge coupling $g_{_{BL}}$
and soft gaugino mass parameters $m_{_{BL}},\;m_{_{1BL}}$. The flavor conservation mixing
between left- and right-handed of the third generation
squarks $(\delta_u^{LR})_{33}=m_{_{\tilde{t}_X}}^2/\Lambda_{_{NP}}^2,\;
(\delta_d^{LR})_{33}=m_{_{\tilde{b}_X}}^2/\Lambda_{_{NP}}^2$ are chosen to give
the lightest Higgs mass in the range $124-126$ GeV, where the concrete expressions
of $m_{_{\tilde{t}_X}}^2,\;m_{_{\tilde{b}_X}}^2$ are presented in appendix \ref{appA}.
The $b\rightarrow s$ transitions are mediated by those flavor changing insertions
$(\delta_{U,D}^{LL})_{23},\;(\delta_{U,D}^{LR})_{23},\;(\delta_{U,D}^{RR})_{23}$,
which are originated from flavour-violating scalar mass terms and trilinear scalar couplings
in soft breaking terms.

%%%%%%%%%%%%%%%%%%%%%%%%%%%%%%%%%%%%%%%%%%%%%%%%%%%%%
\begin{figure}[h]
\setlength{\unitlength}{1cm}
\centering
\vspace{0.0cm}\hspace{-1.5cm}
\includegraphics[height=16cm,width=18.0cm]{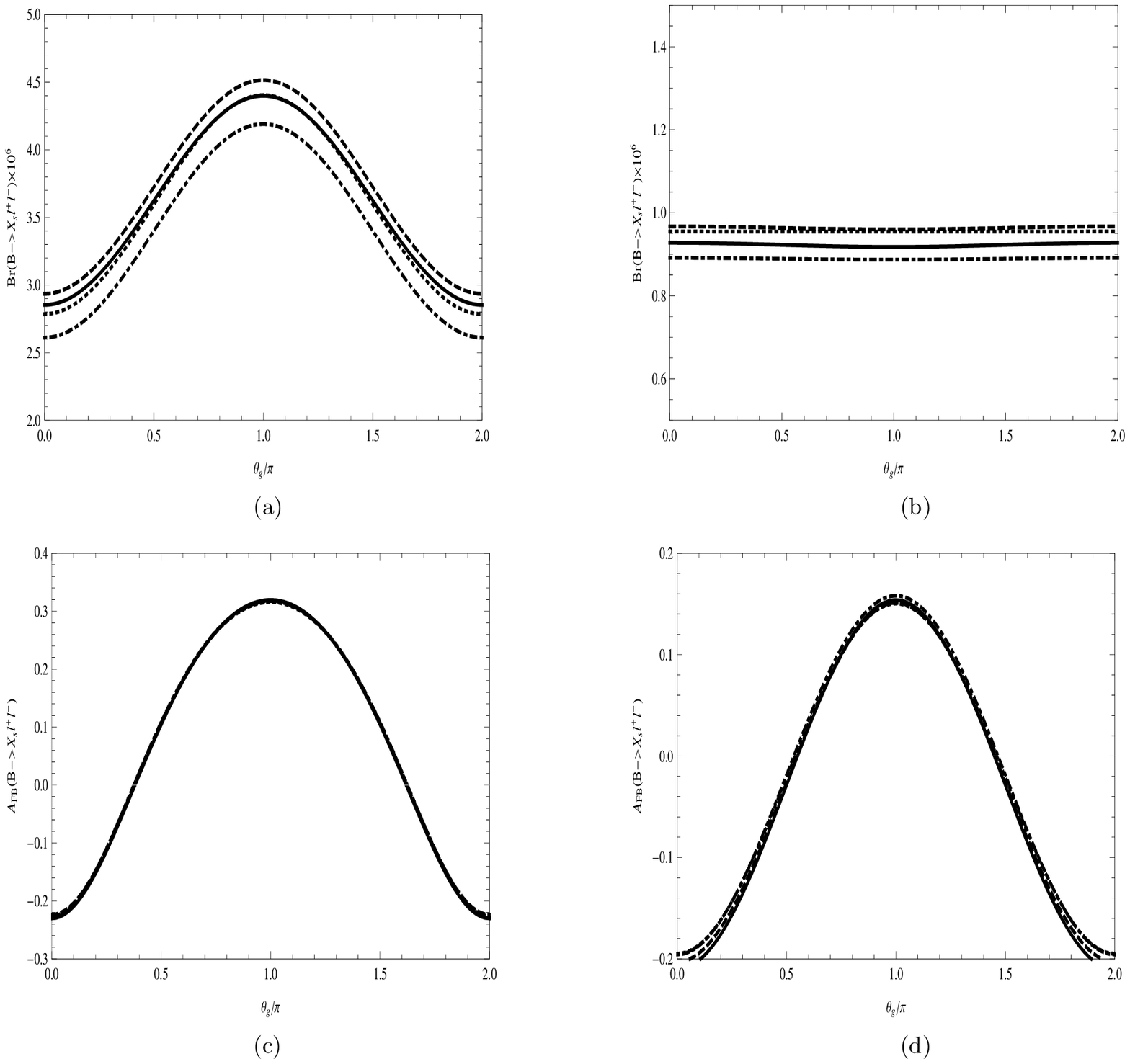}
\vspace{-7cm}
\caption[]{Taking $(\delta_{D}^{LL})_{23}=(\delta_{D}^{RR})_{23}=(\delta_{D}^{LR})_{23}=0.125$,
we plot (a) $BR(B\rightarrow X_{_s}l^+l^-)_{_{q^2\in[1,6]{\rm GeV}^2}}\times10^6$,
(b) $BR(B\rightarrow X_{_s}l^+l^-)_{_{q^2\in[14.4,25.0]{\rm GeV}^2}}\times10^6$
(c) $A_{_{FB}}(B\rightarrow X_{_s}l^+l^-)_{_{q^2\in[1,6]{\rm GeV}^2}}$,
(d) $A_{_{FB}}(B\rightarrow X_{_s}l^+l^-)_{_{q^2\in[14.4,25.0]{\rm GeV}^2}}$,
varying with the $CP$ phase $\theta_{_{\tilde g}}$.
Where the solid lines denote $\tan\beta=5$,
dashed lines denote $\tan\beta=10$, dotted lines denote $\tan\beta=30$,
dashed-dotted lines denote $\tan\beta=50$, respectively.}
\label{fig2}
\end{figure}
%%%%%%%%%%%%%%%%%%%%%%%%%%%%%%%%%%%%%%%%%%%%%%%%%%%%%

The updated bound from ATLAS collaboration on the gluino mass
is $m_{_{\tilde g}}\ge1460\;{\rm GeV}$, and the bound on
the mass of scalar top is $m_{_{\tilde t}}\ge780\;{\rm GeV}$\cite{ATLAS2016}.
To coincide with those experimental data, we always assume
$\Lambda_{_{NP}}=m_{_{{\tilde Q}_{2,3}}}=m_{_{{\tilde U}_{2,3}}}
=m_{_{{\tilde D}_{2,3}}}=2$ TeV, $m_{_{{\tilde L}_{1,2}}}=m_{_{{\tilde E}_{1,2}}}
=m_{_{{\tilde N}_{1,2}}}=1$ TeV, and $m_{_{\tilde g}}\ge1.5$ TeV
in our numerical discussion unless specified.

%%%%%%%%%%%%%%%%%%%%%%%%%%%%%%%%%%%%%%%%%%%%%%%%%%%%%
\begin{figure}[h]
\setlength{\unitlength}{1cm}
\centering
\vspace{0.0cm}\hspace{-1.5cm}
\includegraphics[height=16cm,width=18.0cm]{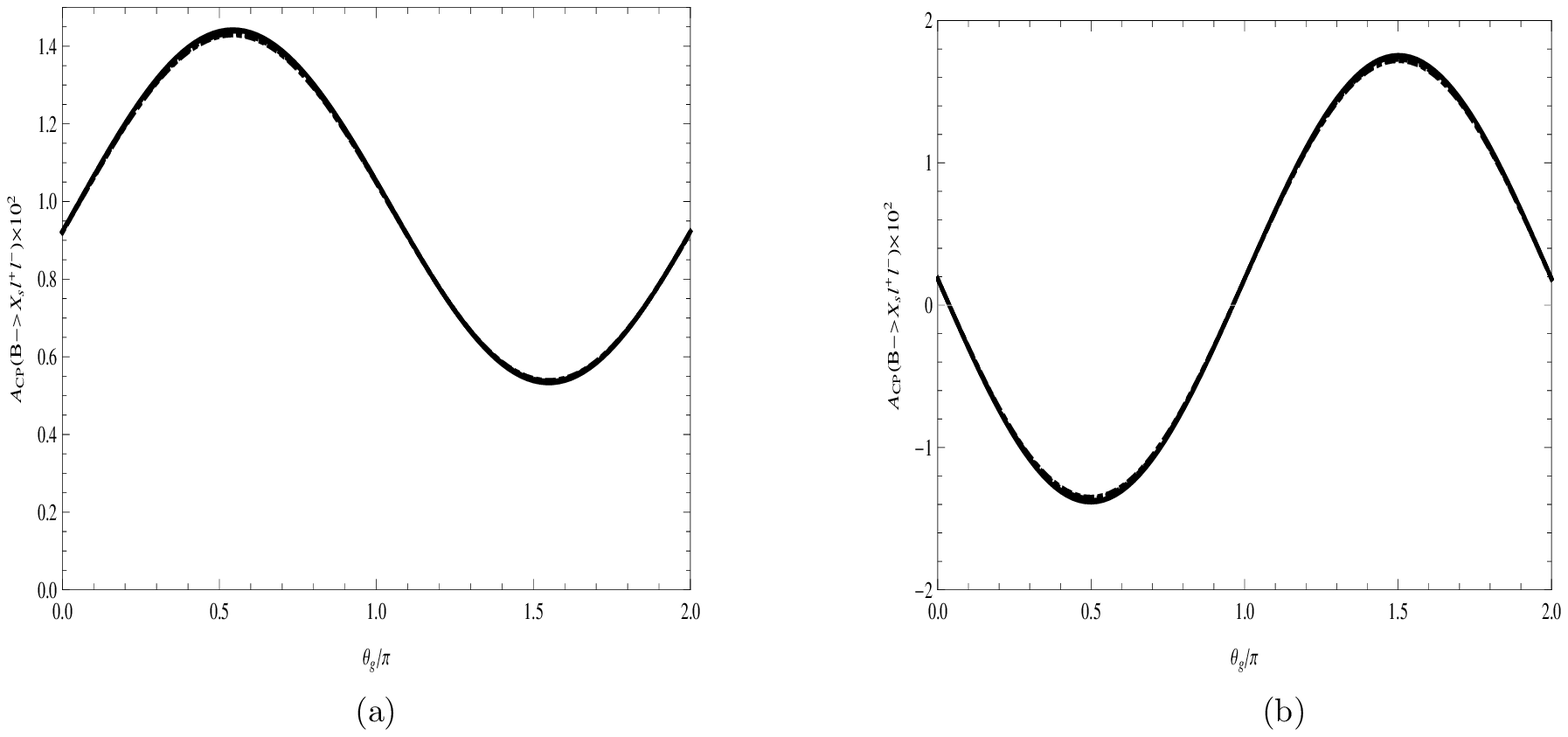}
\vspace{-11cm}
\caption[]{Taking $(\delta_{D}^{LL})_{23}=(\delta_{D}^{RR})_{23}=(\delta_{D}^{LR})_{23}=0.125$,
we plot (a) $A_{_{CP}}(B\rightarrow X_{_s}l^+l^-)_{_{q^2\in[1,6]{\rm GeV}^2}}\times10^2$,
(B) $A_{_{CP}}(B\rightarrow X_{_s}l^+l^-)_{_{q^2\in[14.4,25.0]{\rm GeV}^2}}\times10^2$,
varying with the $CP$ phase $\theta_{_{\tilde g}}$.
Where the solid lines denote $\tan\beta=5$,
dashed lines denote $\tan\beta=10$, dotted lines denote $\tan\beta=30$,
dashed-dotted lines denote $\tan\beta=50$, respectively.}
\label{fig3}
\end{figure}
%%%%%%%%%%%%%%%%%%%%%%%%%%%%%%%%%%%%%%%%%%%%%%%%%%%%%

Certainly the experimental data of $125$ GeV constrain the parameter space
of supersymmetric extension of the SM strongly. The radiative corrections to
mass of the lightest Higgs subtly depend on the parameters $\tan\beta$,
$\mu$, the squark masses of the third generation and relevant trilinear
couplings $A_{_t},\;A_{_b}$ in soft terms.
Furthermore the observed average on branching ratio of $B_{_s}^0\rightarrow\mu^+\mu^-$
is\cite{PDG}
\begin{eqnarray}
%%%%%%%%%%%%%%%%%%%%%%%%%%%%%%%%%%%%%%%%%%%%%%%%%%%
&&BR(B\rightarrow \mu^+\mu^-)^{exp}
\simeq\Big(3.1\pm0.7\Big)\times10^{-9}\;,
%%%%%%%%%%%%%%%%%%%%%%%%%%%%%%%%%%%%%%%%%%%%%%%%%%%
\label{BrBstomumu}
\end{eqnarray}
which coincides with the SM evaluation:
\begin{eqnarray}
%%%%%%%%%%%%%%%%%%%%%%%%%%%%%%%%%%%%%%%%%%%%%%%%%%%
&&BR(B\rightarrow \mu^+\mu^-)^{SM}
\simeq\Big(3.23\pm0.27\Big)\times10^{-9}\;.
%%%%%%%%%%%%%%%%%%%%%%%%%%%%%%%%%%%%%%%%%%%%%%%%%%%
\label{SMBrBstomumu}
\end{eqnarray}
The average experimental data on the branching ratio of the inclusive
$\bar{B}\rightarrow X_{_s}\gamma$ reads
\begin{eqnarray}
%%%%%%%%%%%%%%%%%%%%%%%%%%%%%%%%%%%%%%%%%%%%%%%%%%%
&&BR(\bar{B}\rightarrow X_{_s}\gamma)^{exp}
\simeq\Big(3.40\pm0.21\Big)\times10^{-4}\;,
%%%%%%%%%%%%%%%%%%%%%%%%%%%%%%%%%%%%%%%%%%%%%%%%%%%
\label{BrBstoXsgamma}
\end{eqnarray}
and the corresponding SM prediction at NNLO order is
\begin{eqnarray}
%%%%%%%%%%%%%%%%%%%%%%%%%%%%%%%%%%%%%%%%%%%%%%%%%%%
&&BR(\bar{B}\rightarrow X_{_s}\gamma)^{SM}
\simeq\Big(3.36\pm0.23\Big)\times10^{-4}\;.
%%%%%%%%%%%%%%%%%%%%%%%%%%%%%%%%%%%%%%%%%%%%%%%%%%%
\label{SMBrBstoXsgamma}
\end{eqnarray}
The experimental data from $\bar{B}\rightarrow X_{_s}\gamma$
and $B_{_s}^0\rightarrow\mu^+\mu^-$ also constrain on correlations between
the flavor-changing parameters and energy scale of new physics strongly.

%%%%%%%%%%%%%%%%%%%%%%%%%%%%%%%%%%%%%%%%%%%%%%%%%%%%%
\begin{figure}[h]
\setlength{\unitlength}{1cm}
\centering
\vspace{0.0cm}\hspace{-1.5cm}
\includegraphics[height=16cm,width=18.0cm]{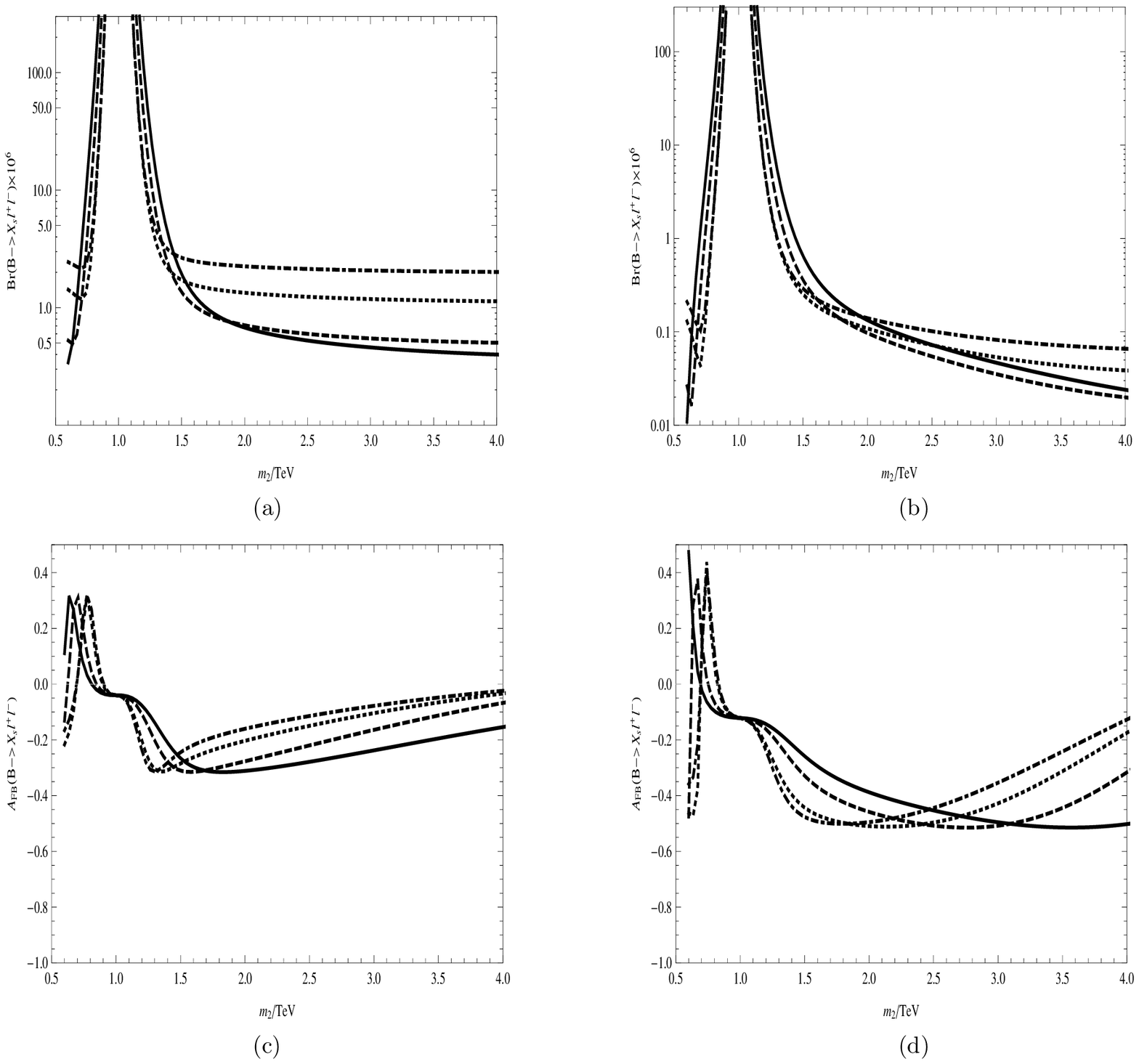}
\vspace{-7cm}
\caption[]{Taking $(\delta_{U}^{LL})_{23}=(\delta_{U}^{RR})_{23}=(\delta_{U}^{LR})_{23}=0.125$,
we plot (a) $BR(B\rightarrow X_{_s}l^+l^-)_{_{q^2\in[1,6]{\rm GeV}^2}}\times10^6$,
(b) $BR(B\rightarrow X_{_s}l^+l^-)_{_{q^2\in[14.4,25.0]{\rm GeV}^2}}\times10^6$
(c) $A_{_{FB}}(B\rightarrow X_{_s}l^+l^-)_{_{q^2\in[1,6]{\rm GeV}^2}}$,
(d) $A_{_{FB}}(B\rightarrow X_{_s}l^+l^-)_{_{q^2\in[14.4,25.0]{\rm GeV}^2}}$,
varying with the $SU(2)$ gaugino mass $|m_{_2}|$.
Where the solid lines denote $\tan\beta=5$,
dashed lines denote $\tan\beta=10$, dotted lines denote $\tan\beta=30$,
dashed-dotted lines denote $\tan\beta=50$, respectively.}
\label{fig4}
\end{figure}
%%%%%%%%%%%%%%%%%%%%%%%%%%%%%%%%%%%%%%%%%%%%%%%%%%%%%

To obtain mass of the lightest Higgs in reasonable range, we further
choose the mass of CP-odd Higgs $m_{_A}=1$ TeV,
and the following assumptions on the parameter space.
\begin{itemize}
\item Taking $\tan\beta=5$, $A_{_t}=1.5$ TeV, $A_{_b}=1$ TeV, $\mu=-500$ GeV,
one gets $m_{_h}\simeq124.6$ GeV correspondingly.
\item Taking $\tan\beta=10$, $A_{_t}=1$ TeV, $A_{_b}=1$ TeV, $\mu=500$ GeV,
one gets $m_{_h}\simeq125.3$ GeV correspondingly.
\item Taking $\tan\beta=30$, $A_{_t}=0.5$ TeV, $A_{_b}=1$ TeV, $\mu=500$ GeV,
one gets $m_{_h}\simeq125.2$ GeV correspondingly.
\item Taking $\tan\beta=50$, $A_{_t}=.5$ TeV, $A_{_b}=0.5$ TeV, $\mu=500$ GeV,
one gets $m_{_h}\simeq125.3$ GeV correspondingly.
\end{itemize}
For the gauge coupling of local $B-L$ symmetry and nonzero VEVs of right-handed
sneutrinos, we take $g_{_{BL}}=0.7$, $\upsilon_{_N}=(0,\;0,\;3)\;{\rm TeV}$ here.
This choice induces the mass of $U(1)_{_{B-L}}$ gauge boson
$m_{_{Z_{BL}}}=2.1\;{\rm TeV}$. Through scanning the parameter space,
we find that the theoretical evaluations depend on the
$U(1)_{_{B-L}}\times U(1)_{_Y}$ gauginos masses $|m_{_1}|$
and $|m_{_{BL}}|$ mildly. In our numerical analysis below, we
choose $m_{_1}=|m_{_{BL}}|=1\;{\rm TeV}$ for simplification.
Furthermore, the parameter $m_{_{1BL}}$ only evokes the mixing between $U(1)_{_{B-L}}$
and $U(1)_{_Y}$ gauginos, and affects our numerical results gently.
Not loss of generality, we also take $m_{_{1BL}}=0$ in our numerical analysis.
Under the assumptions above, the numerical results
are decided by the gaugino masses $|m_{_2}|,\;|m_{_{\tilde g}}|$,
the $CP$ violating phases $\theta_{_{\tilde g}},\;\theta_{_2},
\;\theta_{_{BL}}$, and the corresponding flavor-changing
insertions $(\delta_{U,D}^{LL})_{23},\;(\delta_{U,D}^{RR})_{23},\;
(\delta_{U,D}^{LR})_{23}$.

%%%%%%%%%%%%%%%%%%%%%%%%%%%%%%%%%%%%%%%%%%%%%%%%%%%%%
\begin{figure}[h]
\setlength{\unitlength}{1cm}
\centering
\vspace{0.0cm}%\hspace{-1.5cm}
\includegraphics[height=14.5cm,width=18.0cm]{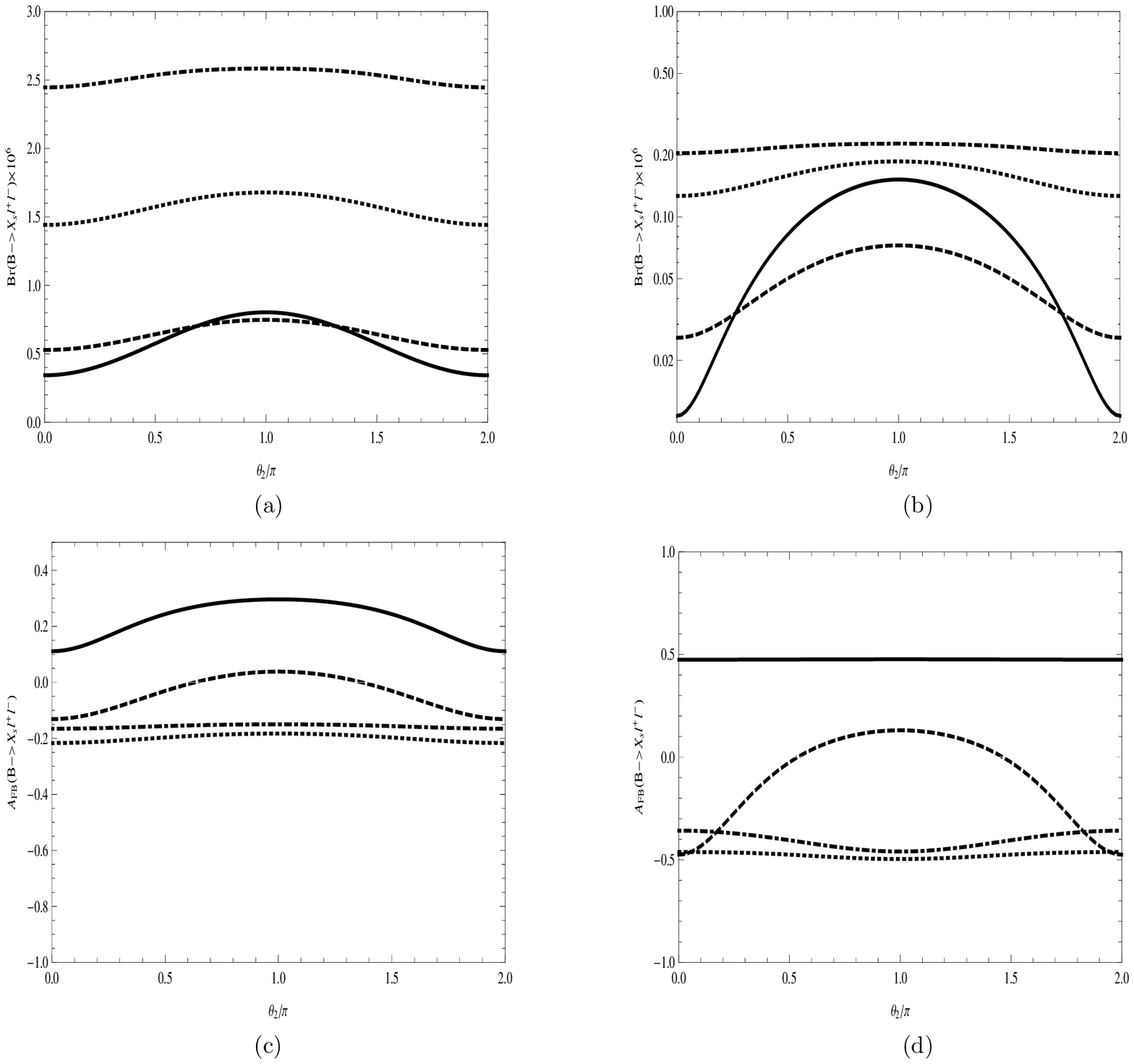}
\vspace{-6cm}
\caption[]{Taking $(\delta_{U}^{LL})_{23}=(\delta_{U}^{RR})_{23}=(\delta_{U}^{LR})_{23}=0.125$,
we plot (a) $BR(B\rightarrow X_{_s}l^+l^-)_{_{q^2\in[1,6]{\rm GeV}^2}}\times10^6$,
(b) $BR(B\rightarrow X_{_s}l^+l^-)_{_{q^2\in[14.4,25.0]{\rm GeV}^2}}\times10^6$
(c) $A_{_{FB}}(B\rightarrow X_{_s}l^+l^-)_{_{q^2\in[1,6]{\rm GeV}^2}}$,
(d) $A_{_{FB}}(B\rightarrow X_{_s}l^+l^-)_{_{q^2\in[14.4,25.0]{\rm GeV}^2}}$,
varying with the $CP$ phase $\theta_{_2}$.
Where the solid lines denote $\tan\beta=5$,
dashed lines denote $\tan\beta=10$, dotted lines denote $\tan\beta=30$,
dashed-dotted lines denote $\tan\beta=50$, respectively.}
\label{fig5}
\end{figure}
%%%%%%%%%%%%%%%%%%%%%%%%%%%%%%%%%%%%%%%%%%%%%%%%%%%%%

Assuming $|m_{_2}|=600\;$GeV, $\theta_{_{\tilde g}}=\theta_{_2}=\theta_{_{BL}}=0$,
we present $BR(B\rightarrow X_{_s}l^+l^-)_{_{q^2\in[1,6]{\rm GeV}^2}}\times10^6$
versus $|m_{_{\tilde g}}|$ in Fig.\ref{fig1}(a),
$BR(B\rightarrow X_{_s}l^+l^-)_{_{q^2\in[14.4,25.0]{\rm GeV}^2}}\times10^6$
versus $|m_{_{\tilde g}}|$ in Fig.\ref{fig1}(b),
$A_{_{FB}}(B\rightarrow X_{_s}l^+l^-)_{_{q^2\in[1,6]{\rm GeV}^2}}$
versus $|m_{_{\tilde g}}|$ in Fig.\ref{fig1}(c),
and $A_{_{FB}}(B\rightarrow X_{_s}l^+l^-)_{_{q^2\in[14.4,25.0]{\rm GeV}^2}}$
versus $|m_{_{\tilde g}}|$ in Fig.\ref{fig1}(d), respectively.
Since the gluino affects our numerical results through the down-type squark sector,
we choose $(\delta_{D}^{LL})_{23}=(\delta_{D}^{RR})_{23}=
(\delta_{D}^{LR})_{23}=0.125$, $(\delta_{U}^{LL})_{23}=(\delta_{U}^{RR})_{23}=
(\delta_{U}^{LR})_{23}=0$ here. As $|m_{_{\tilde g}}|<1.8$ TeV,
the numerical evaluations of $BR(B\rightarrow X_{_s}l^+l^-)_{_{q^2\in[1,6]{\rm GeV}^2}}$
exceed $2\times10^{-6}$. With increasing of $|m_{_{\tilde g}}|$,
the numerical evaluations of $BR(B\rightarrow X_{_s}l^+l^-)_{_{q^2\in[1,6]{\rm GeV}^2}}$
decrease slowly. Meanwhile the corresponding numerical results
depend on the parameter $\tan\beta$ mildly because the main corrections
originate from the operators ${\cal O}_{_{7,9,10}}$ in low $q^2$ region. Similarly
the numerical evaluations of $BR(B\rightarrow X_{_s}l^+l^-)_{_{q^2\in[14.4,25]{\rm GeV}^2}}$
exceed $0.5\times10^{-6}$ as $|m_{_{\tilde g}}|<2$ TeV.
With increasing of $|m_{_{\tilde g}}|$, the numerical evaluations
of $BR(B\rightarrow X_{_s}l^+l^-)_{_{q^2\in[14.4,25]{\rm GeV}^2}}$
decrease mildly. The corresponding numerical results
depend on the parameter $\tan\beta$ subtly because the main corrections
originate from the operators ${\cal O}_{_{S,P}}$ in high $q^2$ region.
In addition the numerical evaluations on $A_{_{FB}}$ in low $q^2$
region is lying in the range
$-0.32\le A_{_{FB}}(B\rightarrow X_{_s}l^+l^-)_{_{q^2\in[1,6]{\rm GeV}^2}}\le -0.18$,
the numerical evaluations on $A_{_{FB}}$ in high $q^2$
region is lying in the range
$0.28\le A_{_{FB}}(B\rightarrow X_{_s}l^+l^-)_{_{q^2\in[14.4,25.0]{\rm GeV}^2}}\le-0.18$
as $1.6\le |m_{_{\tilde g}}|/{\rm TeV}\le4$, which are all coincide
with the experimental data within three standard deviations.
Because the main corrections originate from the operators
${\cal O}_{_{S,P}}$ in high $q^2$ region, the numerical evaluations of
$A_{_{FB}}(B\rightarrow X_{_s}l^+l^-)_{_{q^2\in[14.4,25.0]{\rm GeV}^2}}$
depend on the parameter $\tan\beta$ subtly.

%%%%%%%%%%%%%%%%%%%%%%%%%%%%%%%%%%%%%%%%%%%%%%%%%%%%%
\begin{figure}[h]
\setlength{\unitlength}{1cm}
\centering
\vspace{0.0cm}\hspace{-1.5cm}
\includegraphics[height=16cm,width=18.0cm]{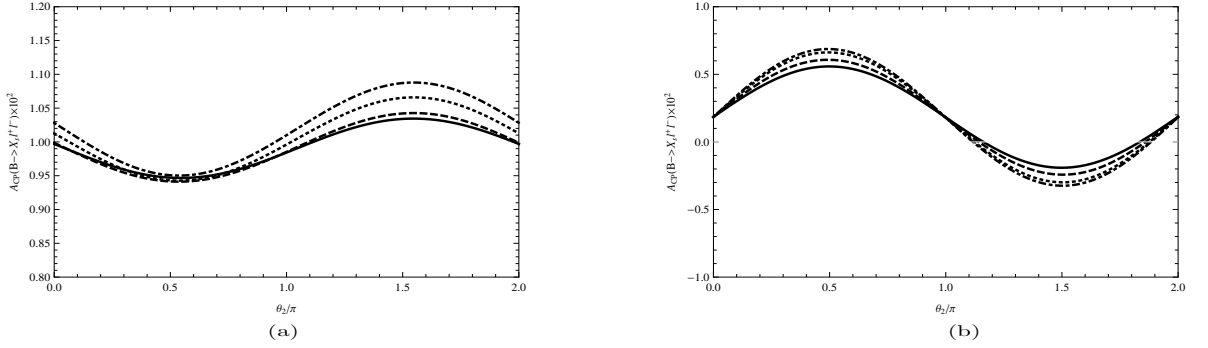}
\vspace{-11cm}
\caption[]{Taking $(\delta_{U}^{LL})_{23}=(\delta_{U}^{RR})_{23}=(\delta_{U}^{LR})_{23}=0.125$,
we plot (a) $A_{_{CP}}(B\rightarrow X_{_s}l^+l^-)_{_{q^2\in[1,6]{\rm GeV}^2}}\times10^2$,
(B) $A_{_{CP}}(B\rightarrow X_{_s}l^+l^-)_{_{q^2\in[14.4,25.0]{\rm GeV}^2}}\times10^2$,
varying with the $CP$ phase $\theta_{_2}$.
Where the solid lines denote $\tan\beta=5$,
dashed lines denote $\tan\beta=10$, dotted lines denote $\tan\beta=30$,
dashed-dotted lines denote $\tan\beta=50$, respectively.}
\label{fig6}
\end{figure}
%%%%%%%%%%%%%%%%%%%%%%%%%%%%%%%%%%%%%%%%%%%%%%%%%%%%%

As the mass of gluino is relatively light, the $CP$ phase of
$m_{_{\tilde g}}$ also affects our final result strongly.
Taking $|m_{_2}|=600\;$GeV, $|m_{_{\tilde g}}|=1.6$ TeV,
and $\theta_{_2}=\theta_{_{BL}}=0$, we plot
$BR(B\rightarrow X_{_s}l^+l^-)_{_{q^2\in[1,6]{\rm GeV}^2}}\times10^6$
varying with $\theta_{_{\tilde g}}$ in Fig.\ref{fig2}(a),
$BR(B\rightarrow X_{_s}l^+l^-)_{_{q^2\in[14.4,25.0]{\rm GeV}^2}}\times10^6$
varying with $\theta_{_{\tilde g}}$ in Fig.\ref{fig2}(b),
$A_{_{FB}}(B\rightarrow X_{_s}l^+l^-)_{_{q^2\in[1,6]{\rm GeV}^2}}$
varying with $\theta_{_{\tilde g}}$ in Fig.\ref{fig2}(c),
and $A_{_{FB}}(B\rightarrow X_{_s}l^+l^-)_{_{q^2\in[14.4,25.0]{\rm GeV}^2}}$
varying with $\theta_{_{\tilde g}}$ in Fig.\ref{fig2}(d), respectively.
In low $q^2$ region, the theoretical prediction on
$BR(B\rightarrow X_{_s}l^+l^-)_{_{q^2\in[1,6]{\rm GeV}^2}}$
increases from $2.5\times10^{-6}$ to $4.5\times10^{-6}$
as the $CP$ phase $\theta_{_{\tilde g}}$ increases from
$0$ to $\pi$, and the forward-backward asymmetry
$A_{_{FB}}(B\rightarrow X_{_s}l^+l^-)_{_{q^2\in[1,6]{\rm GeV}^2}}$
changes from $-0.2$ to $0.3$ when the $CP$ phase $\theta_{_{\tilde g}}$
increases from $0$ to $\pi$, respectively. In high $q^2$ region,
the branching ratio $BR(B\rightarrow X_{_s}l^+l^-)_{_{q^2\in[14.4,25.0]{\rm GeV}^2}}$
varies mildly with increasing of the $CP$ phase $\theta_{_{\tilde g}}$
because the corrections from the operators ${\cal O}_{_{S,P}}$
are important in this region. Nevertheless
$A_{_{FB}}(B\rightarrow X_{_s}l^+l^-)_{_{q^2\in[14.4,25]{\rm GeV}^2}}$
changes from $-0.2$ to $0.18$ when the $CP$ phase $\theta_{_{\tilde g}}$
increases from $0$ to $\pi$.

%%%%%%%%%%%%%%%%%%%%%%%%%%%%%%%%%%%%%%%%%%%%%%%%%%%%%
\begin{figure}[h]
\setlength{\unitlength}{1cm}
\centering
\vspace{0.0cm}\hspace{-1.5cm}
\includegraphics[height=16cm,width=18.0cm]{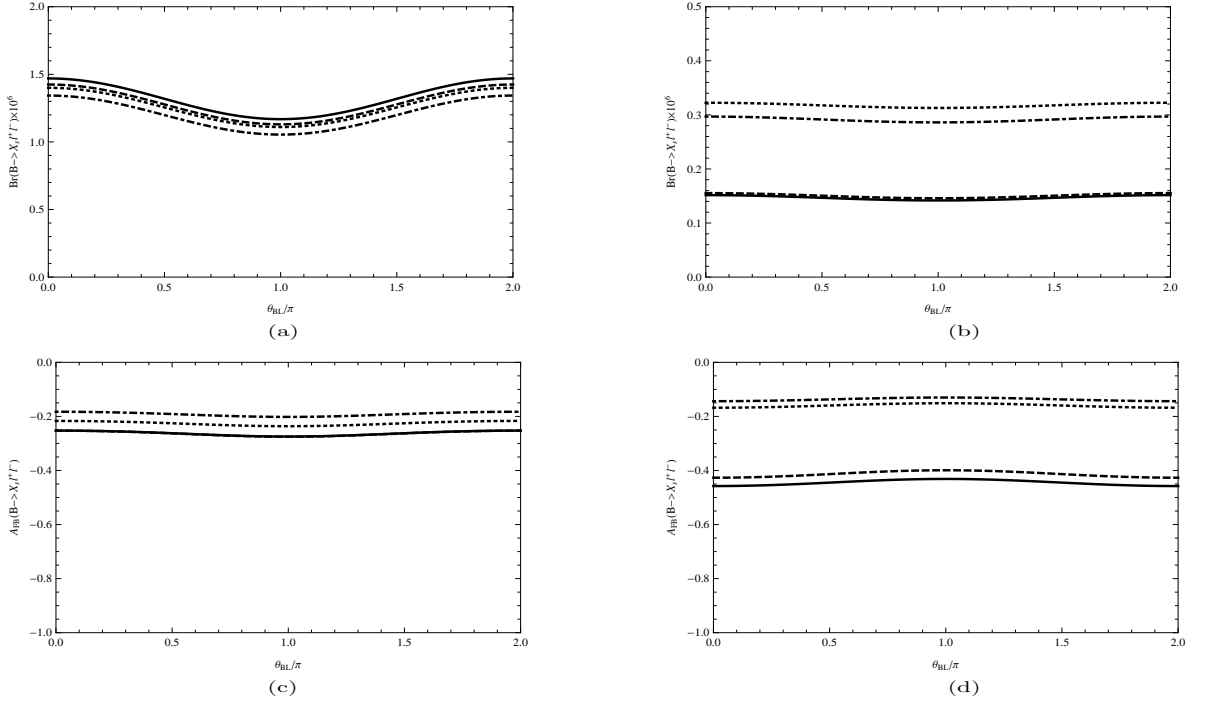}
\vspace{-7cm}
\caption[]{Taking $(\delta_{D}^{LL})_{23}=(\delta_{D}^{RR})_{23}=(\delta_{D}^{LR})_{23}=0.125$,
we plot (a) $BR(B\rightarrow X_{_s}l^+l^-)_{_{q^2\in[1,6]{\rm GeV}^2}}\times10^6$,
(b) $BR(B\rightarrow X_{_s}l^+l^-)_{_{q^2\in[14.4,25.0]{\rm GeV}^2}}\times10^6$
(c) $A_{_{FB}}(B\rightarrow X_{_s}l^+l^-)_{_{q^2\in[1,6]{\rm GeV}^2}}$,
(d) $A_{_{FB}}(B\rightarrow X_{_s}l^+l^-)_{_{q^2\in[14.4,25.0]{\rm GeV}^2}}$,
varying with the $CP$ phase $\theta_{_{\tilde g}}$.
Where the solid lines denote $\tan\beta=5$,
dashed lines denote $\tan\beta=10$, dotted lines denote $\tan\beta=30$,
dashed-dotted lines denote $\tan\beta=50$, respectively.}
\label{fig7}
\end{figure}
%%%%%%%%%%%%%%%%%%%%%%%%%%%%%%%%%%%%%%%%%%%%%%%%%%%%%

Using the inputs presented in Table.(\ref{tab2}), one gets the SM predictions
on the $CP$ asymmetries as
$A_{_{CP}}(B\rightarrow X_{_s}l^+l^-)_{_{q^2\in[1,6]{\rm GeV}^2}}\sim10^{-3}$,
$A_{_{CP}}(B\rightarrow X_{_s}l^+l^-)_{_{q^2\in[14.4,25]{\rm GeV}^2}}<10^{-4}$, respectively.
Taking $(\delta_{D}^{LL})_{23}=(\delta_{D}^{RR})_{23}=(\delta_{D}^{LR})_{23}=0.125$,
we plot $A_{_{CP}}(B\rightarrow X_{_s}l^+l^-)_{_{q^2\in[1,6]{\rm GeV}^2}}\times10^2$
versus $\theta_{_{\tilde g}}$ in Fig.\ref{fig3}(a),
$A_{_{CP}}(B\rightarrow X_{_s}l^+l^-)_{_{q^2\in[14.4,25.0]{\rm GeV}^2}}\times10^2$,
versus the $CP$ phase $\theta_{_{\tilde g}}$  in Fig.\ref{fig3}(b).
Where the solid lines denote $\tan\beta=5$,
dashed lines denote $\tan\beta=10$, dotted lines denote $\tan\beta=30$,
dashed-dotted lines denote $\tan\beta=50$, respectively. The $CP$ asymmetry
$A_{_{CP}}(B\rightarrow X_{_s}l^+l^-)_{_{q^2\in[1,6]{\rm GeV}^2}}$
reaches $1.4\%$ as $\theta_{_{\tilde g}}=\pi/2$, the $CP$ asymmetry
$A_{_{CP}}(B\rightarrow X_{_s}l^+l^-)_{_{q^2\in[14.4,25]{\rm GeV}^2}}$
changes from $-1.2\%$ to $1.8\%$ when the $CP$ phase $\theta_{_{\tilde g}}$
varies from $\pi/2$ to $3\pi/2$. We anticipate that the $CP$ asymmetries
exceeding $0.01$ can be detected in near future.

%%%%%%%%%%%%%%%%%%%%%%%%%%%%%%%%%%%%%%%%%%%%%%%%%%%%%
\begin{figure}[h]
\setlength{\unitlength}{1cm}
\centering
\vspace{0.0cm}\hspace{-1.5cm}
\includegraphics[height=16cm,width=18.0cm]{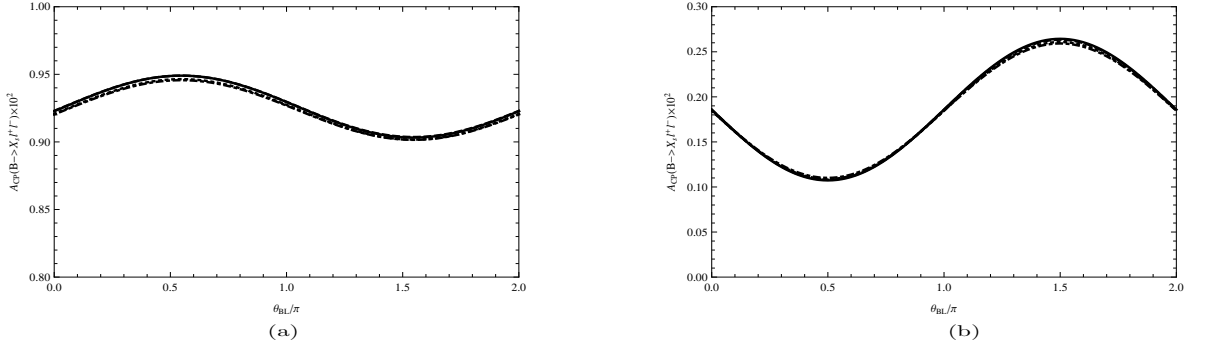}
\vspace{-11cm}
\caption[]{Taking $(\delta_{D}^{LL})_{23}=(\delta_{D}^{RR})_{23}=(\delta_{D}^{LR})_{23}=0.125$,
we plot (a) $A_{_{CP}}(B\rightarrow X_{_s}l^+l^-)_{_{q^2\in[1,6]{\rm GeV}^2}}\times10^2$,
(B) $A_{_{CP}}(B\rightarrow X_{_s}l^+l^-)_{_{q^2\in[14.4,25.0]{\rm GeV}^2}}\times10^2$,
varying with the $CP$ phase $\theta_{_{BL}}$.
Where the solid lines denote $\tan\beta=5$,
dashed lines denote $\tan\beta=10$, dotted lines denote $\tan\beta=30$,
dashed-dotted lines denote $\tan\beta=50$, respectively.}
\label{fig8}
\end{figure}
%%%%%%%%%%%%%%%%%%%%%%%%%%%%%%%%%%%%%%%%%%%%%%%%%%%%%

To investigate the corrections from chargino sector, we choose
$(\delta_{D}^{LL})_{23}=(\delta_{D}^{RR})_{23}=
(\delta_{D}^{LR})_{23}=0$, and $|m_{_{\tilde g}}|=4$ TeV
to suppress the contributions from gluinos and down-type squarks.
So far the experimental data do not exclude relatively light
neutralinos and charginos with several hundred GeV masses yet.
Assuming $\theta_{_{\tilde g}}=\theta_{_2}=\theta_{_{BL}}=0$,
and the insertions $(\delta_{U}^{LL})_{23}=(\delta_{U}^{RR})_{23}=
(\delta_{U}^{LR})_{23}=0.125$,
we present $BR(B\rightarrow X_{_s}l^+l^-)_{_{q^2\in[1,6]{\rm GeV}^2}}\times10^6$
versus the $SU(2)$ gaugino mass $|m_{_2}|$ in Fig.\ref{fig3}(a),
$BR(B\rightarrow X_{_s}l^+l^-)_{_{q^2\in[14.4,25.0]{\rm GeV}^2}}\times10^6$
versus the $SU(2)$ gaugino mass $|m_{_2}|$ in Fig.\ref{fig3}(b),
$A_{_{FB}}(B\rightarrow X_{_s}l^+l^-)_{_{q^2\in[1,6]{\rm GeV}^2}}$
versus the $SU(2)$ gaugino mass $|m_{_2}|$ in Fig.\ref{fig3}(c),
and $A_{_{FB}}(B\rightarrow X_{_s}l^+l^-)_{_{q^2\in[14.4,25.0]{\rm GeV}^2}}$
versus the $SU(2)$ gaugino mass $|m_{_2}|$ in Fig.\ref{fig3}(d), respectively.
As $|m_{_2}|>2$ TeV, $BR(B\rightarrow X_{_s}l^+l^-)_{_{q^2\in[1,6]{\rm GeV}^2}}$
and $BR(B\rightarrow X_{_s}l^+l^-)_{_{q^2\in[14.4,25]{\rm GeV}^2}}$
both coincide with the experimental data in three standard deviations.
The resonance peaks around $1$ TeV originate from
the box diagrams involving sleptons/sneutrinos and neutralinos/charginos under our
assumptions on parameter space. As $|m_{_2}|=1.5$ TeV, the numerical evaluation on
$A_{_{FB}}(B\rightarrow X_{_s}l^+l^-)_{_{q^2\in[1,6]{\rm GeV}^2}}$ is about $-0.4$,
and increases slowly with increasing of $|m_{_2}|$.
In high $q^2$ region the corrections from the operators ${\cal O}_{_{S,P}}$
affect our numerical results heavily.  As $|m_{_2}|=2.5$ TeV, the numerical result indicates
$A_{_{FB}}(B\rightarrow X_{_s}l^+l^-)_{_{q^2\in[14.4,25]{\rm GeV}^2}}\simeq-0.5$.
Along with increasing of $|m_{_2}|$, the numerical evaluations
on $A_{_{FB}}(B\rightarrow X_{_s}l^+l^-)_{_{q^2\in[14.4,25]{\rm GeV}^2}}$
of $\tan\beta=50$ are faster than those of lower of $\tan\beta$.

Under our assumptions on the parameter space, there is a
relatively light chargino with mass in the range $500\;{\rm GeV}
\le m_{_{\chi_1^\pm}}\le1\;{\rm TeV}$. Therefore the $CP$ phase of
$m_{_2}$ also affects our numerical results.
Taking $|m_{_2}|=600\;$GeV, $|m_{_{\tilde g}}|=4$ TeV,
and $\theta_{_{\tilde g}}=\theta_{_{BL}}=0$, we present
$BR(B\rightarrow X_{_s}l^+l^-)_{_{q^2\in[1,6]{\rm GeV}^2}}\times10^6$
versus $\theta_{_2}$ in Fig.\ref{fig5}(a),
$BR(B\rightarrow X_{_s}l^+l^-)_{_{q^2\in[14.4,25.0]{\rm GeV}^2}}\times10^6$
versus $\theta_{_2}$ in Fig.\ref{fig5}(b),
$A_{_{FB}}(B\rightarrow X_{_s}l^+l^-)_{_{q^2\in[1,6]{\rm GeV}^2}}$
versus $\theta_{_2}$ in Fig.\ref{fig5}(c),
and $A_{_{FB}}(B\rightarrow X_{_s}l^+l^-)_{_{q^2\in[14.4,25.0]{\rm GeV}^2}}$
versus $\theta_{_2}$ in Fig.\ref{fig5}(d), respectively.
When $\tan\beta=5$ and $\pi/2\le\theta_{_2}\le3\pi/2$,
we get $0.4\times^{-6}\le BR(B\rightarrow X_{_s}l^+l^-)_{_{q^2\in[1,6]{\rm GeV}^2}}
\le0.8\times10^{-6},\;0.07\times^{-6}\le BR(B\rightarrow X_{_s}l^+l^-)_{_{q^2\in[14.4,25]{\rm GeV}^2}}
\le0.12\times10^{-6}$ respectively,  which coincide with
the experimental data in $3\sigma$ permissions. As $\tan\beta=10$, one
finds the numerical evaluations of $BR(B\rightarrow X_{_s}l^+l^-)_{_{q^2\in[1,6]{\rm GeV}^2}}$
satisfying the experimental constraint within three standard deviations.
However the choice of $\tan\beta=10$ is excluded by experimental observations
because $0.02\times^{-6}\le BR(B\rightarrow X_{_s}l^+l^-)_{_{q^2\in[14.4,25]{\rm GeV}^2}}
\le0.05\times10^{-6}$, which does not satisfy the
experimental data in 3 standard deviations. For large $\tan\beta=30,\;50$,
the corrections from the operators ${\cal O}_{_{S,P}}$ are enhanced drastically.
Correspondingly the theoretical predictions on the branching ratios
in low and high $q^2$ regions all coincide with the experimental data, respectively.
The forward-backward asymmetries both in low and high $q^2$ regions
depend on the $CP$ phase $\theta_{_2}$ smoothly, the absolute values of
corresponding evaluations exceed $0.05$ which can be detected in future.

As mentioned above, the SM predictions on the $CP$ asymmetries are
$A_{_{CP}}(B\rightarrow X_{_s}l^+l^-)_{_{q^2\in[1,6]{\rm GeV}^2}}\sim10^{-3}$,
$A_{_{CP}}(B\rightarrow X_{_s}l^+l^-)_{_{q^2\in[14.4,25]{\rm GeV}^2}}<10^{-4}$,
which are difficult to detected in near future.
Taking $(\delta_{U}^{LL})_{23}=(\delta_{U}^{RR})_{23}=(\delta_{U}^{LR})_{23}=0.125$,
we plot $A_{_{CP}}(B\rightarrow X_{_s}l^+l^-)_{_{q^2\in[1,6]{\rm GeV}^2}}\times10^2$
varying with $\theta_{_2}$ in Fig.\ref{fig6}(a),
$A_{_{CP}}(B\rightarrow X_{_s}l^+l^-)_{_{q^2\in[14.4,25.0]{\rm GeV}^2}}\times10^2$,
varying with the $CP$ phase $\theta_{_2}$ in Fig.\ref{fig6}(b), respectively.
Assuming that the $CP$ violation originates from CKM in the SM, one finds
$A_{_{CP}}(B\rightarrow X_{_s}l^+l^-)_{_{q^2\in[1,6]{\rm GeV}^2}}\sim0.01$,
$A_{_{CP}}(B\rightarrow X_{_s}l^+l^-)_{_{q^2\in[14.4,25]{\rm GeV}^2}}\sim0.002$, respectively.
The theoretical predictions of $A_{_{CP}}(B\rightarrow X_{_s}l^+l^-)_{_{q^2\in[1,6]{\rm GeV}^2}}$
depends on $\theta_{_2}$ gently. Nevertheless, the new $CP$ phase
$\theta_{_2}$ modifies numerical results of
$A_{_{CP}}(B\rightarrow X_{_s}l^+l^-)_{_{q^2\in[14.4,25.0]{\rm GeV}^2}}$ strongly.

When $(\delta_{U}^{LL})_{23}=(\delta_{U}^{RR})_{23}=(\delta_{U}^{LR})_{23}=0$,
and $(\delta_{D}^{LL})_{23}=(\delta_{D}^{RR})_{23}=(\delta_{D}^{LR})_{23}\neq0$,
the $SU(2)$ gaugino mass $m_{_2}$ and $U(1)$ gaugino mass $m_{_1}$
affect our numerical results through the mixing matrix of neutralinos.
The numerical evaluations indicate that those physics quantities depend on
$m_{_2},\;m_{_1}$ mildly in the sectors of parameter space. Similarly the
$U(1)_{_{B-L}}$ gaugino mass $|m_{_{BL}}|$ also affects our results gently.

Under our assumptions on the parameter space, it is interesting to investigate
the effect of $CP$ phase $\theta_{_{BL}}$ on our numerical analyses.
Taking $(\delta_{D}^{LL})_{23}=(\delta_{D}^{RR})_{23}=(\delta_{D}^{LR})_{23}=0.125$,
$(\delta_{U}^{LL})_{23}=(\delta_{U}^{RR})_{23}=(\delta_{U}^{LR})_{23}=0$,
$|m_{_2}|=600\;$GeV, $|m_{_{\tilde g}}|=4$ TeV,
and $\theta_{_{\tilde g}}=\theta_{_2}=0$, we present
$BR(B\rightarrow X_{_s}l^+l^-)_{_{q^2\in[1,6]{\rm GeV}^2}}\times10^6$
versus $\theta_{_{BL}}$ in Fig.\ref{fig7}(a),
$BR(B\rightarrow X_{_s}l^+l^-)_{_{q^2\in[14.4,25.0]{\rm GeV}^2}}\times10^6$
versus $\theta_{_{BL}}$ in Fig.\ref{fig7}(b),
$A_{_{FB}}(B\rightarrow X_{_s}l^+l^-)_{_{q^2\in[1,6]{\rm GeV}^2}}$
versus $\theta_{_2}$ in Fig.\ref{fig7}(c),
and $A_{_{FB}}(B\rightarrow X_{_s}l^+l^-)_{_{q^2\in[14.4,25.0]{\rm GeV}^2}}$
versus $\theta_{_{BL}}$ in Fig.\ref{fig5}(d), respectively.
Our theoretical evaluations depend on the $CP$ phase $\theta_{_{BL}}$
slowly. The numerical results on branching ratios in low $q^2$
and high $q^2$ regions satisfy the experimental data simultaneously
in three standard permissions.
$A_{_{FB}}(B\rightarrow X_{_s}l^+l^-)_{_{q^2\in[1,6]{\rm GeV}^2}}\sim-0.2$
for all $\tan\beta$ chosen. As $\tan\beta=5,\;10$,
$A_{_{FB}}(B\rightarrow X_{_s}l^+l^-)_{_{q^2\in[14.4,25.0]{\rm GeV}^2}}\sim-0.4$,
and $A_{_{FB}}(B\rightarrow X_{_s}l^+l^-)_{_{q^2\in[14.4,25.0]{\rm GeV}^2}}\sim-0.2$
as $\tan\beta=30,\;50$. Additionally
we present $A_{_{CP}}(B\rightarrow X_{_s}l^+l^-)_{_{q^2\in[1,6]{\rm GeV}^2}}\times10^2$
varying with $\theta_{_{BL}}$ in Fig.\ref{fig8}(a),
$A_{_{CP}}(B\rightarrow X_{_s}l^+l^-)_{_{q^2\in[14.4,25.0]{\rm GeV}^2}}\times10^2$,
varying with the $CP$ phase $\theta_{_{BL}}$ in Fig.\ref{fig8}(b), respectively.
Assuming that the $CP$ violation originates from CKM matrix elements, one finds
$A_{_{CP}}(B\rightarrow X_{_s}l^+l^-)_{_{q^2\in[1,6]{\rm GeV}^2}}\sim0.01$,
$A_{_{CP}}(B\rightarrow X_{_s}l^+l^-)_{_{q^2\in[14.4,25]{\rm GeV}^2}}\sim0.002$, respectively.
The theoretical predictions of $A_{_{CP}}(B\rightarrow X_{_s}l^+l^-)_{_{q^2\in[1,6]{\rm GeV}^2}}$
depends on $\theta_{_{BL}}$ gently. Nevertheless, the new $CP$ phase
$\theta_{_{BL}}$ modifies numerical results of
$A_{_{CP}}(B\rightarrow X_{_s}l^+l^-)_{_{q^2\in[14.4,25.0]{\rm GeV}^2}}$ strongly.

In our assumptions on parameter space, the theoretical predictions
on $Br(\bar{B}\rightarrow X_{_s}\gamma)$ and $Br(B_{_s}\rightarrow\mu^+\mu^-)$
all coincide with the experimental observations in three standard permissions.
Obviously our numerical results on branching ratios, forward-backward asymmetries,
and $CP$ asymmetries in $B\rightarrow X_{_s}l^+l^-$ depend on the mass insertions
$(\delta_{U,D}^{LL})_{23},\;(\delta_{U,D}^{RR})_{23},\;(\delta_{U,D}^{LR})_{23}$
and corresponding $CP$ phases subtly. Here we do not present
theoretical evaluations on above quantities versus mass insertions explicitly,
because some similar analyses are given in our previous works\cite{Feng1,Feng2016}.
In addition we take a relatively small coupling of $U(1)_{_{B-L}}$ as
$g_{_{BL}}\le g_{_2}$, this choice avoid the Landau pole of $g_{_{BL}}$ below
the energy scale of grand unified theories.

\section{Summary\label{sec6}}
\indent\indent
Rare $B$-meson decays are very sensitive to new physics beyond the SM
since the theoretical evaluations on corresponding physical quantities
are not seriously affected by the uncertainties originating from unperturbative
QCD effects. Considering the constraint from the observed Higgs signal at the LHC,
we study the supersymmetric corrections to the branching ratios
$BR(B\rightarrow X_{_s}l^+l^-),\;(l=e,\;\mu)$
in the MSSM with local $U(1)_{B-L}$ symmetry\cite{Perez1,Perez2,Perez3,Perez4}
with nonuniversal soft breaking terms. After obtaining the Wilson
coefficients at matching scale, we evolve the Wilson coefficients from the SM
down to hadronic scale at NNLL accuracy, and evolve that from new physics
down to hadronic scale at LL accuracy, respectively.
The lightest neutral Higgs with mass around 125 GeV constrains
the correlation between $\tan\beta$ and the soft Yukawa coupling
$A_{_t},\;A_{_b}$ strongly, nevertheless constrains neutral flavor changing
mass insertions weakly.
Under our assumptions on parameters of the considered model,
the numerical analyses indicate that the branching ratios, and forward-backward asymmetries
depend on the gaugino masses $m_{_{\tilde g}},\;m_{_2}$
strongly, new possible $CP$ phases can enhance the $CP$ asymmetries exceed $1\%$,
which can be detected in near future.

\begin{acknowledgments}
\indent\indent
The work has been supported by the National Natural
Science Foundation of China (NNSFC) with Grant No. 11275036, and No. 11535002,
the Open Project Program of State Key Laboratory of Theoretical Physics, Institute of Theoretical Physics, Chinese
Academy of Sciences, China(No.Y5KF131CJ1), the Natural Science Foundation of
Hebei province with Grant No. A2013201277, No. A2016201010, No. A2016201069, and Natural Science Foundation of
Hebei University with Grant No. 2011JQ05, No. 2012-242.
\end{acknowledgments}

\appendix

\section{The mass squared matrices for squarks\label{appA}}
\indent\indent
With the minimal flavor violation assumption, the $2\times2$ mass squared matrix
for scalar tops is given as
\begin{eqnarray}
&&{\cal Z}_{_t}^\dagger\left(\begin{array}{cc}
m_{_{\tilde{t}_L}}^2\;\;&m_{_{\tilde{t}_X}}^2\\ \\
m_{_{\tilde{t}_X}}^2\;\;&m_{_{\tilde{t}_R}}^2
\end{array}\right){\cal Z}_{_t}=diag\Big(m_{_{\tilde{t}_1}}^2,\;m_{_{\tilde{t}_2}}^2\Big)\;,
\label{UP-squark1}
\end{eqnarray}
with
\begin{eqnarray}
&&m_{_{\tilde{t}_L}}^2={(g_1^2+g_2^2)\upsilon_{_{\rm EW}}^2\over24}
\Big(1-2\cos^2\beta\Big)\Big(1-4c_{_{\rm W}}^2\Big)
\nonumber\\
&&\hspace{1.4cm}
+{g_{_{BL}}^2\over6}\Big(\upsilon_{_N}^2-\upsilon_{_{\rm EW}}^2
+\upsilon_{_{\rm SM}}^2\Big)+m_{_t}^2+m_{_{\tilde{Q}_3}}^2
\;,\nonumber\\
&&m_{_{\tilde{t}_R}}^2=-{g_1^2\upsilon_{_{\rm EW}}^2\over6}
\Big(1-2\cos^2\beta\Big)
\nonumber\\
&&\hspace{1.4cm}
-{g_{_{BL}}^2\over6}\Big(\upsilon_{_N}^2-\upsilon_{_{\rm EW}}^2
+\upsilon_{_{\rm SM}}^2\Big)
+m_{_t}^2+m_{_{\tilde{U}_3}}^2\;,
\nonumber\\
&&m_{_{\tilde{t}_X}}^2=-{\upsilon_{_u}\over\sqrt{2}}A_{_t}Y_{_t}
+{\mu\upsilon_{_d}\over\sqrt{2}}Y_{_t}\;.
\label{UP-squark2}
\end{eqnarray}
Here $Y_{_t},\;A_{_t}$ denote Yukawa coupling
and trilinear soft-breaking parameters in top quark sector, respectively.
In a similar way, the mass-squared matrix for scalar bottoms is
\begin{eqnarray}
&&{\cal Z}_{_b}^\dagger\left(\begin{array}{cc}
m_{_{\tilde{b}_L}}^2\;\;&m_{_{\tilde{b}_X}}^2\\ \\
m_{_{\tilde{b}_X}}^2\;\;&m_{_{\tilde{b}_R}}^2
\end{array}\right){\cal Z}_{_b}=diag\Big(m_{_{\tilde{b}_1}}^2,\;m_{_{\tilde{b}_2}}^2\Big)\;,
\label{DOWN-squark1}
\end{eqnarray}
with
\begin{eqnarray}
&&m_{_{\tilde{b}_L}}^2={(g_1^2+g_2^2)\upsilon_{_{\rm EW}}^2\over24}
\Big(1-2\cos^2\beta\Big)\Big(1+2c_{_{\rm W}}^2\Big)
\nonumber\\
&&\hspace{1.4cm}
+{g_{_{BL}}^2\over6}\Big(\upsilon_{_N}^2-\upsilon_{_{\rm EW}}^2
+\upsilon_{_{\rm SM}}^2\Big)
+m_{_b}^2+m_{_{\tilde{Q}_3}}^2\;,
\nonumber\\
&&m_{_{\tilde{b}_R}}^2={g_1^2\upsilon_{_{\rm EW}}^2\over12}
\Big(1-2\cos^2\beta\Big)
\nonumber\\
&&\hspace{1.4cm}
-{g_{_{BL}}^2\over6}\Big(\upsilon_{_N}^2-\upsilon_{_{\rm EW}}^2
+\upsilon_{_{\rm SM}}^2\Big)
+m_{_b}^2+m_{_{\tilde{D}_3}}^2\;,
\nonumber\\
&&m_{_{\tilde{b}_X}}^2={\upsilon_{_d}\over\sqrt{2}}A_{_b}Y_{_b}
-{\mu\upsilon_{_u}\over\sqrt{2}}Y_{_b}\;,
\label{DOWN-squark2}
\end{eqnarray}
here $Y_{_b},\;A_{_b}$ denote Yukawa couplings
and trilinear soft-breaking parameters in $B$ quark sector, respectively.

\section{The Wilson coefficients from $U(1)_{_{B-L}}$ interaction at electroweak scale\label{appB}}
\indent\indent
Adopting mass insertion approximation, we present the corrections to those Wilson coefficients
from $U(1)_{_{B-L}}$ interaction here
\begin{eqnarray}
%%%%%%%%%%%%%%%%%%%%%%%%%%%%%%%%%%%%%%%%%%%%%%%%%%%
&&C_{_{9,{\tilde Z}_{_{BL}}}}^\gamma(\mu_{_{\rm EW}})=
{Q_{_d}\alpha_{_s}(\mu_{_{\rm EW}})\alpha_{_{BL}}s_{_{\rm W}}^2\over324\pi\alpha_{_{\rm EW}}(\mu_{_{\rm EW}})}
{(\delta^2m_{_{\tilde D}}^{LL})_{_{23}}\over\Lambda_{_{NP}}^2V_{_{tb}}V_{_{ts}}^*}
x_{_{\rm W}}T_3(x_{_{{\tilde Z}_{_{BL}}}},x_{_{{\tilde b}_L}},x_{_{{\tilde s}_L}})
\nonumber\\
&&\hspace{2.0cm}
-{Q_{_d}\alpha_{_s}(\mu_{_{\rm EW}})\alpha_{_{BL}}s_{_{\rm W}}^2\over324\pi\alpha_{_{\rm EW}}(\mu_{_{\rm EW}})}
{(\delta^2m_{_{\tilde D}}^{LR})_{_{23}}(\delta^2m_{_{\tilde D}}^{LR})_{_{33}}^*
\over\Lambda_{_{NP}}^4V_{_{tb}}V_{_{ts}}^*}x_{_{\rm W}}
D_3(x_{_{{\tilde Z}_{_{BL}}}},x_{_{{\tilde b}_L}},x_{_{{\tilde b}_R}},x_{_{{\tilde s}_L}})
\;,\nonumber\\
%%%%%%%%%%%%%%%%%%%%%%%%%%%%%%%%%%%%%%%%%%%%%%%%%%%
&&C_{_{10,{\tilde Z}_{_{BL}}}}^Z(\mu_{_{\rm EW}})=
{\alpha_{_s}(\mu_{_{\rm EW}})\alpha_{_{BL}}\over144\pi\alpha_{_{\rm EW}}(\mu_{_{\rm EW}})}
(1-{2\over3}s_{_{\rm W}}^2)
{(\delta^2m_{_{\tilde D}}^{LL})_{_{23}}\over\Lambda_{_{NP}}^2V_{_{tb}}V_{_{ts}}^*}
\Big[T_{_B}-{\partial\varrho_{_{2,1}}\over\partial x_{_{{\tilde s}_L}}}
-{\partial\varrho_{_{2,1}}\over\partial x_{_{{\tilde b}_L}}}
\Big](x_{_{{\tilde Z}_{_{BL}}}},x_{_{{\tilde b}_L}},x_{_{{\tilde s}_L}})
\nonumber\\
&&\hspace{2.0cm}
-{\alpha_{_s}(\mu_{_{\rm EW}})\alpha_{_{BL}}\over144\pi\alpha_{_{\rm EW}}(\mu_{_{\rm EW}})}
{(\delta^2m_{_{\tilde D}}^{LR})_{_{23}}(\delta^2m_{_{\tilde D}}^{LR})_{_{33}}^*
\over\Lambda_{_{NP}}^4V_{_{tb}}V_{_{ts}}^*}\Big[(1-{2\over3}s_{_{\rm W}}^2)(D_{_B}
-{\partial\varrho_{_{2,1}}\over\partial x_{_{{\tilde s}_L}}}
-{\partial\varrho_{_{2,1}}\over\partial x_{_{{\tilde b}_L}}})
\nonumber\\
&&\hspace{2.0cm}
-{2\over3}s_{_{\rm W}}^2{\partial\varrho_{_{2,1}}\over\partial x_{_{{\tilde b}_L}}}\Big]
(x_{_{{\tilde Z}_{_{BL}}}},x_{_{{\tilde b}_L}},x_{_{{\tilde b}_R}},x_{_{{\tilde s}_L}})
\;,\nonumber\\
%%%%%%%%%%%%%%%%%%%%%%%%%%%%%%%%%%%%%%%%%%%%%%%%%%%
&&C_{_{9,H^\pm}}^{Z_{_{BL}}}(\mu_{_{\rm EW}})=
{\alpha_{_s}(\mu_{_{\rm EW}})\alpha_{_{BL}}\over24\pi
\alpha_{_{\rm EW}}(\mu_{_{\rm EW}})s_{_{\rm W}}^2}{x_{_{\rm W}}\over x_{_{Z_{BL}}}}
\Big\{\Big[-1-2\varrho_{_{1,1}}+2x_{_t}{\partial\varrho_{_{1,1}}\over\partial x_{_t}}
\Big](x_{_t},x_{_{\rm W}})
\nonumber\\
&&\hspace{2.5cm}
+{x_{_t}\over x_{_{\rm W}}s_{_\beta}^2}
\sum\limits_{i=1}^2\Big(Z_{_{H^\pm}}\Big)_{1i}\Big(Z_{_{H^\pm}}\Big)_{1i}^*
\Big[-3+2\varrho_{_{1,1}}-2{\partial\varrho_{_{2,1}}\over\partial x_{_t}}
+2x_{_t}{\partial\varrho_{_{1,1}}\over\partial x_{_t}}\Big](x_{_t},x_{_{H_i^\pm}})
\;,\nonumber\\
%%%%%%%%%%%%%%%%%%%%%%%%%%%%%%%%%%%%%%%%%%%%%%%%%%%
&&C_{_{9,\chi^\pm}}^{Z_{_{BL}}}(\mu_{_{\rm EW}})=
-{\alpha_{_s}(\mu_{_{\rm EW}})\alpha_{_{BL}}\over24\pi\alpha_{_{\rm EW}}(\mu_{_{\rm EW}})}
{(\delta^2m_{_{\tilde U}}^{LL})_{_{23}}\over\Lambda_{_{NP}}^2V_{_{tb}}V_{_{ts}}^*}
(U_+)_{1i}^*(U_+)_{1i}{x_{_{\rm W}}\over x_{_{Z_{BL}}}}
(T_{_B}-2T_{_{BL}})(x_{_{\chi_i^\pm}},x_{_{{\tilde t}_L}},x_{_{{\tilde c}_L}})
\nonumber\\
&&\hspace{2.5cm}
+{\alpha_{_s}(\mu_{_{\rm EW}})\alpha_{_{BL}}m_{_t}\over24\pi\alpha_{_{\rm EW}}(\mu_{_{\rm EW}})m_{_{\rm W}}s_\beta}
{(\delta^2m_{_{\tilde U}}^{LR})_{_{23}}\over\Lambda_{_{NP}}^2V_{_{tb}}V_{_{ts}}^*}
(U_+)_{1i}^*(U_+)_{2i}{x_{_{\rm W}}\over x_{_{Z_{BL}}}}
(T_{_B}-2T_{_{BL}})(x_{_{\chi_i^\pm}},x_{_{{\tilde t}_R}},x_{_{{\tilde c}_L}})
\nonumber\\
&&\hspace{2.5cm}
+{\alpha_{_s}(\mu_{_{\rm EW}})\alpha_{_{BL}}\over24\pi\alpha_{_{\rm EW}}(\mu_{_{\rm EW}})}
{(\delta^2m_{_{\tilde U}}^{LL})_{_{23}}(\delta^2m_{_{\tilde U}}^{LR})_{_{33}}^*
\over\Lambda_{_{NP}}^4V_{_{tb}}V_{_{ts}}^*}(U_+)_{1i}^*(U_+)_{1i}{x_{_{\rm W}}\over x_{_{Z_{BL}}}}
\nonumber\\
&&\hspace{2.5cm}\times
(D_{_B}-2D_{_{BL}})(x_{_{\chi_i^\pm}},x_{_{{\tilde t}_L}},x_{_{{\tilde t}_R}},x_{_{{\tilde c}_L}})
\nonumber\\
&&\hspace{2.5cm}
-{\alpha_{_s}(\mu_{_{\rm EW}})\alpha_{_{BL}}m_{_t}\over24\pi\alpha_{_{\rm EW}}(\mu_{_{\rm EW}})m_{_{\rm W}}s_\beta}
{(\delta^2m_{_{\tilde U}}^{LR})_{_{23}}(\delta^2m_{_{\tilde U}}^{LR})_{_{33}}^*
\over\Lambda_{_{NP}}^4V_{_{tb}}V_{_{ts}}^*}(U_+)_{1i}^*(U_+)_{2i}{x_{_{\rm W}}\over x_{_{Z_{BL}}}}
\nonumber\\
&&\hspace{2.5cm}\times
(D_{_B}-2D_{_{BL}})(x_{_{\chi_i^\pm}},x_{_{{\tilde t}_R}},x_{_{{\tilde t}_L}},x_{_{{\tilde c}_L}})
\;,\nonumber\\
%%%%%%%%%%%%%%%%%%%%%%%%%%%%%%%%%%%%%%%%%%%%%%%%%%%
&&C_{_{9,\chi^0}}^{Z_{_{BL}}}(\mu_{_{\rm EW}})=
-{\alpha_{_s}(\mu_{_{\rm EW}})\alpha_{_{BL}}\over24\pi\alpha_{_{\rm EW}}(\mu_{_{\rm EW}})c_{_{\rm W}}^2}
{(\delta^2m_{_{\tilde D}}^{LL})_{_{23}}\over\Lambda_{_{NP}}^2V_{_{tb}}V_{_{ts}}^*}
\Big|{\cal N}_d^i\Big|^2{x_{_{\rm W}}\over x_{_{Z_{BL}}}}(T_{_B}-{1\over2}T_{_{BL}})
(x_{_{\chi_i^0}},x_{_{{\tilde b}_L}},x_{_{{\tilde s}_L}})
\nonumber\\
&&\hspace{2.5cm}
-{\alpha_{_s}(\mu_{_{\rm EW}})\alpha_{_{BL}}m_{_b}m_{_s}\over24\pi\alpha_{_{\rm EW}}(\mu_{_{\rm EW}})
m_{_{\rm W}}^2c_\beta^2}{(\delta^2m_{_{\tilde D}}^{RR})_{_{23}}\over\Lambda_{_{NP}}^2V_{_{tb}}V_{_{ts}}^*}
\Big|(U_{_N})_{3i}\Big|^2{x_{_{\rm W}}\over x_{_{Z_{BL}}}}
(T_{_B}-{1\over2}T_{_{BL}})(x_{_{\chi_i^0}},x_{_{{\tilde b}_R}},x_{_{{\tilde s}_R}})
\nonumber\\
&&\hspace{2.5cm}
-{\alpha_{_s}(\mu_{_{\rm EW}})\alpha_{_{BL}}m_{_b}\over24\pi\alpha_{_{\rm EW}}(\mu_{_{\rm EW}})
m_{_{\rm W}}c_{_{\rm W}}c_\beta}{(\delta^2m_{_{\tilde D}}^{LR})_{_{23}}
\over\Lambda_{_{NP}}^2V_{_{tb}}V_{_{ts}}^*}(U_{_N})_{3i}{\cal N}_d^{i*}
{x_{_{\rm W}}\over x_{_{Z_{BL}}}}(T_{_B}-{1\over2}T_{_{BL}})(x_{_{\chi_i^0}},x_{_{{\tilde b}_R}},x_{_{{\tilde s}_L}})
\nonumber\\
&&\hspace{2.5cm}
-{\alpha_{_s}(\mu_{_{\rm EW}})\alpha_{_{BL}}m_{_s}\over24\pi\alpha_{_{\rm EW}}(\mu_{_{\rm EW}})
m_{_{\rm W}}c_{_{\rm W}}c_\beta}{(\delta^2m_{_{\tilde D}}^{LR})_{_{23}}
\over\Lambda_{_{NP}}^2V_{_{tb}}V_{_{ts}}^*}(U_{_N})_{3i}^*{\cal N}_d^i
{x_{_{\rm W}}\over x_{_{Z_{BL}}}}
(T_{_B}-{1\over2}T_{_{BL}})(x_{_{\chi_i^0}},x_{_{{\tilde b}_L}},x_{_{{\tilde s}_R}})
\nonumber\\
&&\hspace{2.5cm}
+{\alpha_{_s}(\mu_{_{\rm EW}})\alpha_{_{BL}}\over24\pi\alpha_{_{\rm EW}}(\mu_{_{\rm EW}})c_{_{\rm W}}^2}
{(\delta^2m_{_{\tilde D}}^{LR})_{_{23}}(\delta^2m_{_{\tilde D}}^{LR})_{_{33}}^*
\over\Lambda_{_{NP}}^4V_{_{tb}}V_{_{ts}}^*}\Big|{\cal N}_d^i\Big|^2
{x_{_{\rm W}}\over x_{_{Z_{BL}}}}
\nonumber\\
&&\hspace{2.5cm}\times
(D_{_B}-{1\over2}D_{_{BL}})(x_{_{\chi_i^0}},x_{_{{\tilde b}_L}},x_{_{{\tilde b}_R}},x_{_{{\tilde s}_L}})
\nonumber\\
&&\hspace{2.5cm}
+{\alpha_{_s}(\mu_{_{\rm EW}})\alpha_{_{BL}}m_{_b}m_{_s}\over24\pi\alpha_{_{\rm EW}}(\mu_{_{\rm EW}})
m_{_{\rm W}}^2c_\beta^2}{(\delta^2m_{_{\tilde D}}^{LR})_{_{23}}^*
(\delta^2m_{_{\tilde D}}^{LR})_{_{33}}\over\Lambda_{_{NP}}^4V_{_{tb}}V_{_{ts}}^*}
\Big|(U_{_N})_{3i}\Big|^2{x_{_{\rm W}}\over x_{_{Z_{BL}}}}
\nonumber\\
&&\hspace{2.5cm}\times
(D_{_B}-{1\over2}D_{_{BL}})(x_{_{\chi_i^0}},x_{_{{\tilde b}_R}},x_{_{{\tilde b}_L}},x_{_{{\tilde s}_R}})
\nonumber\\
&&\hspace{2.5cm}
+{\alpha_{_s}(\mu_{_{\rm EW}})\alpha_{_{BL}}m_{_b}\over24\pi\alpha_{_{\rm EW}}(\mu_{_{\rm EW}})
m_{_{\rm W}}c_{_{\rm W}}c_\beta}{(\delta^2m_{_{\tilde D}}^{LL})_{_{23}}
(\delta^2m_{_{\tilde D}}^{LR})_{_{33}}\over\Lambda_{_{NP}}^4V_{_{tb}}V_{_{ts}}^*}(U_{_N})_{3i}
{\cal N}_d^{i*}{x_{_{\rm W}}\over x_{_{Z_{BL}}}}
\nonumber\\
&&\hspace{2.5cm}\times
(D_{_B}-{1\over2}D_{_{BL}})(x_{_{\chi_i^0}},x_{_{{\tilde b}_R}},x_{_{{\tilde b}_L}},x_{_{{\tilde s}_L}})
\nonumber\\
&&\hspace{2.5cm}
+{\alpha_{_s}(\mu_{_{\rm EW}})\alpha_{_{BL}}m_{_s}\over24\pi\alpha_{_{\rm EW}}(\mu_{_{\rm EW}})
m_{_{\rm W}}c_{_{\rm W}}c_\beta}{(\delta^2m_{_{\tilde D}}^{RR})_{_{23}}
(\delta^2m_{_{\tilde D}}^{LR})_{_{33}}^*\over\Lambda_{_{NP}}^4V_{_{tb}}V_{_{ts}}^*}
(U_{_N})_{3i}^*{\cal N}_d^{i}
\nonumber\\
&&\hspace{2.5cm}\times
{x_{_{\rm W}}\over x_{_{Z_{BL}}}}
(D_{_B}-{1\over2}D_{_{BL}})(x_{_{\chi_i^0}},x_{_{{\tilde b}_L}},x_{_{{\tilde b}_R}},x_{_{{\tilde s}_R}})
\;,\nonumber\\
%%%%%%%%%%%%%%%%%%%%%%%%%%%%%%%%%%%%%%%%%%%%%%%%%%%
&&C_{_{9,\tilde{g}}}^{Z_{_{BL}}}(\mu_{_{\rm EW}})=
-{\alpha_{_s}^2(\mu_{_{\rm EW}})\alpha_{_{BL}}s_{_{\rm W}}^2\over9\pi\alpha_{_{\rm EW}}^2(\mu_{_{\rm EW}})}
{(\delta^2m_{_{\tilde D}}^{LL})_{_{23}}\over\Lambda_{_{NP}}^2V_{_{tb}}V_{_{ts}}^*}
{x_{_{\rm W}}\over x_{_{Z_{BL}}}}(T_{_B}-T_{_{BL}})(x_{_{\tilde g}},x_{_{{\tilde b}_L}},x_{_{{\tilde s}_L}})
\nonumber\\
&&\hspace{2.5cm}
+{\alpha_{_s}^2(\mu_{_{\rm EW}})\alpha_{_{BL}}s_{_{\rm W}}^2\over9\pi\alpha_{_{\rm EW}}^2(\mu_{_{\rm EW}})}
{(\delta^2m_{_{\tilde D}}^{LR})_{_{23}}(\delta^2m_{_{\tilde D}}^{LR})_{_{33}}^*
\over\Lambda_{_{NP}}^4V_{_{tb}}V_{_{ts}}^*}{x_{_{\rm W}}\over x_{_{Z_{BL}}}}
(D_{_B}-D_{_{BL}})(x_{_{\tilde g}},x_{_{{\tilde b}_L}},x_{_{{\tilde b}_R}},x_{_{{\tilde s}_L}})
\;,\nonumber\\
%%%%%%%%%%%%%%%%%%%%%%%%%%%%%%%%%%%%%%%%%%%%%%%%%%%
&&C_{_{9,{\tilde Z}_{_{BL}}}}^{Z_{_{BL}}}(\mu_{_{\rm EW}})=
-{\alpha_{_s}(\mu_{_{\rm EW}})\alpha_{_{BL}}^2s_{_{\rm W}}^2\over54\pi\alpha_{_{\rm EW}}^2(\mu_{_{\rm EW}})}
{(\delta^2m_{_{\tilde D}}^{LL})_{_{23}}\over\Lambda_{_{NP}}^2V_{_{tb}}V_{_{ts}}^*}
{x_{_{\rm W}}\over x_{_{Z_{BL}}}}
(T_{_B}-{1\over2}T_{_{BL}})(x_{_{{\tilde Z}_{_{BL}}}},x_{_{{\tilde b}_L}},x_{_{{\tilde s}_L}})
\nonumber\\
&&\hspace{2.5cm}
+{\alpha_{_s}(\mu_{_{\rm EW}})\alpha_{_{BL}}^2s_{_{\rm W}}^2\over54\pi\alpha_{_{\rm EW}}^2(\mu_{_{\rm EW}})}
{(\delta^2m_{_{\tilde D}}^{LR})_{_{23}}(\delta^2m_{_{\tilde D}}^{LR})_{_{33}}^*
\over\Lambda_{_{NP}}^4V_{_{tb}}V_{_{ts}}^*}{x_{_{\rm W}}\over x_{_{Z_{BL}}}}
(D_{_B}-{1\over2}D_{_{BL}})(x_{_{{\tilde Z}_{_{BL}}}},x_{_{{\tilde b}_L}},x_{_{{\tilde b}_R}},x_{_{{\tilde s}_L}})
\;,\nonumber\\
%%%%%%%%%%%%%%%%%%%%%%%%%%%%%%%%%%%%%%%%%%%%%%%%%%%
&&C_{_{S,{\tilde Z}_{_{BL}}}}^{H_k^0}(\mu_{_{\rm EW}})=
-{\alpha_{_{BL}}m_{_\mu}\over18\alpha_{_{\rm EW}}(\mu_{_{\rm EW}})m_{_{\rm W}}^2c_\beta^2}
{(\delta^2m_{_{\tilde D}}^{RR})_{_{23}}\over\Lambda_{_{NP}}^2V_{_{tb}}V_{_{ts}}^*}
(Z_{_H})_{2k}^2{x_{_{\rm W}}\over x_{_{H_k^0}}}
T_{_B}(x_{_{{\tilde Z}_{_{BL}}}},x_{_{{\tilde b}_R}},x_{_{{\tilde s}_R}})
\nonumber\\
&&\hspace{2.5cm}
+{\alpha_{_{BL}}m_{_\mu}\over18\alpha_{_{\rm EW}}(\mu_{_{\rm EW}})m_{_{\rm W}}^2c_\beta^2}
{(\delta^2m_{_{\tilde D}}^{LR})_{_{23}}^*(\delta^2m_{_{\tilde D}}^{LR})_{_{33}}
\over\Lambda_{_{NP}}^4V_{_{tb}}V_{_{ts}}^*}
(Z_{_H})_{2k}^2{x_{_{\rm W}}\over x_{_{H_k^0}}}
D_{_B}(x_{_{{\tilde Z}_{_{BL}}}},x_{_{{\tilde b}_R}},
x_{_{{\tilde b}_L}},x_{_{{\tilde s}_R}})
\nonumber\\
&&\hspace{2.5cm}
+{\alpha_{_{BL}}m_{_\mu}e^{i\theta_{_{BL}}}\over9\alpha_{_{\rm EW}}(\mu_{_{\rm EW}})m_{_{\rm W}}m_{_b}c_{_{\rm W}}^2c_\beta}
{(\delta^2m_{_{\tilde D}}^{LR})_{_{23}}\over\Lambda_{_{NP}}^2V_{_{tb}}V_{_{ts}}^*}
(Z_{_H})_{2k}{(x_{_{\rm W}}^3x_{_{{\tilde Z}_{_{BL}}}})^{1/2}\over x_{_{H_k^0}}}
\nonumber\\
&&\hspace{2.5cm}\times
\Big[(\zeta_{_{LL}}^s)_k{\partial\varrho_{_{1,1}}\over\partial x_{_{{\tilde s}_L}}}
+{2\over3}s_{_{\rm W}}^2(\zeta_{_{RR}}^b)_k
{\partial\varrho_{_{1,1}}\over\partial x_{_{{\tilde b}_R}}}\Big](x_{_{{\tilde Z}_{_{BL}}}},
x_{_{{\tilde b}_R}},x_{_{{\tilde s}_L}})
\nonumber\\
&&\hspace{2.5cm}
+{2\alpha_{_{BL}}m_{_\mu}A_{_{b\mu}}^ke^{i\theta_{_{BL}}}\over
9\alpha_{_{\rm EW}}(\mu_{_{\rm EW}})m_{_{\rm W}}^3c_\beta^2}
{(\delta^2m_{_{\tilde D}}^{LL})_{_{23}}\over\Lambda_{_{NP}}^2V_{_{tb}}V_{_{ts}}^*}
(Z_{_H})_{2k}{(x_{_{\rm W}}^3x_{_{{\tilde Z}_{_{BL}}}})^{1/2}\over x_{_{H_k^0}}}
\nonumber\\
&&\hspace{2.5cm}\times
\varrho_{_{1,1}}(x_{_{{\tilde Z}_{_{BL}}}},
x_{_{{\tilde b}_R}},x_{_{{\tilde b}_L}},x_{_{{\tilde s}_L}})
\nonumber\\
&&\hspace{2.5cm}
-{\alpha_{_{BL}}m_{_\mu}e^{i\theta_{_{BL}}}\over9\alpha_{_{\rm EW}}(\mu_{_{\rm EW}})m_{_{\rm W}}m_{_b}c_{_{\rm W}}^2c_\beta}
{(\delta^2m_{_{\tilde D}}^{LL})_{_{23}}(\delta^2m_{_{\tilde D}}^{LR})_{_{33}}
\over\Lambda_{_{NP}}^4V_{_{tb}}V_{_{ts}}^*}
(Z_{_H})_{2k}{(x_{_{\rm W}}^3x_{_{{\tilde Z}_{_{BL}}}})^{1/2}\over x_{_{H_k^0}}}
\nonumber\\
&&\hspace{2.5cm}\times
\Big[(\zeta_{_{LL}}^s)_k{\partial\varrho_{_{1,1}}\over\partial x_{_{{\tilde s}_L}}}
+(\zeta_{_{LL}}^b)_k{\partial\varrho_{_{1,1}}\over\partial x_{_{{\tilde b}_L}}}
+{2\over3}s_{_{\rm W}}^2(\zeta_{_{RR}}^b)_k
{\partial\varrho_{_{1,1}}\over\partial x_{_{{\tilde b}_R}}}\Big](x_{_{{\tilde Z}_{_{BL}}}},
x_{_{{\tilde b}_R}},x_{_{{\tilde b}_L}},x_{_{{\tilde s}_L}})
\nonumber\\
&&\hspace{2.5cm}
-{2\alpha_{_{BL}}m_{_\mu}e^{i\theta_{_{BL}}}\over
9\alpha_{_{\rm EW}}(\mu_{_{\rm EW}})m_{_{\rm W}}^3c_\beta^2}
{(\delta^2m_{_{\tilde D}}^{LL})_{_{23}}\Re[A_{_{b\mu}}^k(\delta^2m_{_{\tilde D}}^{LR})_{_{33}}]
\over\Lambda_{_{NP}}^4V_{_{tb}}V_{_{ts}}^*}
(Z_{_H})_{2k}{(x_{_{\rm W}}^3x_{_{{\tilde Z}_{_{BL}}}})^{1/2}\over x_{_{H_k^0}}}
\nonumber\\
&&\hspace{2.5cm}\times
{\partial\varrho_{_{1,1}}\over\partial x_{_{{\tilde b}_R}}}(x_{_{{\tilde Z}_{_{BL}}}},
x_{_{{\tilde b}_R}},x_{_{{\tilde b}_L}},x_{_{{\tilde s}_L}})
\;,\nonumber\\
%%%%%%%%%%%%%%%%%%%%%%%%%%%%%%%%%%%%%%%%%%%%%%%%%%%
&&C_{_{P,{\tilde Z}_{_{BL}}}}^{A_k^0}(\mu_{_{\rm EW}})=
{\alpha_{_{BL}}m_{_\mu}\over18\alpha_{_{\rm EW}}(\mu_{_{\rm EW}})m_{_{\rm W}}^2c_\beta^2}
{(\delta^2m_{_{\tilde D}}^{RR})_{_{23}}\over\Lambda_{_{NP}}^2V_{_{tb}}V_{_{ts}}^*}
(Z_{_{H^\pm}})_{2k}^2{x_{_{\rm W}}\over x_{_{A_k^0}}}
T_{_B}(x_{_{{\tilde Z}_{_{BL}}}},x_{_{{\tilde b}_R}},x_{_{{\tilde s}_R}})
\nonumber\\
&&\hspace{2.5cm}
-{\alpha_{_{BL}}m_{_\mu}\over18\alpha_{_{\rm EW}}(\mu_{_{\rm EW}})m_{_{\rm W}}^2c_\beta^2}
{(\delta^2m_{_{\tilde D}}^{LR})_{_{23}}^*(\delta^2m_{_{\tilde D}}^{LR})_{_{33}}
\over\Lambda_{_{NP}}^4V_{_{tb}}V_{_{ts}}^*}
(Z_{_{H^\pm}})_{2k}^2{x_{_{\rm W}}\over x_{_{A_k^0}}}
D_{_B}(x_{_{{\tilde Z}_{_{BL}}}},x_{_{{\tilde b}_R}},
x_{_{{\tilde b}_L}},x_{_{{\tilde s}_R}})
\nonumber\\
&&\hspace{2.5cm}
-{2\alpha_{_{BL}}m_{_\mu}P_{_{b\mu}}^ke^{i\theta_{_{BL}}}\over
9\alpha_{_{\rm EW}}(\mu_{_{\rm EW}})m_{_{\rm W}}^3c_\beta^2}
{(\delta^2m_{_{\tilde D}}^{LL})_{_{23}}\over\Lambda_{_{NP}}^2V_{_{tb}}V_{_{ts}}^*}
(Z_{_{H^\pm}})_{2k}{(x_{_{\rm W}}^3x_{_{{\tilde Z}_{_{BL}}}})^{1/2}\over x_{_{A_k^0}}}
\nonumber\\
&&\hspace{2.5cm}\times
\varrho_{_{1,1}}(x_{_{{\tilde Z}_{_{BL}}}},
x_{_{{\tilde b}_R}},x_{_{{\tilde b}_L}},x_{_{{\tilde s}_L}})
\nonumber\\
&&\hspace{2.5cm}
+{2\alpha_{_{BL}}m_{_\mu}e^{i\theta_{_{BL}}}\over
9\alpha_{_{\rm EW}}(\mu_{_{\rm EW}})m_{_{\rm W}}^3c_\beta^2}
{(\delta^2m_{_{\tilde D}}^{LL})_{_{23}}\Re[P_{_{b\mu}}^k(\delta^2m_{_{\tilde D}}^{LR})_{_{33}}]
\over\Lambda_{_{NP}}^4V_{_{tb}}V_{_{ts}}^*}
(Z_{_{H^\pm}})_{2k}{(x_{_{\rm W}}^3x_{_{{\tilde Z}_{_{BL}}}})^{1/2}\over x_{_{A_k^0}}}
\nonumber\\
&&\hspace{2.5cm}\times
{\partial\varrho_{_{1,1}}\over\partial x_{_{{\tilde b}_R}}}(x_{_{{\tilde Z}_{_{BL}}}},
x_{_{{\tilde b}_R}},x_{_{{\tilde b}_L}},x_{_{{\tilde s}_L}})
\;,\nonumber\\
%%%%%%%%%%%%%%%%%%%%%%%%%%%%%%%%%%%%%%%%%%%%%%%%%%%
%%%%%%%%%%%%%%%%%%%%%%%%%%%%%%%%%%%%%%%%%%%%%%%%%%%
%%%%%%%%%%%%%%%%%%%%%%%%%%%%%%%%%%%%%%%%%%%%%%%%%%%
%%%%%%%%%%%%%%%%%%%%%%%%%%%%%%%%%%%%%%%%%%%%%%%%%%%
&&C_{_{9,{\tilde Z}_{_{BL}}}}^{box}(\mu_{_{\rm EW}})=
-{\alpha_{_s}(\mu_{_{\rm EW}})\alpha_{_{BL}}^2s_{_{\rm W}}^2\over36\pi\alpha_{_{\rm EW}}^2(\mu_{_{\rm EW}})}
{(\delta^2m_{_{\tilde D}}^{LL})_{_{23}}\over\Lambda_{_{NP}}^2V_{_{tb}}V_{_{ts}}^*}x_{_{\rm W}}
\nonumber\\
&&\hspace{2.5cm}\times
\Big[{\partial\varrho_{_{2,1}}\over\partial x_{_{{\tilde Z}_{_{BL}}}}}
-2x_{_{{\tilde Z}_{_{BL}}}}{\partial\varrho_{_{1,1}}\over\partial x_{_{{\tilde Z}_{_{BL}}}}}
\Big](x_{_{{\tilde Z}_{_{BL}}}},x_{_{{\tilde b}_L}},x_{_{{\tilde s}_L}},x_{_{{\tilde e}_L}})
\nonumber\\
&&\hspace{2.5cm}
+{\alpha_{_s}(\mu_{_{\rm EW}})\alpha_{_{BL}}^2s_{_{\rm W}}^2\over36\pi\alpha_{_{\rm EW}}^2(\mu_{_{\rm EW}})}
{(\delta^2m_{_{\tilde D}}^{LR})_{_{23}}(\delta^2m_{_{\tilde D}}^{LR})_{_{33}}^*
\over\Lambda_{_{NP}}^4V_{_{tb}}V_{_{ts}}^*}x_{_{\rm W}}
\nonumber\\
&&\hspace{2.5cm}\times
\Big[{\partial\varrho_{_{2,1}}\over\partial x_{_{{\tilde Z}_{_{BL}}}}}
-2x_{_{{\tilde Z}_{_{BL}}}}{\partial\varrho_{_{1,1}}\over\partial x_{_{{\tilde Z}_{_{BL}}}}}
\Big](x_{_{{\tilde Z}_{_{BL}}}},x_{_{{\tilde b}_L}},x_{_{{\tilde b}_R}},
x_{_{{\tilde s}_L}},x_{_{{\tilde e}_L}})
\;,\nonumber\\
%%%%%%%%%%%%%%%%%%%%%%%%%%%%%%%%%%%%%%%%%%%%%%%%%%%
&&C_{_{9,{\chi^0\tilde Z}_{_{BL}}}}^{box}(\mu_{_{\rm EW}})=
-{\alpha_{_s}(\mu_{_{\rm EW}})\alpha_{_{BL}}\over24\pi\alpha_{_{\rm EW}}(\mu_{_{\rm EW}})c_{_{\rm W}}^2}
{(\delta^2m_{_{\tilde D}}^{LL})_{_{23}}\over\Lambda_{_{NP}}^2V_{_{tb}}V_{_{ts}}^*}x_{_{\rm W}}
\Big[\Re\Big({\cal N}_d^{i*}{\cal N}_l^i\Big)\varrho_{_{2,1}}
\nonumber\\
&&\hspace{2.5cm}
-{4s_{_{\rm W}}\over c_{_{\rm W}}}\Re\Big(e^{i\theta_{_{BL}}}(U_{_N})_{1i}{\cal N}_d^i\Big)
(x_{_{\chi_i^0}}x_{_{{\tilde Z}_{_{BL}}}})^{1/2}
\varrho_{_{1,1}}\Big](x_{_{\chi_i^0}},x_{_{{\tilde Z}_{_{BL}}}},x_{_{{\tilde b}_L}},
x_{_{{\tilde s}_L}},x_{_{{\tilde e}_L}})
\nonumber\\
&&\hspace{2.5cm}
-{\alpha_{_s}(\mu_{_{\rm EW}})\alpha_{_{BL}}m_{_\mu}m_{_s}\over48\pi
\alpha_{_{\rm EW}}(\mu_{_{\rm EW}})m_{_{\rm W}}^2c_\beta^2}
{(\delta^2m_{_{\tilde D}}^{RR})_{_{23}}\over\Lambda_{_{NP}}^2V_{_{tb}}V_{_{ts}}^*}
(U_{_N})_{3i}^*(U_{_N})_{3i}x_{_{\rm W}}
\nonumber\\
&&\hspace{2.5cm}\times
\varrho_{_{2,1}}(x_{_{\chi_i^0}},x_{_{{\tilde Z}_{_{BL}}}},x_{_{{\tilde b}_R}}
,x_{_{{\tilde s}_R}},x_{_{{\tilde e}_R}})
\nonumber\\
&&\hspace{2.5cm}
-{\alpha_{_s}(\mu_{_{\rm EW}})\alpha_{_{BL}}\over48\pi\alpha_{_{\rm EW}}(\mu_{_{\rm EW}})}
{(\delta^2m_{_{\tilde D}}^{LR})_{_{23}}\over\Lambda_{_{NP}}^2V_{_{tb}}V_{_{ts}}^*}
x_{_{\rm W}}\Big[{m_{_b}\over m_{_{\rm W}}c_{_{\rm W}}c_\beta}{\cal N}_l^{i*}(U_{_N})_{3i}
\varrho_{_{2,1}}
+{m_{_\mu}\over m_{_{\rm W}}c_\beta}{\cal N}_d^{i*}(U_{_N})_{3i}\varrho_{_{2,1}}
\nonumber\\
&&\hspace{2.5cm}
-{4m_{_b}s_{_{\rm W}}e^{i\theta_{_{BL}}}\over m_{_{\rm W}}c_\beta}
(U_{_N})_{1i}(U_{_N})_{3i}(x_{_{\chi_i^0}}x_{_{{\tilde Z}_{_{BL}}}})^{1/2}\varrho_{_{1,1}}\Big]
(x_{_{\chi_i^0}},x_{_{{\tilde Z}_{_{BL}}}},x_{_{{\tilde b}_R}},
x_{_{{\tilde s}_L}},x_{_{{\tilde e}_L}})
\nonumber\\
&&\hspace{2.5cm}
-{\alpha_{_s}(\mu_{_{\rm EW}})\alpha_{_{BL}}\over24\pi\alpha_{_{\rm EW}}(\mu_{_{\rm EW}})
m_{_{\rm W}}c_{_{\rm W}}^2c_\beta}{(\delta^2m_{_{\tilde D}}^{LR})_{_{23}}^*\over
\Lambda_{_{NP}}^2V_{_{tb}}V_{_{ts}}^*}x_{_{\rm W}}
\Big[(U_{_N})_{3i}^*{\cal N}_l^i\varrho_{_{2,1}}
\nonumber\\
&&\hspace{2.5cm}
-4s_{_{\rm W}}e^{i\theta_{_{BL}}}(U_{_N})_{3i}^*(U_{_N})_{1i}^*(x_{_{\chi_i^0}}x_{_{{\tilde Z}_{_{BL}}}})^{1/2}
\varrho_{_{1,1}}\Big](x_{_{\chi_i^0}},x_{_{{\tilde Z}_{_{BL}}}},x_{_{{\tilde b}_L}},
x_{_{{\tilde s}_R}},x_{_{{\tilde e}_R}})
\nonumber\\
&&\hspace{2.5cm}
+{\alpha_{_s}(\mu_{_{\rm EW}})\alpha_{_{BL}}\over24\pi\alpha_{_{\rm EW}}(\mu_{_{\rm EW}})c_{_{\rm W}}^2}
{(\delta^2m_{_{\tilde D}}^{LR})_{_{23}}(\delta^2m_{_{\tilde D}}^{LR})_{_{33}}^*
\over\Lambda_{_{NP}}^4V_{_{tb}}V_{_{ts}}^*}x_{_{\rm W}}\Big[\Re\Big(e^{i\theta_{_{BL}}}{\cal N}_d^{i*}
{\cal N}_l^i\Big)\varrho_{_{2,1}}
\nonumber\\
&&\hspace{2.5cm}
-{4s_{_{\rm W}}\over c_{_{\rm W}}}\Re\Big((U_{_N})_{1i}{\cal N}_d^i\Big)
(x_{_{\chi_i^0}}x_{_{{\tilde Z}_{_{BL}}}})^{1/2}
\varrho_{_{1,1}}\Big](x_{_{\chi_i^0}},x_{_{{\tilde Z}_{_{BL}}}},x_{_{{\tilde b}_L}},
x_{_{{\tilde b}_R}},x_{_{{\tilde s}_L}},x_{_{{\tilde e}_L}})
\nonumber\\
&&\hspace{2.5cm}
+{\alpha_{_s}(\mu_{_{\rm EW}})\alpha_{_{BL}}m_{_\mu}m_{_s}\over48\pi
\alpha_{_{\rm EW}}(\mu_{_{\rm EW}})m_{_{\rm W}}^2c_\beta^2}
{(\delta^2m_{_{\tilde D}}^{LR})_{_{23}}^*(\delta^2m_{_{\tilde D}}^{LR})_{_{33}}
\over\Lambda_{_{NP}}^4V_{_{tb}}V_{_{ts}}^*}
(U_{_N})_{3i}^*(U_{_N})_{3i}x_{_{\rm W}}
\nonumber\\
&&\hspace{2.5cm}\times
\varrho_{_{2,1}}(x_{_{\chi_i^0}},x_{_{{\tilde Z}_{_{BL}}}},x_{_{{\tilde b}_R}},
x_{_{{\tilde b}_L}},x_{_{{\tilde s}_R}},x_{_{{\tilde e}_R}})
\nonumber\\
&&\hspace{2.5cm}
+{\alpha_{_s}(\mu_{_{\rm EW}})\alpha_{_{BL}}\over48\pi\alpha_{_{\rm EW}}(\mu_{_{\rm EW}})}
{(\delta^2m_{_{\tilde D}}^{LL})_{_{23}}(\delta^2m_{_{\tilde D}}^{LR})_{_{33}}
\over\Lambda_{_{NP}}^4V_{_{tb}}V_{_{ts}}^*}x_{_{\rm W}}
\Big[{m_{_b}\over m_{_{\rm W}}c_{_{\rm W}}c_\beta}{\cal N}_l^{i*}(U_{_N})_{3i}\varrho_{_{2,1}}
\nonumber\\
&&\hspace{2.5cm}
+{m_{_\mu}\over m_{_{\rm W}}c_\beta}{\cal N}_d^{i*}(U_{_N})_{3i}\varrho_{_{2,1}}
\nonumber\\
&&\hspace{2.5cm}
-{4m_{_b}s_{_{\rm W}}\over m_{_{\rm W}}c_\beta}e^{i\theta_{_{BL}}}
(U_{_N})_{1i}(U_{_N})_{3i}(x_{_{\chi_i^0}}x_{_{{\tilde Z}_{_{BL}}}})^{1/2}\varrho_{_{1,1}}\Big]
(x_{_{\chi_i^0}},x_{_{{\tilde Z}_{_{BL}}}},x_{_{{\tilde b}_R}},x_{_{{\tilde b}_L}},
x_{_{{\tilde s}_L}},x_{_{{\tilde e}_L}})
\nonumber\\
&&\hspace{2.5cm}
+{\alpha_{_s}(\mu_{_{\rm EW}})\alpha_{_{BL}}\over24\pi\alpha_{_{\rm EW}}(\mu_{_{\rm EW}})
m_{_{\rm W}}c_{_{\rm W}}^2c_\beta}
{(\delta^2m_{_{\tilde D}}^{RR})_{_{23}}(\delta^2m_{_{\tilde D}}^{LR})_{_{33}}^*
\over\Lambda_{_{NP}}^4V_{_{tb}}V_{_{ts}}^*}x_{_{\rm W}}
\Big[(U_{_N})_{3i}^*{\cal N}_l^i\varrho_{_{2,1}}
\nonumber\\
&&\hspace{2.5cm}
-4s_{_{\rm W}}e^{i\theta_{_{BL}}}(U_{_N})_{3i}^*(U_{_N})_{1i}^*(x_{_{\chi_i^0}}x_{_{{\tilde Z}_{_{BL}}}})^{1/2}
\varrho_{_{1,1}}\Big](x_{_{\chi_i^0}},x_{_{{\tilde Z}_{_{BL}}}},x_{_{{\tilde b}_L}},
x_{_{{\tilde b}_R}},x_{_{{\tilde s}_R}},x_{_{{\tilde e}_R}})
\;,\nonumber\\
%%%%%%%%%%%%%%%%%%%%%%%%%%%%%%%%%%%%%%%%%%%%%%%%%%%
&&C_{_{10,{\tilde Z}_{_{BL}}}}^{box}(\mu_{_{\rm EW}})=
{\alpha_{_s}(\mu_{_{\rm EW}})\alpha_{_{BL}}^2s_{_{\rm W}}^2\over36\pi\alpha_{_{\rm EW}}^2(\mu_{_{\rm EW}})}
{(\delta^2m_{_{\tilde D}}^{LL})_{_{23}}\over\Lambda_{_{NP}}^2V_{_{tb}}V_{_{ts}}^*}x_{_{\rm W}}
\nonumber\\
&&\hspace{2.5cm}\times
\Big[{\partial\varrho_{_{2,1}}\over\partial x_{_{{\tilde Z}_{_{BL}}}}}
+2x_{_{{\tilde Z}_{_{BL}}}}{\partial\varrho_{_{1,1}}\over\partial x_{_{{\tilde Z}_{_{BL}}}}}
\Big](x_{_{{\tilde Z}_{_{BL}}}},x_{_{{\tilde b}_L}},x_{_{{\tilde s}_L}},x_{_{{\tilde e}_L}})
\nonumber\\
&&\hspace{2.5cm}
-{\alpha_{_s}(\mu_{_{\rm EW}})\alpha_{_{BL}}^2s_{_{\rm W}}^2\over36\pi\alpha_{_{\rm EW}}^2(\mu_{_{\rm EW}})}
{(\delta^2m_{_{\tilde D}}^{LR})_{_{23}}(\delta^2m_{_{\tilde D}}^{LR})_{_{33}}^*
\over\Lambda_{_{NP}}^4V_{_{tb}}V_{_{ts}}^*}x_{_{\rm W}}
\nonumber\\
&&\hspace{2.5cm}\times
\Big[{\partial\varrho_{_{2,1}}\over\partial x_{_{{\tilde Z}_{_{BL}}}}}
+2x_{_{{\tilde Z}_{_{BL}}}}{\partial\varrho_{_{1,1}}\over\partial x_{_{{\tilde Z}_{_{BL}}}}}
\Big](x_{_{{\tilde Z}_{_{BL}}}},x_{_{{\tilde b}_L}},x_{_{{\tilde b}_R}},
x_{_{{\tilde s}_L}},x_{_{{\tilde e}_L}})
\;,\nonumber\\
%%%%%%%%%%%%%%%%%%%%%%%%%%%%%%%%%%%%%%%%%%%%%%%%%%%
&&C_{_{10,\chi^0{\tilde Z}_{_{BL}}}}^{box}(\mu_{_{\rm EW}})=
{\alpha_{_s}(\mu_{_{\rm EW}})\alpha_{_{BL}}\over24\pi\alpha_{_{\rm EW}}(\mu_{_{\rm EW}})c_{_{\rm W}}^2}
{(\delta^2m_{_{\tilde D}}^{LL})_{_{23}}\over\Lambda_{_{NP}}^2V_{_{tb}}V_{_{ts}}^*}x_{_{\rm W}}
\Big[\Re\Big({\cal N}_d^{i*}{\cal N}_l^i\Big)\varrho_{_{2,1}}
\nonumber\\
&&\hspace{2.5cm}
+{4s_{_{\rm W}}\over c_{_{\rm W}}}\Re\Big(e^{i\theta_{_{BL}}}(U_{_N})_{1i}{\cal N}_d^i\Big)
(x_{_{\chi_i^0}}x_{_{{\tilde Z}_{_{BL}}}})^{1/2}
\varrho_{_{1,1}}\Big](x_{_{\chi_i^0}},x_{_{{\tilde Z}_{_{BL}}}},x_{_{{\tilde b}_L}},
x_{_{{\tilde s}_L}},x_{_{{\tilde e}_L}})
\nonumber\\
&&\hspace{2.5cm}
+{\alpha_{_s}(\mu_{_{\rm EW}})\alpha_{_{BL}}m_{_\mu}m_{_s}\over48\pi
\alpha_{_{\rm EW}}(\mu_{_{\rm EW}})m_{_{\rm W}}^2c_\beta^2}
{(\delta^2m_{_{\tilde D}}^{RR})_{_{23}}\over\Lambda_{_{NP}}^2V_{_{tb}}V_{_{ts}}^*}
(U_{_N})_{3i}^*(U_{_N})_{3i}x_{_{\rm W}}
\nonumber\\
&&\hspace{2.5cm}\times
\varrho_{_{2,1}}(x_{_{\chi_i^0}},x_{_{{\tilde Z}_{_{BL}}}},x_{_{{\tilde b}_R}}
,x_{_{{\tilde s}_R}},x_{_{{\tilde e}_R}})
\nonumber\\
&&\hspace{2.5cm}
+{\alpha_{_s}(\mu_{_{\rm EW}})\alpha_{_{BL}}\over48\pi\alpha_{_{\rm EW}}(\mu_{_{\rm EW}})}
{(\delta^2m_{_{\tilde D}}^{LR})_{_{23}}\over\Lambda_{_{NP}}^2V_{_{tb}}V_{_{ts}}^*}
x_{_{\rm W}}\Big[{m_{_b}\over m_{_{\rm W}}c_{_{\rm W}}c_\beta}{\cal N}_l^{i*}(U_{_N})_{3i}
\varrho_{_{2,1}}
+{m_{_\mu}\over m_{_{\rm W}}c_\beta}{\cal N}_d^{i*}(U_{_N})_{3i}\varrho_{_{2,1}}
\nonumber\\
&&\hspace{2.5cm}
+{4m_{_b}s_{_{\rm W}}\over m_{_{\rm W}}c_\beta}e^{i\theta_{_{BL}}}
(U_{_N})_{1i}(U_{_N})_{3i}(x_{_{\chi_i^0}}x_{_{{\tilde Z}_{_{BL}}}})^{1/2}\varrho_{_{1,1}}\Big]
(x_{_{\chi_i^0}},x_{_{{\tilde Z}_{_{BL}}}},x_{_{{\tilde b}_R}},
x_{_{{\tilde s}_L}},x_{_{{\tilde e}_L}})
\nonumber\\
&&\hspace{2.5cm}
+{\alpha_{_s}(\mu_{_{\rm EW}})\alpha_{_{BL}}\over24\pi\alpha_{_{\rm EW}}(\mu_{_{\rm EW}})
m_{_{\rm W}}c_{_{\rm W}}^2c_\beta}{(\delta^2m_{_{\tilde D}}^{LR})_{_{23}}^*\over
\Lambda_{_{NP}}^2V_{_{tb}}V_{_{ts}}^*}x_{_{\rm W}}\Big[(U_{_N})_{3i}^*{\cal N}_l^i
\varrho_{_{2,1}}
\nonumber\\
&&\hspace{2.5cm}
+4s_{_{\rm W}}e^{i\theta_{_{BL}}}(U_{_N})_{3i}^*(U_{_N})_{1i}^*(x_{_{\chi_i^0}}x_{_{{\tilde Z}_{_{BL}}}})^{1/2}
\varrho_{_{1,1}}\Big](x_{_{\chi_i^0}},x_{_{{\tilde Z}_{_{BL}}}},x_{_{{\tilde b}_L}},
x_{_{{\tilde s}_R}},x_{_{{\tilde e}_R}})
\nonumber\\
&&\hspace{2.5cm}
-{\alpha_{_s}(\mu_{_{\rm EW}})\alpha_{_{BL}}\over24\pi\alpha_{_{\rm EW}}(\mu_{_{\rm EW}})c_{_{\rm W}}^2}
{(\delta^2m_{_{\tilde D}}^{LR})_{_{23}}(\delta^2m_{_{\tilde D}}^{LR})_{_{33}}^*
\over\Lambda_{_{NP}}^4V_{_{tb}}V_{_{ts}}^*}x_{_{\rm W}}\Big[\Re\Big({\cal N}_d^{i*}{\cal N}_l^i\Big)
\varrho_{_{2,1}}
\nonumber\\
&&\hspace{2.5cm}
+{4s_{_{\rm W}}\over c_{_{\rm W}}}\Re\Big(e^{i\theta_{_{BL}}}(U_{_N})_{1i}{\cal N}_d^i\Big)
(x_{_{\chi_i^0}}x_{_{{\tilde Z}_{_{BL}}}})^{1/2}
\varrho_{_{1,1}}\Big](x_{_{\chi_i^0}},x_{_{{\tilde Z}_{_{BL}}}},x_{_{{\tilde b}_L}},
x_{_{{\tilde b}_R}},x_{_{{\tilde s}_L}},x_{_{{\tilde e}_L}})
\nonumber\\
&&\hspace{2.5cm}
-{\alpha_{_s}(\mu_{_{\rm EW}})\alpha_{_{BL}}m_{_\mu}m_{_s}\over48\pi
\alpha_{_{\rm EW}}(\mu_{_{\rm EW}})m_{_{\rm W}}^2c_\beta^2}
{(\delta^2m_{_{\tilde D}}^{LR})_{_{23}}^*(\delta^2m_{_{\tilde D}}^{LR})_{_{33}}
\over\Lambda_{_{NP}}^4V_{_{tb}}V_{_{ts}}^*}
(U_{_N})_{3i}^*(U_{_N})_{3i}x_{_{\rm W}}
\nonumber\\
&&\hspace{2.5cm}\times
\varrho_{_{2,1}}(x_{_{\chi_i^0}},x_{_{{\tilde Z}_{_{BL}}}},x_{_{{\tilde b}_R}},
x_{_{{\tilde b}_L}},x_{_{{\tilde s}_R}},x_{_{{\tilde e}_R}})
\nonumber\\
&&\hspace{2.5cm}
-{\alpha_{_s}(\mu_{_{\rm EW}})\alpha_{_{BL}}\over48\pi\alpha_{_{\rm EW}}(\mu_{_{\rm EW}})}
{(\delta^2m_{_{\tilde D}}^{LL})_{_{23}}(\delta^2m_{_{\tilde D}}^{LR})_{_{33}}
\over\Lambda_{_{NP}}^4V_{_{tb}}V_{_{ts}}^*}x_{_{\rm W}}
\Big[{m_{_b}\over m_{_{\rm W}}c_{_{\rm W}}c_\beta}{\cal N}_l^{i*}(U_{_N})_{3i}\varrho_{_{2,1}}
\nonumber\\
&&\hspace{2.5cm}
+{m_{_\mu}\over m_{_{\rm W}}c_\beta}{\cal N}_d^{i*}(U_{_N})_{3i}\varrho_{_{2,1}}
\nonumber\\
&&\hspace{2.5cm}
+{4m_{_b}s_{_{\rm W}}\over m_{_{\rm W}}c_\beta}e^{i\theta_{_{BL}}}
(U_{_N})_{1i}(U_{_N})_{3i}(x_{_{\chi_i^0}}x_{_{{\tilde Z}_{_{BL}}}})^{1/2}\varrho_{_{1,1}}\Big]
(x_{_{\chi_i^0}},x_{_{{\tilde Z}_{_{BL}}}},x_{_{{\tilde b}_R}},x_{_{{\tilde b}_L}},
x_{_{{\tilde s}_L}},x_{_{{\tilde e}_L}})
\nonumber\\
&&\hspace{2.5cm}
-{\alpha_{_s}(\mu_{_{\rm EW}})\alpha_{_{BL}}\over24\pi\alpha_{_{\rm EW}}(\mu_{_{\rm EW}})
m_{_{\rm W}}c_{_{\rm W}}^2c_\beta}
{(\delta^2m_{_{\tilde D}}^{RR})_{_{23}}(\delta^2m_{_{\tilde D}}^{LR})_{_{33}}^*
\over\Lambda_{_{NP}}^4V_{_{tb}}V_{_{ts}}^*}x_{_{\rm W}}\Big[(U_{_N})_{3i}^*{\cal N}_l^i
\varrho_{_{2,1}}
\nonumber\\
&&\hspace{2.5cm}
+4s_{_{\rm W}}e^{i\theta_{_{BL}}}(U_{_N})_{3i}^*(U_{_N})_{1i}^*(x_{_{\chi_i^0}}x_{_{{\tilde Z}_{_{BL}}}})^{1/2}
\varrho_{_{1,1}}\Big](x_{_{\chi_i^0}},x_{_{{\tilde Z}_{_{BL}}}},x_{_{{\tilde b}_L}},
x_{_{{\tilde b}_R}},x_{_{{\tilde s}_R}},x_{_{{\tilde e}_R}})
\;,\nonumber\\
%%%%%%%%%%%%%%%%%%%%%%%%%%%%%%%%%%%%%%%%%%%%%%%%%%%
&&C_{_{S,\chi^0{\tilde Z}_{_{BL}}}}^{box}(\mu_{_{\rm EW}})=
-{\alpha_{_{BL}}m_{_\mu}\over6\alpha_{_{\rm EW}}(\mu_{_{\rm EW}})m_{_{\rm W}}^2c_\beta^2}
{(\delta^2m_{_{\tilde D}}^{LL})_{_{23}}\over\Lambda_{_{NP}}^2V_{_{tb}}V_{_{ts}}^*}
x_{_{\rm W}}\Big[(U_{_N})_{3i}(U_{_N})_{3i}^*\varrho_{_{2,1}}
\nonumber\\
&&\hspace{2.5cm}
-{e^{i\theta_{_{BL}}}\over c_{_{\rm W}}}(U_{_N})_{3i}^*(U_{_N})_{3i}^*(x_{_{\chi_i^0}}x_{_{{\tilde Z}_{_{BL}}}})^{1/2}
\varrho_{_{1,1}}\Big](x_{_{\chi_i^0}},x_{_{{\tilde Z}_{_{BL}}}},x_{_{{\tilde b}_L}},
x_{_{{\tilde s}_L}},x_{_{{\tilde e}_L}})
\nonumber\\
&&\hspace{2.5cm}
-{\alpha_{_{BL}}m_{_\mu}m_{_s}e^{i\theta_{_{BL}}}\over6\alpha_{_{\rm EW}}(\mu_{_{\rm EW}})m_{_{\rm W}}^2m_{_b}c_\beta^2}
{(\delta^2m_{_{\tilde D}}^{RR})_{_{23}}\over\Lambda_{_{NP}}^2V_{_{tb}}V_{_{ts}}^*}
x_{_{\rm W}}(U_{_N})_{3i}^*(U_{_N})_{3i}^*
(x_{_{\chi_i^0}}x_{_{{\tilde Z}_{_{BL}}}})^{1/2}
\nonumber\\
&&\hspace{2.5cm}\times
\varrho_{_{1,1}}(x_{_{\chi_i^0}},x_{_{{\tilde Z}_{_{BL}}}},x_{_{{\tilde b}_R}}
,x_{_{{\tilde s}_R}},x_{_{{\tilde e}_L}})
\nonumber\\
&&\hspace{2.5cm}
-{\alpha_{_{BL}}m_{_\mu}\over18\alpha_{_{\rm EW}}(\mu_{_{\rm EW}})m_{_{\rm W}}m_{_b}c_{_{\rm W}}c_\beta}
{(\delta^2m_{_{\tilde D}}^{LR})_{_{23}}\over\Lambda_{_{NP}}^2V_{_{tb}}V_{_{ts}}^*}
x_{_{\rm W}}\Big[2s_{_{\rm W}}(U_{_N})_{3i}(U_{_N})_{1i}^*\varrho_{_{2,1}}
\nonumber\\
&&\hspace{2.5cm}
+3c_{_{\rm W}}e^{i\theta_{_{BL}}}{\cal N}_d^{i*}(U_{_N})_{3i}^*
(x_{_{\chi_i^0}}x_{_{{\tilde Z}_{_{BL}}}})^{1/2}\varrho_{_{1,1}}
\nonumber\\
&&\hspace{2.5cm}
-2s_{_{\rm W}}e^{i\theta_{_{BL}}}(U_{_N})_{3i}^*(U_{_N})_{1i}^*(x_{_{\chi_i^0}}x_{_{{\tilde Z}_{_{BL}}}})^{1/2}
\varrho_{_{1,1}}\Big](x_{_{\chi_i^0}},x_{_{{\tilde Z}_{_{BL}}}},x_{_{{\tilde b}_R}},
x_{_{{\tilde s}_L}},x_{_{{\tilde e}_R}})
\nonumber\\
&&\hspace{2.5cm}
+{\alpha_{_{BL}}m_{_\mu}\over6\alpha_{_{\rm EW}}(\mu_{_{\rm EW}})m_{_{\rm W}}^2c_\beta^2}
{(\delta^2m_{_{\tilde D}}^{LR})_{_{23}}(\delta^2m_{_{\tilde D}}^{LR})_{_{33}}^*
\over\Lambda_{_{NP}}^4V_{_{tb}}V_{_{ts}}^*}x_{_{\rm W}}\Big[(U_{_N})_{3i}(U_{_N})_{3i}^*\varrho_{_{2,1}}
\nonumber\\
&&\hspace{2.5cm}
-{e^{i\theta_{_{BL}}}\over c_{_{\rm W}}}(U_{_N})_{3i}^*(U_{_N})_{3i}^*(x_{_{\chi_i^0}}x_{_{{\tilde Z}_{_{BL}}}})^{1/2}
\varrho_{_{1,1}}\Big](x_{_{\chi_i^0}},x_{_{{\tilde Z}_{_{BL}}}},x_{_{{\tilde b}_L}},x_{_{{\tilde b}_R}},
x_{_{{\tilde s}_L}},x_{_{{\tilde e}_L}})
\nonumber\\
&&\hspace{2.5cm}
+{\alpha_{_{BL}}m_{_\mu}m_{_s}e^{i\theta_{_{BL}}}\over6\alpha_{_{\rm EW}}(\mu_{_{\rm EW}})m_{_{\rm W}}^2m_{_b}c_\beta^2}
{(\delta^2m_{_{\tilde D}}^{LR})_{_{23}}^*(\delta^2m_{_{\tilde D}}^{LR})_{_{33}}
\over\Lambda_{_{NP}}^4V_{_{tb}}V_{_{ts}}^*}x_{_{\rm W}}(U_{_N})_{3i}^*(U_{_N})_{3i}^*
(x_{_{\chi_i^0}}x_{_{{\tilde Z}_{_{BL}}}})^{1/2}
\nonumber\\
&&\hspace{2.5cm}\times
\varrho_{_{1,1}}(x_{_{\chi_i^0}},x_{_{{\tilde Z}_{_{BL}}}},x_{_{{\tilde b}_R}},
x_{_{{\tilde b}_L}},x_{_{{\tilde s}_R}},x_{_{{\tilde e}_L}})
\nonumber\\
&&\hspace{2.5cm}
+{\alpha_{_{BL}}m_{_\mu}\over18\alpha_{_{\rm EW}}(\mu_{_{\rm EW}})m_{_{\rm W}}m_{_b}c_{_{\rm W}}c_\beta}
{(\delta^2m_{_{\tilde D}}^{LL})_{_{23}}(\delta^2m_{_{\tilde D}}^{LR})_{_{33}}
\over\Lambda_{_{NP}}^4V_{_{tb}}V_{_{ts}}^*}x_{_{\rm W}}
\nonumber\\
&&\hspace{2.5cm}\times
\Big[2s_{_{\rm W}}(U_{_N})_{3i}(U_{_N})_{1i}^*\varrho_{_{2,1}}
+3c_{_{\rm W}}e^{i\theta_{_{BL}}}{\cal N}_d^{i*}(U_{_N})_{3i}^*
(x_{_{\chi_i^0}}x_{_{{\tilde Z}_{_{BL}}}})^{1/2}\varrho_{_{1,1}}
\nonumber\\
&&\hspace{2.5cm}
-2s_{_{\rm W}}e^{i\theta_{_{BL}}}(U_{_N})_{3i}^*(U_{_N})_{1i}^*(x_{_{\chi_i^0}}x_{_{{\tilde Z}_{_{BL}}}})^{1/2}
\varrho_{_{1,1}}\Big](x_{_{\chi_i^0}},x_{_{{\tilde Z}_{_{BL}}}},x_{_{{\tilde b}_R}},
x_{_{{\tilde b}_L}},x_{_{{\tilde s}_L}},x_{_{{\tilde e}_R}})
\;,\nonumber\\
%%%%%%%%%%%%%%%%%%%%%%%%%%%%%%%%%%%%%%%%%%%%%%%%%%%
&&C_{_{P,\chi^0{\tilde Z}_{_{BL}}}}^{box}(\mu_{_{\rm EW}})=
{\alpha_{_{BL}}m_{_\mu}\over6\alpha_{_{\rm EW}}(\mu_{_{\rm EW}})m_{_{\rm W}}^2c_\beta^2}
{(\delta^2m_{_{\tilde D}}^{LL})_{_{23}}\over\Lambda_{_{NP}}^2V_{_{tb}}V_{_{ts}}^*}
x_{_{\rm W}}\Big[(U_{_N})_{3i}(U_{_N})_{3i}^*\varrho_{_{2,1}}
\nonumber\\
&&\hspace{2.5cm}
+{e^{i\theta_{_{BL}}}\over c_{_{\rm W}}}(U_{_N})_{3i}^*(U_{_N})_{3i}^*(x_{_{\chi_i^0}}x_{_{{\tilde Z}_{_{BL}}}})^{1/2}
\varrho_{_{1,1}}\Big](x_{_{\chi_i^0}},x_{_{{\tilde Z}_{_{BL}}}},x_{_{{\tilde b}_L}},
x_{_{{\tilde s}_L}},x_{_{{\tilde e}_L}})
\nonumber\\
&&\hspace{2.5cm}
-{\alpha_{_{BL}}m_{_\mu}m_{_s}e^{i\theta_{_{BL}}}\over6\alpha_{_{\rm EW}}(\mu_{_{\rm EW}})m_{_{\rm W}}^2m_{_b}c_\beta^2}
{(\delta^2m_{_{\tilde D}}^{RR})_{_{23}}\over\Lambda_{_{NP}}^2V_{_{tb}}V_{_{ts}}^*}
(U_{_N})_{3i}^*(U_{_N})_{3i}^*x_{_{\rm W}}(x_{_{\chi_i^0}}x_{_{{\tilde Z}_{_{BL}}}})^{1/2}
\nonumber\\
&&\hspace{2.5cm}\times
\varrho_{_{1,1}}(x_{_{\chi_i^0}},x_{_{{\tilde Z}_{_{BL}}}},x_{_{{\tilde b}_R}}
,x_{_{{\tilde s}_R}},x_{_{{\tilde e}_L}})
\nonumber\\
&&\hspace{2.5cm}
+{\alpha_{_{BL}}m_{_\mu}\over18\alpha_{_{\rm EW}}(\mu_{_{\rm EW}})m_{_{\rm W}}m_{_b}c_{_{\rm W}}c_\beta}
{(\delta^2m_{_{\tilde D}}^{LR})_{_{23}}\over\Lambda_{_{NP}}^2V_{_{tb}}V_{_{ts}}^*}
x_{_{\rm W}}
\nonumber\\
&&\hspace{2.5cm}\times
\Big[2s_{_{\rm W}}(U_{_N})_{3i}(U_{_N})_{1i}^*\varrho_{_{2,1}}
-3c_{_{\rm W}}e^{i\theta_{_{BL}}}{\cal N}_d^{i*}(U_{_N})_{3i}^*
(x_{_{\chi_i^0}}x_{_{{\tilde Z}_{_{BL}}}})^{1/2}\varrho_{_{1,1}}
\nonumber\\
&&\hspace{2.5cm}
+2s_{_{\rm W}}e^{i\theta_{_{BL}}}(U_{_N})_{3i}^*(U_{_N})_{1i}^*(x_{_{\chi_i^0}}x_{_{{\tilde Z}_{_{BL}}}})^{1/2}
\varrho_{_{1,1}}\Big](x_{_{\chi_i^0}},x_{_{{\tilde Z}_{_{BL}}}},x_{_{{\tilde b}_R}},
x_{_{{\tilde s}_L}},x_{_{{\tilde e}_R}})
\nonumber\\
&&\hspace{2.5cm}
-{\alpha_{_{BL}}m_{_\mu}\over6\alpha_{_{\rm EW}}(\mu_{_{\rm EW}})m_{_{\rm W}}^2c_\beta^2}
{(\delta^2m_{_{\tilde D}}^{LR})_{_{23}}(\delta^2m_{_{\tilde D}}^{LR})_{_{33}}^*
\over\Lambda_{_{NP}}^4V_{_{tb}}V_{_{ts}}^*}x_{_{\rm W}}\Big[(U_{_N})_{3i}(U_{_N})_{3i}^*\varrho_{_{2,1}}
\nonumber\\
&&\hspace{2.5cm}
+{e^{i\theta_{_{BL}}}\over c_{_{\rm W}}}(U_{_N})_{3i}^*(U_{_N})_{3i}^*(x_{_{\chi_i^0}}x_{_{{\tilde Z}_{_{BL}}}})^{1/2}
\varrho_{_{1,1}}\Big](x_{_{\chi_i^0}},x_{_{{\tilde Z}_{_{BL}}}},x_{_{{\tilde b}_L}},
x_{_{{\tilde b}_R}},x_{_{{\tilde s}_L}},x_{_{{\tilde e}_L}})
\nonumber\\
&&\hspace{2.5cm}
+{\alpha_{_{BL}}m_{_\mu}m_{_s}e^{i\theta_{_{BL}}}\over6\alpha_{_{\rm EW}}(\mu_{_{\rm EW}})m_{_{\rm W}}^2m_{_b}c_\beta^2}
{(\delta^2m_{_{\tilde D}}^{LR})_{_{23}}^*(\delta^2m_{_{\tilde D}}^{LR})_{_{33}}
\over\Lambda_{_{NP}}^4V_{_{tb}}V_{_{ts}}^*}(U_{_N})_{3i}^*(U_{_N})_{3i}^*x_{_{\rm W}}
(x_{_{\chi_i^0}}x_{_{{\tilde Z}_{_{BL}}}})^{1/2}
\nonumber\\
&&\hspace{2.5cm}\times
\varrho_{_{1,1}}(x_{_{\chi_i^0}},x_{_{{\tilde Z}_{_{BL}}}},x_{_{{\tilde b}_R}}
,x_{_{{\tilde b}_L}},x_{_{{\tilde s}_R}},x_{_{{\tilde e}_L}})
\nonumber\\
&&\hspace{2.5cm}
-{\alpha_{_{BL}}m_{_\mu}\over18\alpha_{_{\rm EW}}(\mu_{_{\rm EW}})m_{_{\rm W}}m_{_b}c_{_{\rm W}}c_\beta}
{(\delta^2m_{_{\tilde D}}^{LL})_{_{23}}(\delta^2m_{_{\tilde D}}^{LR})_{_{33}}
\over\Lambda_{_{NP}}^4V_{_{tb}}V_{_{ts}}^*}x_{_{\rm W}}
\nonumber\\
&&\hspace{2.5cm}\times
\Big[2s_{_{\rm W}}(U_{_N})_{3i}(U_{_N})_{1i}^*\varrho_{_{2,1}}
-3c_{_{\rm W}}e^{i\theta_{_{BL}}}{\cal N}_d^{i*}(U_{_N})_{3i}^*
(x_{_{\chi_i^0}}x_{_{{\tilde Z}_{_{BL}}}})^{1/2}\varrho_{_{1,1}}
\nonumber\\
&&\hspace{2.5cm}
+2s_{_{\rm W}}e^{i\theta_{_{BL}}}(U_{_N})_{3i}^*(U_{_N})_{1i}^*(x_{_{\chi_i^0}}x_{_{{\tilde Z}_{_{BL}}}})^{1/2}
\varrho_{_{1,1}}\Big](x_{_{\chi_i^0}},x_{_{{\tilde Z}_{_{BL}}}},x_{_{{\tilde b}_R}},
x_{_{{\tilde b}_L}},x_{_{{\tilde s}_L}},x_{_{{\tilde e}_R}})\;,
%%%%%%%%%%%%%%%%%%%%%%%%%%%%%%%%%%%%%%%%%%%%%%%%%%%
\label{appB-1}
\end{eqnarray}
where the couplings in the expression are
\begin{eqnarray}
%%%%%%%%%%%%%%%%%%%%%%%%%%%%%%%%%%%%%%%%%%%%%%%%%%%%%%%%%%%%%%%%%%%%%%%%%%%%%
&&{\cal N}_{_d}^i={1\over3}(U_{_N})_{1i}s_{_{\rm W}}-(U_{_N})_{2i}c_{_{\rm W}}
\;,\nonumber\\
%%%%%%%%%%%%%%%%%%%%%%%%%%%%%%%%%%%%%%%%%%%%%%%%%%%%%%%%%%%%%%%%%%%%%%%%%%%%%
&&{\cal N}_{_l}^i=(U_{_N})_{1i}s_{_{\rm W}}+(U_{_N})_{2i}c_{_{\rm W}}
\;,\nonumber\\
%%%%%%%%%%%%%%%%%%%%%%%%%%%%%%%%%%%%%%%%%%%%%%%%%%%%%%%%%%%%%%%%%%%%%%%%%%%%%
&&{\cal C}_{ij}^{L}=2(c_{_{\rm W}}^2-s_{_{\rm W}}^2)\delta_{ij}+(U_-)_{1i}^*(U_-)_{1j}
\;,\nonumber\\
%%%%%%%%%%%%%%%%%%%%%%%%%%%%%%%%%%%%%%%%%%%%%%%%%%%%%%%%%%%%%%%%%%%%%%%%%%%%%
&&{\cal C}_{ij}^{R}=2(c_{_{\rm W}}^2-s_{_{\rm W}}^2)\delta_{ij}+(U_+)_{1i}(U_+)_{1j}^*
\;,\nonumber\\
%%%%%%%%%%%%%%%%%%%%%%%%%%%%%%%%%%%%%%%%%%%%%%%%%%%%%%%%%%%%%%%%%%%%%%%%%%%%%
&&B_{_H}^k=\cos\beta(Z_{_H})_{1k}-\sin\beta(Z_{_H})_{2k}
\;,\nonumber\\
%%%%%%%%%%%%%%%%%%%%%%%%%%%%%%%%%%%%%%%%%%%%%%%%%%%%%%%%%%%%%%%%%%%%%%%%%%%%%
&&(\zeta_{_{LL}}^u)_k=(1-{4\over3}s_{_{\rm W}}^2)B_{_H}^k
+{2m_{_u}^2c_{_{\rm W}}^2\over m_{_{\rm W}}^2s_\beta}(Z_{_H})_{1k}
\;,\nonumber\\
%%%%%%%%%%%%%%%%%%%%%%%%%%%%%%%%%%%%%%%%%%%%%%%%%%%%%%%%%%%%%%%%%%%%%%%%%%%%%
&&(\zeta_{_{RR}}^u)_k=B_{_H}^k
+{3m_{_u}^2c_{_{\rm W}}^2\over2m_{_{\rm W}}^2s_{_{\rm W}}^2s_\beta}(Z_{_H})_{1k}
\;,\nonumber\\
%%%%%%%%%%%%%%%%%%%%%%%%%%%%%%%%%%%%%%%%%%%%%%%%%%%%%%%%%%%%%%%%%%%%%%%%%%%%%
&&(\zeta_{_{LL}}^d)_k=(1-{2\over3}s_{_{\rm W}}^2)B_{_H}^k
-{2m_{_d}^2c_{_{\rm W}}^2\over m_{_{\rm W}}^2c_\beta}(Z_{_H})_{2k}
\;,\nonumber\\
%%%%%%%%%%%%%%%%%%%%%%%%%%%%%%%%%%%%%%%%%%%%%%%%%%%%%%%%%%%%%%%%%%%%%%%%%%%%%
&&(\zeta_{_{RR}}^d)_k=B_{_H}^k
-{3m_{_d}^2c_{_{\rm W}}^2\over m_{_{\rm W}}^2s_{_{\rm W}}^2c_\beta}(Z_{_H})_{2k}
\;,\nonumber\\
%%%%%%%%%%%%%%%%%%%%%%%%%%%%%%%%%%%%%%%%%%%%%%%%%%%%%%%%%%%%%%%%%%%%%%%%%%%%%
&&(\xi_{_k}^{\chi^\pm})_{_{ij}}=
(Z_{_H})_{1k}(U_-)_{1i}(U_+)_{2j}+(Z_{_H})_{2k}(U_-)_{2i}(U_+)_{1j}
\;,\nonumber\\
%%%%%%%%%%%%%%%%%%%%%%%%%%%%%%%%%%%%%%%%%%%%%%%%%%%%%%%%%%%%%%%%%%%%%%%%%%%%%
&&(\xi_{_k}^{\chi^0})_{_{ij}}=
\Big[(Z_{_H})_{_{1k}}(U_{_N})_{3j}-(Z_{_H})_{_{2k}}(U_{_N})_{4j}\Big]
\Big[(U_{_N})_{1i}s_{_{\rm W}}-(U_{_N})_{2i}c_{_{\rm W}}\Big]
\;,\nonumber\\
%%%%%%%%%%%%%%%%%%%%%%%%%%%%%%%%%%%%%%%%%%%%%%%%%%%%%%%%%%%%%%%%%%%%%%%%%%%%%
&&(\eta_{_k}^{\chi^\pm})_{_{ij}}=
(Z_{_{H^\pm}})_{1k}(U_-)_{1i}(U_+)_{2j}+(Z_{_{H^\pm}})_{2k}(U_-)_{2i}(U_+)_{1j}
\;,\nonumber\\
%%%%%%%%%%%%%%%%%%%%%%%%%%%%%%%%%%%%%%%%%%%%%%%%%%%%%%%%%%%%%%%%%%%%%%%%%%%%%
&&(\eta_{_k}^{\chi^0})_{_{ij}}=
\Big[(Z_{_{H^\pm}})_{_{1k}}(U_{_N})_{3j}-(Z_{_{H^\pm}})_{_{2k}}(U_{_N})_{4j}\Big]
\Big[(U_{_N})_{1i}s_{_{\rm W}}-(U_{_N})_{2i}c_{_{\rm W}}\Big]
\;,\nonumber\\
%%%%%%%%%%%%%%%%%%%%%%%%%%%%%%%%%%%%%%%%%%%%%%%%%%%%%%%%%%%%%%%%%%%%%%%%%%%%%
&&A_{_M}^{ki}=(Z_{_H})_{1k}(Z_{_{H^\pm}})_{1i}-(Z_{_H})_{2k}(Z_{_{H^\pm}})_{2i}
\;,\nonumber\\
%%%%%%%%%%%%%%%%%%%%%%%%%%%%%%%%%%%%%%%%%%%%%%%%%%%%%%%%%%%%%%%%%%%%%%%%%%%%%
&&A_{_{u\mu}}^k=A_{_u}(Z_{_H})_{1k}+\mu^*(Z_{_H})_{2k}
\;,\nonumber\\
%%%%%%%%%%%%%%%%%%%%%%%%%%%%%%%%%%%%%%%%%%%%%%%%%%%%%%%%%%%%%%%%%%%%%%%%%%%%%
&&A_{_{d\mu}}^k=A_{_d}(Z_{_H})_{2k}+\mu^*(Z_{_H})_{1k}
\;,\nonumber\\
%%%%%%%%%%%%%%%%%%%%%%%%%%%%%%%%%%%%%%%%%%%%%%%%%%%%%%%%%%%%%%%%%%%%%%%%%%%%%
&&P_{_{u\mu}}^k=A_{_u}(Z_{_{H^\pm}})_{1k}-\mu^*(Z_{_{H^\pm}})_{2k}
\;,\nonumber\\
%%%%%%%%%%%%%%%%%%%%%%%%%%%%%%%%%%%%%%%%%%%%%%%%%%%%%%%%%%%%%%%%%%%%%%%%%%%%%
&&P_{_{d\mu}}^k=A_{_d}(Z_{_{H^\pm}})_{2k}-\mu^*(Z_{_{H^\pm}})_{1k}\;.
%%%%%%%%%%%%%%%%%%%%%%%%%%%%%%%%%%%%%%%%%%%%%%%%%%%%%%%%%%%%%%%%%%%%%%%%%%%%%
\label{appB-2}
\end{eqnarray}
Here $U_{_N},\;U_{_\pm}$ are the mixing matrices of neutralinos and charginos
in the MSSM, respectively. $Z_{_H}$ is the $2\times2$ mixing matrix of CP-even
Higgs, and
\begin{eqnarray}
%%%%%%%%%%%%%%%%%%%%%%%%%%%%%%%%%%%%%%%%%%%%%%%%%%%%%%%%%%%%%%%%%%%%%%%%%%%%%
&&Z_{_{H^\pm}}=\left(\begin{array}{ll}c_{_\beta}&-s_{_\beta}\\
s_{_\beta}&c_{_\beta}\end{array}\right)
%%%%%%%%%%%%%%%%%%%%%%%%%%%%%%%%%%%%%%%%%%%%%%%%%%%%%%%%%%%%%%%%%%%%%%%%%%%%%
\label{appB-3}
\end{eqnarray}
is the mixing matrix between charged Higgs and Goldstone. Furthermore, we adopt
the shorten-cutting notations as $c_{_\beta}=\cos\beta,\;s_{_\beta}=\sin\beta,\;
c_{_{\rm W}}=\cos\theta_{_{\rm W}},\;s_{_{\rm W}}=\sin\theta_{_{\rm W}}$.

\section{The functions\label{appC}}
\indent\indent
The functions in the wilson coefficients of $\gamma-$ and $g-$ penguin operators
are
\begin{eqnarray}
%%%%%%%%%%%%%%%%%%%%%%%%%%%%%%%%%%%%%%%%%%%%%%%%%%%%%%%%%%%%%%%%%%%%%%%%%%%%%
&&T_{_B}(x,y,z)=\Big[2\varrho_{_{1,1}}
-{\partial\varrho_{_{2,1}}\over\partial x}\Big](x,y,z)
\;,\nonumber\\
%%%%%%%%%%%%%%%%%%%%%%%%%%%%%%%%%%%%%%%%%%%%%%%%%%%%%%%%%%%%%%%%%%%%%%%%%%%%%
&&T_{_{BL}}(x,y,z)={\partial\varrho_{_{2,1}}\over\partial y}
+{\partial\varrho_{_{2,1}}\over\partial z}
\;,\nonumber\\
%%%%%%%%%%%%%%%%%%%%%%%%%%%%%%%%%%%%%%%%%%%%%%%%%%%%%%%%%%%%%%%%%%%%%%%%%%%%%
&&T_1(x,y,z)=\Big[{\partial\varrho_{_{1,1}}\over\partial y}
+{\partial\varrho_{_{1,1}}\over\partial z}\Big](x,y,z)
\;,\nonumber\\
%%%%%%%%%%%%%%%%%%%%%%%%%%%%%%%%%%%%%%%%%%%%%%%%%%%%%%%%%%%%%%%%%%%%%%%%%%%%%
&&T_2(x,y,z)=\Big[{\partial^2\varrho_{_{2,1}}\over\partial y^2}
+2{\partial^2\varrho_{_{2,1}}\over\partial y\partial z}
+{\partial^2\varrho_{_{2,1}}\over\partial z^2}\Big](x,y,z)
\;,\nonumber\\
%%%%%%%%%%%%%%%%%%%%%%%%%%%%%%%%%%%%%%%%%%%%%%%%%%%%%%%%%%%%%%%%%%%%%%%%%%%%%
&&T_3(x,y,z)=\Big[{\partial^3\varrho_{_{3,1}}\over\partial y^3}
+3{\partial^3\varrho_{_{3,1}}\over\partial y^2\partial z}
+3{\partial^3\varrho_{_{3,1}}\over\partial y\partial z^2}
+{\partial^3\varrho_{_{3,1}}\over\partial z^3}\Big](x,y,z)
\;,\nonumber\\
%%%%%%%%%%%%%%%%%%%%%%%%%%%%%%%%%%%%%%%%%%%%%%%%%%%%%%%%%%%%%%%%%%%%%%%%%%%%%
\label{appC-1}
\end{eqnarray}
and
\begin{eqnarray}
%%%%%%%%%%%%%%%%%%%%%%%%%%%%%%%%%%%%%%%%%%%%%%%%%%%%%%%%%%%%%%%%%%%%%%%%%%%%%
&&D_{_B}(x,y,z)=\Big[2\varrho_{_{1,1}}
-{\partial\varrho_{_{2,1}}\over\partial x}\Big](x,y,z,u)
\;,\nonumber\\
%%%%%%%%%%%%%%%%%%%%%%%%%%%%%%%%%%%%%%%%%%%%%%%%%%%%%%%%%%%%%%%%%%%%%%%%%%%%%
&&D_{_{BL}}(x,y,z,u)={\partial\varrho_{_{2,1}}\over\partial y}
+{\partial\varrho_{_{2,1}}\over\partial z}+{\partial\varrho_{_{2,1}}\over\partial u}
\;,\nonumber\\
%%%%%%%%%%%%%%%%%%%%%%%%%%%%%%%%%%%%%%%%%%%%%%%%%%%%%%%%%%%%%%%%%%%%%%%%%%%%%
&&D_1(x,y,z,u)=\Big[{\partial\varrho_{_{1,1}}\over\partial y}
+{\partial\varrho_{_{1,1}}\over\partial z}
+{\partial\varrho_{_{1,1}}\over\partial u}\Big](x,y,z,u)
\;,\nonumber\\
%%%%%%%%%%%%%%%%%%%%%%%%%%%%%%%%%%%%%%%%%%%%%%%%%%%%%%%%%%%%%%%%%%%%%%%%%%%%%
&&D_2(x,y,z,u)=\Big[{\partial^2\varrho_{_{2,1}}\over\partial y^2}
+2{\partial^2\varrho_{_{2,1}}\over\partial y\partial z}
+2{\partial^2\varrho_{_{2,1}}\over\partial y\partial u}
+{\partial^2\varrho_{_{2,1}}\over\partial z^2}
+2{\partial^2\varrho_{_{2,1}}\over\partial z\partial u}
+{\partial^2\varrho_{_{2,1}}\over\partial u^2}\Big](x,y,z,u)
\;,\nonumber\\
%%%%%%%%%%%%%%%%%%%%%%%%%%%%%%%%%%%%%%%%%%%%%%%%%%%%%%%%%%%%%%%%%%%%%%%%%%%%%
&&D_3(x,y,z,u)=\Big[{\partial^3\varrho_{_{3,1}}\over\partial y^3}
+3{\partial^3\varrho_{_{3,1}}\over\partial y^2\partial z}
+3{\partial^3\varrho_{_{3,1}}\over\partial y^2\partial u}
+3{\partial^3\varrho_{_{3,1}}\over\partial y\partial z^2}
+6{\partial^3\varrho_{_{3,1}}\over\partial y\partial z\partial u}
+3{\partial^3\varrho_{_{3,1}}\over\partial y\partial u^2}
\nonumber\\
&&\hspace{3.0cm}
+{\partial^3\varrho_{_{3,1}}\over\partial z^3}
+3{\partial^3\varrho_{_{3,1}}\over\partial z^2\partial u}
+3{\partial^3\varrho_{_{3,1}}\over\partial z\partial u^2}
+{\partial^3\varrho_{_{3,1}}\over\partial u^3}\Big](x,y,z,u)
%%%%%%%%%%%%%%%%%%%%%%%%%%%%%%%%%%%%%%%%%%%%%%%%%%%%%%%%%%%%%%%%%%%%%%%%%%%%%
\label{appC-2}
\end{eqnarray}
with
\begin{eqnarray}
%%%%%%%%%%%%%%%%%%%%%%%%%%%%%%%%%%%%%%%%%%%%%%%%%%%%%%%%%%%%%%%%%%%%%%%%%%%%%
&&\varrho_{_{m,n}}(x_{_1},x_{_2},\cdots,x_{_N})=\sum\limits_{i=1}^N{x_{_i}^m\ln^nx_{_i}
\over\prod\limits_{j\neq i}^N(x_{_i}-x_{_j})}\;.
%%%%%%%%%%%%%%%%%%%%%%%%%%%%%%%%%%%%%%%%%%%%%%%%%%%%%%%%%%%%%%%%%%%%%%%%%%%%%
\label{appC-3}
\end{eqnarray}

\end{document}